\documentclass{jpp}
\usepackage{graphicx}
\usepackage{color}

\usepackage[utf8]{inputenc}
\usepackage[T1]{fontenc}
\usepackage{amsmath}

\newcommand{\imag}{{\rm i}}
\newcommand{\rmd}{{\rm d}}

\newcommand{\D}[2]{\frac{\rmd #2}{\rmd #1}}

\newcommand\bb[1]{\mbox{\boldmath{$#1$}}}
\newcommand\grad{\bb{\nabla}}
\renewcommand\bcdot{\,\bb{\cdot}\,}

\newcommand{\bdbldot}{\,\bb{:}\,}
\newcommand\btimes{\,\bb{\times}\,}

\newcommand{\msb}[1]{\mathsfbi{#1}}

\newcommand{\ek}{\hat{\bb{k}}}

\newcommand{\ex}{\hat{\bb{x}}}

\newcommand{\metric}{\bb{\mathsf{\rmLambda}}}
\newcommand{\besselJ}{{\rm J}}
\newcommand{\zeroquote}{{`0\textrm{'}}}

\shorttitle{Thermodynamics and collisionality in firehose-susceptible high-$\beta$ plasmas}
\shortauthor{A.~F.~A.~Bott and others}

\title{Thermodynamics and collisionality in firehose-susceptible high-$\beta$ plasmas}

\author{Archie F.~A.~Bott\aff{1,2}
  \corresp{\email{archie.bott@physics.ox.ac.uk}},
  Matthew W.~Kunz\aff{1,3}, 
 Eliot Quataert\aff{1}, \\
 Jonathan Squire\aff{4},
 \and Lev Arzamasskiy\aff{5}}

\affiliation{\aff{1}Department of Astrophysical Sciences, Princeton University, 4 Ivy Lane, Princeton, NJ 08544, USA;
\aff{2}Department of Physics, University of Oxford, Parks Road, Oxford, OX1 3PU, UK; 
\aff{3}Princeton Plasma Physics Laboratory, PO~Box 451, Princeton, NJ 08543, USA
\aff{4}Department of Physics, University of Otago, 730 Cumberland Street, Dunedin 9016, New Zealand
\aff{5}Institute for Advanced Study, 1 Einstein Drive, Princeton, NJ 08540, USA}

\begin{document}

\maketitle

\begin{abstract}
We study the evolution of collisionless plasmas that, due to their macroscopic evolution, are susceptible to the firehose instability, using both analytic theory and hybrid-kinetic particle-in-cell simulations. We establish that, depending on the relative magnitude of 
 the plasma $\beta$, the characteristic timescale of macroscopic evolution, and the
  ion-Larmor frequency, the saturation of the firehose instability in high-$\beta$ plasmas 
  can result in three qualitatively distinct thermodynamic (and electromagnetic) states. By contrast 
  with the previously identified `ultra-high-beta' and `Alfv\'en-inhibiting' states, the newly identified 
  `Alfv\'en-enabling' state, which is realised when the macroscopic evolution time $\tau$ exceeds the ion-Larmor frequency by a $\beta$-dependent critical parameter, can support linear Alfv\'en waves 
  and Alfv\'enic turbulence because the magnetic tension associated with the plasma's 
  macroscopic magnetic field is never completely negated by anisotropic pressure forces. 
  We characterise these states in detail, including their saturated magnetic-energy spectra. 
  The effective collision operator associated with the firehose fluctuations is also 
  described; we find it to be
  well approximated in the Alfv\'en-enabling state by a simple quasilinear pitch-angle scattering operator. The box-averaged collision frequency is $\nu_{\rm eff} \sim \beta/\tau$, in agreement with previous results, but 
  certain sub-populations of particles scatter at a much larger (or smaller) rate 
  depending on their velocity in the direction parallel to the magnetic field.
  Our findings are essential for understanding low-collisionality astrophysical plasmas including the solar wind, the intracluster medium of galaxy clusters and black-hole accretion flows. We show that all three of these plasmas are in the Alfv\'en-enabling regime of firehose saturation and discuss the implications of this result.
  
\end{abstract}

\section{Introduction}\label{sec:introduction}

Over the last decade, numerous studies
have provided compelling evidence that kinetic instabilities play a key role in 
determining many of the basic physical properties of collisionless (or weakly collisional), magnetised 
plasma. These instabilities, which are driven by gradients in macroscopic properties of the plasma such as bulk fluid velocity or temperature, can amplify `microscopic' electromagnetic 
fluctuations in the plasma exponentially at a rate that is 
generically much greater than the plasma's macroscopic evolution rate. 
The fluctuations are microscopic in the sense that their 
characteristic length scales, which are generically related to the Larmor or inertial scales of the plasma's constituent ions and electrons, are 
much smaller than both the plasma's macroscopic length scales and the Coulomb mean free paths of particles. 
Once these electromagnetic fluctuations attain sufficient amplitudes, feedback mechanisms
are thought to affect various features of the plasma in which they are present. 
These features include the plasma's microphysics [e.g., `anomalous' scattering of
particles at a rate much greater than would na\"{i}vely be 
expected given the plasma's Coulomb collisionality~\citep{Kunz2014_b,Riquelme_2015,Melville_2016,Riquelme_2018}], thermodynamics [e.g.,
regulation of pressure anisotropies~\citep{Hellinger_2008,Camporeale_2010} and heating~\citep{Sharma_2007,Lyutikov_2007,Kunz_2011,Sironi_2015}], transport properties [e.g., suppression of heat transport~\citep{Komarov_2016,RobergClark_2018,Komarov_2018,Yerger_2024}], 
and macroscopic dynamics [e.g., wave propagation~\citep{Squire_2016,Squire_2017,Kunz_2020,Majeski_2023} and turbulence~\citep{Hellinger_2015b,Hellinger_2019,Markovskii_2019,Squire_2019,Bott_2021,Squire_2022,Squire_2023,Arzamasskiy_2022,Majeski_2024}]. Because many astrophysical plasma environments -- including the solar wind~\citep{Alexandrova_2013}, black-hole accretion flows~\citep{Yuan_2014}, and the intracluster medium (ICM) of galaxy clusters~\citep{Schekochihin_2006,Simionescu2019} -- are either collisionless or weakly collisional, understanding these types of plasma is vital for obtaining even a rudimentary understanding of these systems.  

Despite the significant progress that has been made towards understanding the
feedback of kinetic instabilities on the macroscopic evolution of collisionless plasmas, 
a comprehensive theoretical framework for this 
phenomenon has not yet been established. There are two current barriers to the completion 
of such a framework. 
First, many different types of kinetic instability can arise~\citep{Bott_2023}. 
For example, bulk fluid motions and temperature gradients can generate pressure
anisotropies, in turn driving kinetic instabilities [e.g., the mirror instability~\citep{Barnes_1966,Hasegawa_1969} and ion-cyclotron instability~\citep{Sagdeev_1960}]. Other instabilities
-- for example, the whistler heat-flux instability~\citep{Levinson_1992} -- are driven directly by temperature gradients. 
Because the precise mechanism of the feedback depends on the properties of electromagnetic fluctuations 
associated with each instability (for example, its scale and/or polarisation), all of these 
kinetic instabilities need to be studied independently, and then their interplay explored 
subsequently. This (frankly Herculean) task has not yet been completed.
Secondly, previous studies have shown that the fundamental nature of kinetic 
instabilities can depend qualitatively on certain parameters including, but not limited to, the plasma $\beta \equiv 8 \upi 
p/B^2$
(defined as the ratio of the thermal pressure $p$ to the magnetic 
pressure), the ion-cyclotron frequency $\Omega_i$, and the macroscopic 
evolution time $\tau$. For many instabilities,
their behaviour over the full extent of parameter space that is relevant to astrophysical 
systems has not yet been explored systematically.  

In this paper, we address this second barrier for \emph{ion firehose instabilities}. These instabilities 
arise when the macroscopic evolution of a collisionless plasma gives rise to an excess of 
parallel ion pressure $p_{\| i}$ as compared with the perpendicular 
 ion pressure $p_{\perp i}$. The development of such a state follows the generic prescription 
 given above: for ion pressure anisotropies $\Delta_i \equiv p_{\perp i}/p_{\| i} - 1 \lesssim -1.35/\beta_{\|i}$ (where $\beta_{\|i} \equiv 8 \upi p_{\| i}/B^2$), 
 ion-Larmor-scale electromagnetic modes becomes unstable\footnote{The precise value of this threshold has a weak 
 dependence on $\beta_{i}$ -- see section \ref{sec:theory}.}, while 
 for $\Delta_i < -2/\beta_{\|i}$, a broad spectrum (from macroscopic scales down to ion-Larmor 
 scales) of Alfv\'enic modes is destabilised. A closely related class of instabilities, \emph{electron firehose instabilities}, can be driven by electron pressure anisotropy~\citep[see, e.g.,][]{Hollweg_1970,Paesold_1999,Li_2000,Gary_2003}. However, for the sake of simplicity, we do not treat these here, and hereafter refer to the ion firehose instability as just {\em the} `firehose instability'.
 
A new study of firehose instabilities in collisionless, $\beta_{i} \gtrsim 1$ plasma is timely, because the plasma's properties after the firehose instability's saturation 
depend on plasma parameters in a manner that remains unclear from previous studies. These prior studies do concur that, once firehose modes are destabilised, 
 they grow, backreact on the evolution of $\Delta_i$, 
 and then regulate it, with this regulation being maintained via an anomalous 
 collisionality $\nu_{\rm eff}$. 
However, several key results change significantly depending on $\beta_i$ and $\tau \Omega_i$, including the specific value $(\Delta_i)_{\rm sat}$ at which the pressure anisotropy is regulated, the specific value of $\nu_{\rm eff}$, as well as the characteristic energy $\delta B^2/B_0^2$ and spectrum of the magnetic-field perturbations. For example, using two-dimensional (2-D) hybrid-kinetic particle-in-cell (PIC) simulations of shearing plasmas with $\beta_{i} = 200$ and $\tau \Omega_i \sim 10^{3}$--$\; 3 \times 10^{4}$ (where $\Omega_i$ is 
 the ion Larmor frequency),~\citet{Kunz2014_b} found that $(\Delta_i)_{\rm sat} \simeq -2/\beta_{i}$ for all the 
 shear rates that were studied, $\nu_{\rm eff} \sim 10^{-2}$--$10^{-1} \Omega_i$, $\delta B^2/B_0^2 \sim 0.07$--$0.3$, and a magnetic-energy spectrum peaked at wavelengths much greater that $\rho_i$. By contrast, hybrid-kinetic PIC simulations of expanding, magnetized plasmas 
  at $\beta_{i} \sim 1$ and $\tau \Omega_i \sim 10^{3}$--$10^{4}$ in both 2-D and three-dimensional (3-D) geometries~\citep{Hellinger_2008,Hellinger_2015a,Hellinger_2019,Bott_2021} found 
  tighter regulation of pressure anisotropy [$(\Delta_i)_{\rm sat} \simeq -1.4/\beta_{\| i}$], much smaller values of the effective collisionality ($\nu_{\rm eff}\lesssim 10^{-3}$) that were time-dependent, $\delta B^2/B_0^2 \lesssim 10^{-2}$, and fluctuations with wavelengths not much larger that $\rho_i$. 
  \citet{Melville_2016}, which performed similar simulations to those of~\citet{Kunz2014_b} with characteristically smaller shearing timescales ($\tau \Omega_i \sim 10^{2}$--$10^{4}$) and larger values of the beta parameter ($\beta_{i} = 10^2$--$10^3$), made some progress on this problem, identifying the \textit{ultra-high-beta} regime ($\beta_i \gg \tau \Omega_i$) in which the regulation of the pressure anisotropy was less efficient ($\Delta_i \lesssim -2/\beta_{\|i}$) than for smaller $\beta_i$. Yet the full range of plasma parameters realised in firehose-susceptible astrophysical plasmas of interest has not been comprehensively explored. 
  
  Understanding quantitatively the thermodynamics and collisionality of firehose-susceptible high-$\beta$ plasmas as a function of $\beta$, $\tau$, and $\Omega_i$ is necessary because these properties can have dramatic implications for the macroscopic 
 dynamics of the plasma in which the firehose instability is operating. 
For example, the discrepancy in the specific value of $(\Delta_i)_{\rm sat}$ ($-2/\beta_{\|i}$~vs.~$-1.4/\beta_{\|i}$), which might na\"{i}vely seem to be a 
  numerical triviality of little consequence, is in fact qualitatively significant, because the effective Alfv\'en speed
\begin{equation}\label{eqn:vAeff}
    v_{\rm A,eff} \equiv v_{\rm A} \left(1+\frac{\beta_{\|i} \Delta_i}{2}\right)^{1/2}
\end{equation}
at which Alfv\'en waves propagate in a pressure-anisotropic plasma decreases as $\Delta_i$ does, with it tending to zero 
 as $\Delta_i\rightarrow-2/\beta_{\|i}$. In a plasma with $\Delta_i = -2/\beta_{\|i}$, the Alfv\'enic 
 restoring force is exactly cancelled out by anisotropic pressure forces, a state we 
  identify as `{\it Alfv\'en inhibiting}' because linear Alfv\'en waves cannot longer propagate. If instead the feedback of the firehose instability regulates the pressure anisotropy such that $\Delta_i \simeq -2/\beta_{\|i}$, an  `{\it Alfv\'en-enabling}' state would result, in which linear Alfv\'en waves would still be able to propagate (albeit with a lower parallel phase speed). Thus, both wave and 
  turbulent dynamics should be profoundly different in a plasma whose firehose-regulated pressure anisotropy satisfies $\Delta_i \simeq -2/\beta_{\|i}$, than in a plasma with {$\Delta_i \simeq -1.4/\beta_{\|i}$}.

\begin{figure}
  \centering
  \includegraphics[width=\linewidth]{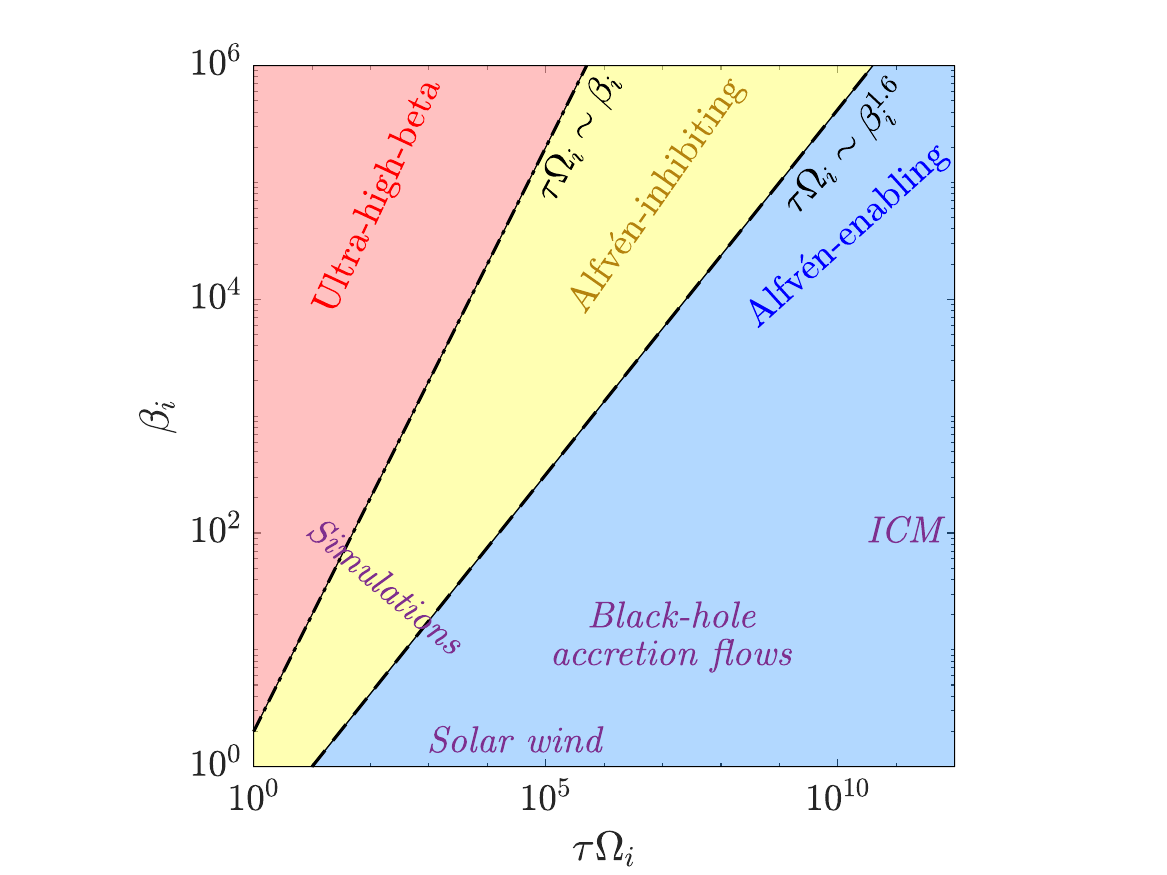}
  \caption{Phase-space map of high-$\beta_{i}$ firehose-susceptible plasmas in $\beta_i$ and $\tau \Omega_i$.}
\label{fig:sims_phasespacegen}
\end{figure}

In this paper, we put forward a comprehensive theory for how the firehose instability grows, saturates and then affects the thermodynamics and collisionality of high-$\beta$ plasma. We claim that, depending on the relative magnitude of $\beta_i$ and $\tau \Omega_i$, there are three qualitatively distinct regimes: ultra-high-$\beta$, Alfv\'en-inhibiting and Alfv\'en-enabling. For each of these regimes, we provide estimates of $(\Delta_i)_{\rm sat}$, $\nu_{\rm eff}$, and $\delta B^2/B_0^2$. We also describe characteristic properties of the wavevectors of firehose modes and various features that emerge in the ion distribution function. A key pillar of our theory, supported by linear calculations and nonlinear simulations, is a complete explanation for when a high-$\beta$ firehose-susceptible plasma attains an Alfv\'en-enabling or Alfv\'en-inhibiting state. We find that, at fixed $\beta_i$, an Alfv\'en-enabling state is attained if $\tau$ exceeds some $\beta_i$-dependent critical value $\tau_{\rm cr} \sim \Omega_i^{-1} \beta_i^{1.6}$.  Figure~\ref{fig:sims_phasespacegen} illustrates which state is realised as a function of $(\tau \Omega_i,\beta_i)$, with some astrophysical high-$\beta$ plasma environments of interest placed in this parameter space. 
Because $\tau\Omega_i$ is very large in most high-$\beta$ astrophysical plasmas, the Alfv\'en-enabling state is the more relevant one (see section~\ref{sec:applications}). We also propose and test a model for an effective firehose collision operator, which we use to understand better certain key properties of plasmas in  Alfv\'en-enabling states (e.g., the saturation energy of the firehose fluctuations and the velocity-space anisotropy of the ion distribution function). 

This paper is organised as follows. In section~\ref{sec:theory}, we outline the linear theory of firehose instabilities in high-$\beta$ plasmas. In section~\ref{propfirehosesat_overview} we describe qualitatively the ultra-high-$\beta$, Alfv\'en-inhibiting and Alfv\'en-enabling states, and account for why they arise. 
In particular, we explain with recourse to the theory outlined in section~\ref{sec:theory} why it is 
 that, for $\tau > \tau_{\rm cr}(\beta_i)$, the
  minimum value of $\Delta_i$ attained during the plasma's evolution obeys $(\Delta_i)_{\rm min} > -2/\beta_{i}$, and thereby why Alfv\'en-enabling states are realised. 
  We then corroborate this theory with a series of simulations of expanding plasmas
   (section \ref{sec:sims}), which we also use to characterise the 
   `saturated' state of the firehose instability in Alfv\'en-inhibiting and Alfv\'en-enabling states. In section~\ref{sec:simtheoryinterp}, we interpret the results of these simulations in detail, and in particular provide further analysis about the more subtle features of the Alfv\'en-enabling state. Of these features, understanding the saturated amplitude of the firehose fluctuations naturally motivates consideration of the effective firehose collision operator that arises in the Alfv\'en-enabling state (see section~\ref{sec:collisionality}). In section~\ref{sec:applications}, we situate our theory with respect to prior studies of firehose instabilities, and also discuss
     their ramifications for various different astrophysical systems.  Finally, in section \ref{sec:summary}, we provide a summary of our key 
    results.

\section{The linear theory of firehose instabilities in high-$\beta$ plasmas} \label{sec:theory}

\subsection{Overview} \label{linfiretheory_overview}

The existence of qualitatively distinct states in firehose-susceptible, high-$\beta$ plasmas stems in part  from 
properties of the instability in its linear stage. In this section, we therefore 
describe the linear theory of the firehose 
instability.
Though the linear theory of firehose instabilities has been discussed extensively in prior research (which we review in section~\ref{linfiretheory_review}), previously reported results do not completely account for the instability's properties in high-$\beta$ plasmas. We therefore report a new analytical and numerical linear study in this regime (section~\ref{linfiretheory_numres}). We find that oblique firehose modes are dominant for $\beta_i \gg 1$, with parallel ion-Larmor-scale firehose modes always having a smaller growth rate, in contrast to plasmas with $\beta_i \sim 1$. Furthermore, the value of the pressure anisotropy at which ion-Larmor-scale oblique firehose modes are destabilised ($\Delta_i = p_{\perp i}/p_{\| i} - 1 \simeq -1.35/\beta_{\|i}$) is less negative than  
that for longer-wavelength firehose modes at fixed $\beta_i$, and is similar to the threshold value in $\beta_i \sim 1$ plasma.
Aided by analytic theory, we explain these results in sections~\ref{linfiretheory_threshexp} and~\ref{linfiretheory_resparfire}, respectively. 

\subsection{A review of the firehose instability's linear theory} \label{linfiretheory_review}

Although a comprehensive understanding of the linear theory of the firehose instability (including at kinetic scales)
was only obtained in the last few decades, the instability itself was first identified well over sixty years ago. The 
first studies of the firehose instability~\citep{Chandrasekhar_1958,Parker_1958,Vedenov_1958} showed that the dispersion relation of
long-wavelength Alfv\'en 
waves ({\em viz.}, those modes with frequency $\omega$ whose parallel and perpendicular wavenumbers satisfy $k_{\|} \rho_i \ll |\Delta_i + 2/\beta_{\|i}|^{-1/2} \lesssim 1$
and $k_{\perp} \rho_i \ll 1$, respectively) is
\begin{equation}
\omega^2 = k_{\|}^2 v_{\rm A}^2 \left(1+\frac{\Delta_i \beta_{\|i}}{2}\right) = k_{\|}^2 v_{\rm A,eff}^2 
. \label{Alfven_wave_longwv_disprel}
\end{equation}
These modes become linearly unstable if the ion pressure anisotropy $\Delta_i$ 
satisfies $\Delta_i < -2/\beta_{\|i}$ (or, equivalently, if $v^2_{\rm A,eff} < 0$). 
We identify this condition as the `fluid' firehose instability threshold, and the resulting instability as the non-resonant (or fluid) firehose instability. 
The instability of these modes can be understood physically as follows: once the 
parallel ion pressure exceeds the perpendicular pressure by an amount equal to twice the magnetic energy, 
parallel pressure forces on an Alfv\'enic perturbation can overpower the restoring 
magnetic tension that, in a pressure-isotropic plasma, is responsible for the wave's propagation.

It is immediately clear from (\ref{Alfven_wave_longwv_disprel}) that shorter-wavelength perturbations grow more rapidly than longer-wavelength ones, implying 
that the scale of the fastest-growing firehose modes must be determined by 
finite-Larmor-radius (FLR) effects. For non-resonant parallel firehose modes, these FLR 
effects can be characterised analytically~\citep{Shapiro_1963,Kennel_1967,Davidson_1968}, with the parallel wavenumber $k_{\|,\mathrm{peak}} \sim |\Delta_i + 2/\beta_{\|i}|^{-1/2} \rho_i^{-1}$ at which
peak growth occurs being determined by gyroviscosity, i.e., the off-diagonal components of the pressure tensor associated with agyrotropy in the distribution function~\citep[e.g.,][]{Schekochihin_2010}. Whenever $|\Delta_i + 2/\beta_{\|i}| \ll 1$, which is either achieved near threshold (that is, when $|\Delta_i + 2/\beta_{\|i}| \ll 1$ in plasmas with $\beta_i \sim 1$, or whenever $|\Delta_i| \ll 1$ in high-$\beta_i$ plasmas), the wavelength of the fastest-growing non-resonant mode is much larger than the ion-Larmor scale.

More recent studies that solved the 
hot-plasma dispersion relation numerically for a bi-Maxwellian plasma discovered the existence of two kinetic variants of the firehose instability: 
the \emph{resonant parallel firehose instability}~\citep{Gary_1998}, and the 
 \emph{oblique firehose instability}~\citep{Yoon_1993,Hellinger_2000}. 
 Modes of the resonant parallel firehose instability are destabilised by gyroresonant interactions with suprathermal ions having parallel velocities $v_{\|} = (\varpi + \Omega_i)/k_{\|}$ (where $\varpi$ is the real frequency of the mode). These modes have a characteristic parallel wavenumber $k_{\|} \rho_i \sim 1$ when $\beta_{\|i} \sim \Delta_i \sim 1$, are circularly polarised, right-handed and propagating. Although the instability is technically 
 thresholdless [\citet{Sagdeev_1960}; see also appendix~\ref{append_resparfiregrowthrate}], previous numerical studies found that such modes only attain growth rates $\gamma_{\|\rm f}$
 at ion-Larmor scales that are not infinitesimal 
 fractions of the ion-Larmor frequency when $\Delta_i$ exceeds some $\beta_{\|i}$-dependent threshold. For example, \citet{Matteini_2006} report that, in order for $\gamma_{\|\rm f} \gtrsim 5 \times 10^{-3} \Omega_i$, one requires that $\Delta_i \lesssim -0.6 (\beta_{\|i}-0.63)^{-0.58}$. By contrast,  oblique firehose modes 
are non-propagating and linearly polarised, with $k_{\|} \rho_i \sim k_{\perp} 
\rho_i \sim 0.5$. Studies with $\beta_{\|i} \gtrsim 1$ identified a threshold that 
scales with $\beta_{\|i}$ in the same way as the fluid firehose threshold, but 
with a less negative numerical prefactor: $\Delta_i \lesssim -1.4
 \beta_{\|i}^{-1}$~\citep{Hellinger_2000,Hellinger_2001}. These conditions together imply that, when $\beta_{\| i} \sim 
 1$, the resonant parallel firehose instability tends to dominate, but that the 
  oblique firehose instability should become dominant when $\beta_{\| i} \gg 
 1$. 
 
 The less negative values of the pressure anisotropy required for the resonant and oblique firehose instabilities to operate { linearly} at 
 ion-Larmor scales have been considered and discussed extensively for collisionless, $\beta_{\|i} \gtrsim 
 1$ plasma similar to the solar wind~\citep{Hellinger_2006,Matteini_2007,Matteini_2012,Matteini_2013}. { Several studies present results of direct relevance to high-$\beta$ plasmas. For example, in addition to identifying the existence of the resonant parallel firehose instability, \citet{Gary_1998} characterise its linear threshold for $\beta_i \leq 10$. \citet{Hellinger_2006} compute linear instability thresholds for both the resonant parallel and oblique firehose instabilities in a bi-Maxwellian plasma for $\beta_i \leq 30$. However, complementary  
 results for linear firehose instability thresholds in plasmas with larger $\beta_i$ have not been the focus of any previous published 
 studies, nor have results for peak growth rates.} We therefore report these 
 results in the next section. 
 
\subsection{The kinetic firehose instability at $\beta_{i} \gg 1$} \label{linfiretheory_numres}

To determine the linear thresholds and growth rates of firehose-unstable modes as a 
function of wavenumber in a plasma with bi-Maxwellian ions and Maxwellian electrons\footnote{As discussed in the Introduction, bi-Maxwellian electron distributions with $\Delta_e$ < 0 are associated with electron firehose instabilities; to focus exclusively on ion firehose instabilities, we therefore choose $\Delta_e = 0$. Physically, this simplification is appropriate in plasmas whose electron population is not `too' collisionless~\citep[e.g., the ICM -- see][]{KJZ22}: more specifically, if $\nu_e \gg \beta/\tau$, where $\nu_e$ is the rate of electron Coulomb collisions, then $|\Delta_e| \ll 1/\beta$, and so the electron population can be treated as Maxwellian from the standpoint of their stability.}, 
we solve their linear dispersion relation numerically. We take the electric 
field $\delta \bb{E}$ and magnetic field $\delta \bb{B}$ associated with such perturbations to be of the form
\begin{equation}
   \delta \bb{E} \propto \exp\bigl[\imag\bigl(\bb{k} \bcdot \bb{r} - \omega t\bigr)\bigr]  , 
   \quad 
   \delta \bb{B} \propto \exp\bigl[\imag\bigl(\bb{k} \bcdot \bb{r} - \omega t\bigr)\bigr] \label{FourierEBfields} 
   ,
\end{equation}
where $\bb{k}$ and $\omega$ are the wavevector and (complex) frequency of the perturbation. The dispersion relation of firehose perturbations having arbitrary $\bb{k}$
is the hot-plasma dispersion relation~\citep{Davidson_1983}, which we provide for a plasma with 
arbitrary distribution functions in Appendix~\ref{append:hotplasmadispreldef}. For a hydrogenic plasma with 
bi-Maxwellian ions (with parallel temperature $T_{\|i}$ and perpendicular temperature $T_{\perp i}$) and Maxwellian electrons (with temperature $T_e$), the dispersion 
relation simplifies to
\begin{equation}
   \mbox{det}\Bigg\{\frac{k^2 \rho_i^2}{\beta_{\perp i}} \Bigg[\ek\ek - \msb{I}\left(1-\frac{\omega^2}{k_\|^2 v_{\mathrm{th}\perp i}^2} \frac{v_{\mathrm{th}\perp i}^2}{c^2}\right)\Bigg] 
 + \widetilde{\bb{\sigma}}_{\textrm{bi-M}} \Bigg\}=0 
   ,
   \label{hotplasmadisprel_bimax}
\end{equation}
where $v_{\mathrm{th}\perp i} \equiv \sqrt{2 T_{\perp i}/m_i}$, and $\widetilde{\bb{\sigma}}_{\textrm{bi-M}} = \widetilde{\bb{\sigma}}_{\textrm{bi-M}} (k_{\|} \rho_i,k_{\perp} \rho_i,{\omega/k_\| v_{\mathrm{th}\perp i}},{m_e/m_i},{T_e/T_{\|i}},\Delta_i )$ 
is a dimensionless rank-three tensor that can be written in terms of the 
plasma dispersion function and sums of modified Bessel functions (see Appendix~\ref{append:hotplasmadispreldef}). 
To find the complex frequency $\omega$ of firehose-unstable modes 
at fixed values of $\beta_i$, $m_e/m_i$, 
$T_e/T_{\|i}$, $v_{\mathrm{th}\perp i}/c$ and $\Delta_i$, we choose values of $k_{\|} \rho_i$ and $k_{\perp} \rho_i$ at which such modes are expected to be realisable, 
and then solve for the roots ${\omega}/{k_{\|} v_{\mathrm{th}\perp i}}$ of 
(\ref{hotplasmadisprel_bimax}). Numerically, this is carried out using the secant 
method, with the initial guesses inputted into the algorithm being determined 
by an analytical approximation to the hot-plasma dispersion relation that is valid when $\beta_i, \beta_e \gg 1$~\citep[taken from][]{Bott_2023}.

\begin{figure}
  \centering
  \includegraphics[width=\linewidth]{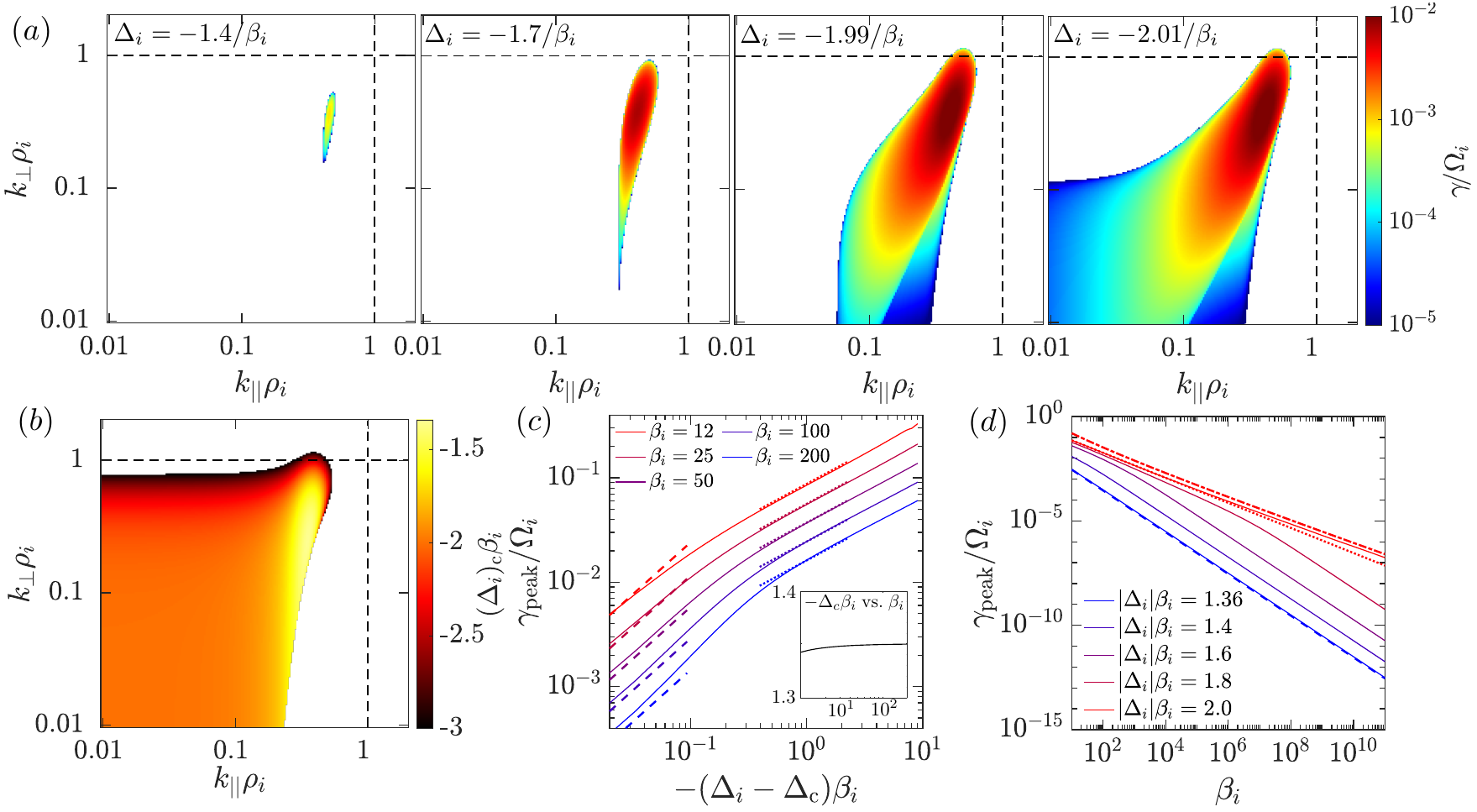}
  \caption{($a$) Linear growth rate $\gamma$ of firehose-unstable modes as a function of parallel and perpendicular wavenumber for a range of different $\Delta_i$
  at $\beta_{i} = 200$, $m_i/m_e = 1836$, $T_e = T_{\|i}$, and $v_{\mathrm{th}e}/c = 0.05$. 
  The growth rates are calculated on a $400^2$ grid in ($k_{\|} \rho_i, k_{\perp} \rho_i$), with equal logarithmic spacing in both directions. 
  ($b$) Critical value of $\Delta_i$ below which firehose instability onsets, $\Delta_{\rm cr}$, as a function of parallel and perpendicular wavenumber at 
  $\beta_{i} = 200$. ($c$) Peak growth rate $\gamma_{\rm peak}$ of the firehose instability as a function of $(\Delta_i-\Delta_{\rm cr}) \beta_i$ for a range of $\beta_i$ (solid lines). The dashed lines shows the semi-analytic result (\ref{firehosescaling_thresholdreg}), the red dotted line shows the power-law scaling (\ref{firehosescaling_growthratefluidthreshold_approx}) that empirically is a good fit for moderately large $\beta_i$.
  ($d$) $\gamma_{\rm peak}$ as a function of $\beta_i$ for a range of $|\Delta_i|\beta_i$. The blue-dashed line shows  (\ref{firehosescaling_thresholdreg}); 
  the red dot-dashed line shows the analytic result (\ref{firehosescaling_fluidthresholdreg_exact_highbeta}); and the red-dotted line shows the power law (\ref{firehosescaling_fluidthresholdreg_approx}).}
\label{fig:lintheory}
\end{figure}

Figure~\ref{fig:lintheory} shows the growth rate $\gamma = \mathrm{Im}(\omega)$ of firehose-unstable modes in a plasma with $\beta_{i} = 200$ as a 
function of $k_{\|} \rho_i$ and $k_{\perp} \rho_i$
for representative choices of the other parameters.
Similarly to prior numerical studies in $\beta_i \gtrsim 1$ plasma, we observe 
that, as $\Delta_i$ is decreased from zero towards the (negative) instability thresholds, modes whose growth rates are not infinitesimally small first emerge only at ion-Larmor scales, at a critical value of the ion pressure anistropy, $\Delta_{\rm cr} \simeq -1.35/\beta_{i}$, that is less negative than the fluid firehose threshold 
$\Delta_{i} = -2/\beta_{i}$ at fixed $\beta_i$ (figure~\ref{fig:lintheory}($a$)). Of these modes, the fastest growing ones are oblique firehose modes ($k_{\|} \rho_i \simeq 0.45$, $k_{\perp} \rho_i \simeq 0.35$) with zero real frequency. 
As $\Delta_i$ is decreased further, the region of $(k_{\|} \rho_i,k_{\perp} 
\rho_i)$-space over which the firehose instability operates extends, with long-wavelength modes becoming unstable once $\Delta_i < -2/\beta_i$ (see figure~\ref{fig:lintheory}($b$)). 

The growth rate $\gamma_{\rm peak}$ of the 
fastest-growing firehose-unstable modes
is an increasing function of $-(\Delta_i-\Delta_{\rm cr}) \beta_{i}$ (see figure~\ref{fig:lintheory}($c$)) and a decreasing function of 
$\beta_i$ at fixed $\Delta_i \beta_i$ (see figure~\ref{fig:lintheory}($d$)).
The particular scaling of $\gamma_{\rm peak}$ with $\beta_i$ 
at fixed $\Delta_i \beta_i$ depends on the latter's exact value. 
When $|\Delta_i - \Delta_{\rm cr}| \beta_i \ll 1$, we find that
\begin{equation}
\gamma_{\rm peak} \approx 2.9 |\Delta_i-\Delta_{\rm cr}| \Omega_i \quad (\mathrm{for} \;|\Delta_i - \Delta_{\rm cr}| \beta_i \ll 1) \, . \label{firehosescaling_thresholdreg}
\end{equation} 
When $\Delta_i = -2/\beta_i$, a different scaling can be derived analytically~\citep{Bott_2023}:
 \begin{equation}
\gamma_{\rm peak} \approx \frac{5^{1/2}}{2^{3/2}} \left(k_{\|} \rho_i\right)_\mathrm{peak}^2 \frac{\Omega_i}{\beta_i^{1/2}} \quad (\mathrm{for} \;\Delta_i = -2/\beta_i) \, , \label{firehosescaling_fluidthresholdreg_exact}
\end{equation} 
where the characteristic parallel wavenumber of the fastest growing mode, $ \left(k_{\|} \rho_i\right)_\mathrm{peak}$, is a weakly varying function of $\beta_i$ that is given by special mathematical functions. For values of $\beta_i$ that are very large ($\beta_i \gg 10^{6}$), the following simple expression for $\left(k_{\|} \rho_i\right)_\mathrm{peak}$ (and therefore $\gamma_{\rm peak}$) can be found through an asymptotic analysis~\citep{Bott_2023}:
 \begin{eqnarray}
\left(k_{\|} \rho_i\right)_\mathrm{peak} \approx \left[\frac{2}{\log{(27 \upi \beta_i/10)}}\right]^{1/2} , \nonumber \\  \gamma_{\rm peak}  \approx \frac{5^{1/2}}{2^{1/2}} \frac{\Omega_i}{\beta_i^{1/2} \log{(27 \upi \beta_i/10)}} & \quad (\mathrm{for} \;\Delta_i = -2/\beta_i, \; \beta_i \mathrm{\; very \; large}) \, . \label{firehosescaling_fluidthresholdreg_exact_highbeta}
\end{eqnarray}
By contrast, for values of $\beta_i$ that are only moderately large ($\beta_i \in [10,10^4]$), it can be shown empirically via fitting to the direct numerical solution of the linearised Vlasov equation that $\gamma_{\rm peak}$ at $\Delta_i = -2/\beta_{\|i}$ has, to a very good approximation, a simple power-law dependence on $\beta_{i}$: 
\begin{equation}
\gamma_{\rm peak} \approx 0.3 \Omega_i \beta_i^{-0.6} \quad (\mathrm{for} \;\Delta_i = -2/\beta_i, \; 1 \ll \beta_i \ll 10^{5}) \, . \label{firehosescaling_fluidthresholdreg_approx}
\end{equation}
This result can be extended to $\Delta_i$ close to (but not exactly equal to) $-2/\beta_{\|i}$, for which we find that the peak growth rate is related to the pressure anisotropy via a simple power-law scaling: 
\begin{equation}
\gamma_{\rm peak} \approx 0.4 |\Delta_i-\Delta_{\rm cr}|^{0.6} \Omega_i \quad (\mathrm{for} \;\Delta_i \approx -2/\beta_i, \; 1 \ll \beta_i \ll 10^{5}) \, . \label{firehosescaling_growthratefluidthreshold_approx}
\end{equation}

The validity of these asymptotic 
approximations is tested in figures~\ref{fig:lintheory}($c$) and~\ref{fig:lintheory}($d$). Figure~\ref{fig:lintheory}($c$) confirms that, in the relevant regime, the expressions (\ref{firehosescaling_thresholdreg}) and (\ref{firehosescaling_growthratefluidthreshold_approx}) are good approximations to the numerically determined growth rate as a function of $\Delta_i - \Delta_{\rm cr}$. Furthermore, figure~\ref{fig:lintheory}($d$) shows that the decrease of $\gamma_{\rm peak}$ with increasing $\beta_i$ is primarily
accounted for by the $\beta_i^{-0.6}$ dependence included in  
(\ref{firehosescaling_fluidthresholdreg_approx}) for $\beta_i \ll 10^5$. For quantitative agreement
over a larger range of $\beta_i$, an exact power-law fit is an oversimplification, as shown by the better agreement of the numerically determined growth rate with (\ref{firehosescaling_fluidthresholdreg_exact_highbeta}).

In summary, we find that, in plasma with $\beta_{\|i} \gg 1$, the fastest-growing unstable modes are oblique firehose modes, and that these modes emerge at less negative pressure anisotropies ($\Delta_i\lesssim -1.35/\beta_{\|i}$) than fluid firehose modes ($\Delta_i < -2/\beta_{\|i}$). Furthermore, the resonant parallel firehose instability does not feature significantly in our numerical solution of the dispersion relation, seeming to imply that it is subdominant to the oblique instability in $\beta_{\|i} \gg 1$ plasma. We account for both of these findings in sections~\ref{linfiretheory_threshexp} and~\ref{linfiretheory_resparfire}, respectively.

\subsection{Why the threshold of the oblique firehose instability is larger than $-2/\beta_{\parallel i}$} \label{linfiretheory_threshexp}

The numerical result that kinetic-scale oblique firehoses in a bi-Maxwellian, ${\beta_i \gg 1}$ plasma are destabilised at a less negative value of the pressure anisotropy can be elucidated by physical arguments and additional mathematical analysis. 

The physical basis for a reduced threshold arises from modifications to the effective parallel pressure force acting on a magnetic-field perturbation when the thermal ion-Larmor radius is only a finite fraction of that perturbation's wavelength. As explained in section~\ref{linfiretheory_review}, the fluid firehose instability is an instability of Alfv\'en waves in which, due to an excess of parallel pressure compared to perpendicular pressure, parallel pressure forces on the perturbed volume of plasma associated with the Alfv\'en wave become sufficiently large to overcome the restorative perpendicular pressure  
and magnetic-tension forces. When the scale of the perturbation is not much larger than $\rho_i$, thermal ions are less well `tied' to the field line's trajectory because of their gyromotion. This results in an additional contribution to the net flux of perpendicular momentum into the perturbed volume of plasma, and therefore enhanced parallel pressure forces. 

That the reduced threshold is a FLR effect can be proven analytically by taking advantage of the numerical observation that marginally unstable oblique firehose modes have no real frequency. Using this fact, we can derive a somewhat simplified (but still transcendental) equation for the threshold condition of the instability for a plasma with arbitrary ion and electron distribution functions. Via a subsidiary expansion in $k_{\|} \rho_i \sim k_{\perp} \rho_i \ll 1$, we can then write down a simple expression for the threshold condition that includes the leading-order FLR corrections (see Appendix~\ref{append_firehose_threshold_condition}). These corrections are proportional to high-order moments of the distribution function (specifically, fourth order or higher). For a plasma with bi-Maxwellian ions and Maxwellian electrons, we deduce that the threshold condition is
\begin{equation}
\Delta_i \left(1+\frac{3}{2} k_{\|}^2 {\rho}_i^2 - \frac{3}{8} k_{\perp}^2 {\rho}_i^2\right)+ \frac{2}{\beta_{\|i}} = \mathcal{O}(\Delta_i k^4 
{\rho}_i^4) . \label{threshold_bimax_main}
\end{equation}
For any perturbation having $k_{\|} < 2 k_{\perp}$, the condition (\ref{threshold_bimax_main}) implies that the value of $\Delta_i$ required for instability is less negative than the fluid firehose threshold $\Delta_i = -2/\beta_{\|i}$. For modes with the same wavevector as those oblique modes that we observe numerically to become unstable at $\Delta_i \approx \Delta_c \simeq -1.35/\beta_i$ (that is, $k_{\|} \rho_i \simeq 0.45$, $k_{\perp} \rho_i \simeq 0.35$), equation (\ref{threshold_bimax_main}) implies $\Delta_i \approx -1.6/\beta_{\|i}$, which is (to the order of accuracy of the subsidiary expansion) not too dissimilar to the numerically determined result. This agreement supports the conjecture that FLR effects are responsible for the observed weakening in the instability threshold. 

\subsection{Why the resonant parallel firehose instability is subdominant in $\beta_{\|i} \gg 1$ plasma} \label{linfiretheory_resparfire}

 The apparent unimportance of the resonant parallel firehose instability when ${\beta_{\|i} \gg 1}$, a finding consistent with previous numerical results (see section \ref{linfiretheory_review}), can be proven analytically. We show in Appendix~\ref{append_resparfiregrowthrate} that, when $\Delta_i \simeq -1.35/\beta_{\|i}$, the fastest-growing resonant parallel firehose modes (which, in contrast to plasma with $\Delta_i \sim \beta_{\|i} \sim 1$, satisfy $k_{\|} \rho_i \ll 1$) have a growth rate that is exponentially small in $1/\beta_{\|i}$, {\em viz.}~$\gamma_{\|\mathrm{f}} \sim \beta_{\|i}^{-1/2} \exp{(-0.74\beta_{\|i})}$. 
By comparison, the peak growth rate $\gamma_{\rm \perp f}$ of the resonant oblique firehose instability satisfies ${\gamma_{\rm \perp f}} \sim \left|\Delta_i -\Delta_{\rm cr}\right| {\Omega_i}$ when $\Delta_i$ is close to the instability's threshold anisotropy, $\Delta_{\rm cr} \simeq -1.35/\beta_{\|i}$ [cf. (\ref{firehosescaling_thresholdreg})]. Assuming that $|\Delta_i - \Delta_{\rm cr}| \ll 
 1/\beta_{\|i}$, it can be shown that $\gamma_{\rm \perp f}$ greatly exceeds $\gamma_{\rm \|f}$ 
  when 
 \begin{equation}
   (\Delta_{\rm cr}-\Delta_i) \beta_{\|i} \gg \beta_{\|i}^{1/2} \exp{\left(-0.74 \beta_{\|i}\right)} 
   \, . \label{bound}
 \end{equation} 
 For $\beta_{\|i} \gtrsim 4$, the right-hand side of (\ref{bound}) is at least an order of magnitude below unity. We 
 conclude that in a high-$\beta_{\|i}$ plasma with an increasingly negative pressure anisotropy, the resonant oblique firehose instability 
becomes much faster growing than its parallel counterpart  
 once $\Delta_i < \Delta_{\rm cr}$. 

 Perhaps more surprisingly, the resonant parallel firehose instability also becomes subdominant to the \emph{non-resonant} (fluid) parallel firehose instability in $\beta_{\|i} \gg 1$, bi-Maxwellian plasma at pressure anisotropies not much more negative than the fluid firehose threshold, $\Delta_i = -2/\beta_{\|i}$. If $\Delta_i < -2/\beta_{\|i}$, then the non-resonant parallel firehose operates at all parallel wavenumbers that satisfy~\citep{Schekochihin_2010}
 \begin{equation}
   k_{\|} \rho_{\|i} < 4 \left|\frac{1}{\beta_{\|i}} + \frac{\Delta_i}{2}\right|^{1/2} 
   \, ,
 \end{equation}
 and the peak instability growth rate is
 \begin{equation}
   \gamma_{\rm \|f,nr} \equiv \left|\frac{2}{\beta_{\|i}} + \Delta_i\right|\Omega_{\rm i} \quad \mathrm{at} \quad (k_{\|} \rho_i)_{\rm \|f,nr} \equiv 2 \left|\frac{2}{\beta_{\|i}} + \Delta_i\right|^{1/2} 
   \, .
 \end{equation}
If we assume that $|{2}/{\beta_{\|i}} + \Delta_i| \ll 2/\beta_{\|i}$, it then follows that $\gamma_{\rm \|f,nr} \gtrsim \gamma_{\rm \|f}$ is equivalent to the condition\footnote{Using the estimate $\gamma_{\|\rm f} \sim |\Delta_i|^{1/2} \exp{(-1/|\Delta_i|)}$ derived in Appendix~\ref{append_resparfiregrowthrate} here with $\Delta_i \approx -2/\beta_{\|i}$ is valid, because the wavenumber $ (k_{\|} \rho_i)_{\rm \|f} \approx 2 \beta_{\|i}^{-1/2}$ at which peak growth of the resonant 
instability of the right-handed mode is attained is much larger than the wavenumber of the smallest-scale mode that becomes unstable to the non-resonant instability: $(k_{\|} \rho_i)_{\rm \|f}/(k_{\|} \rho_i)_{\rm \|f,nr} \approx \left|{2}+ \Delta_i \beta_{\|i}\right|^{-1/2} 
  \gg 1$. Therefore, the fastest-growing resonant parallel firehose mode is still propagating, and its real frequency is still much greater than its growth rate.}
 \begin{equation}
   |2+\Delta_i \beta_{\|i}| \gg \beta_{\|i}^{1/2} \exp{\left(-\frac{\beta_{\|i}}{2}\right)} 
   \, . \label{bound_nr}
 \end{equation}
This bound can be satisfied near marginality of the non-resonant firehose instability provided that $\beta_{\|i} \gtrsim 7$. Thus, in stark contrast to plasmas with $\beta_{\|i} \sim 1$~\citep[see, e.g,][]{Hellinger_2001}, the resonant parallel firehose instability is unimportant in $\beta_{\|i} \gg 1$ plasma. 

The relative inefficacy of the resonant parallel firehose instability in high-$\beta$ plasma has a simple physical explanation. As was mentioned in section~\ref{linfiretheory_review}, the instability is driven by 
 resonant wave-particle interactions: specifically, right-handed circularly polarised hydromagnetic waves drain  
 energy from gyroresonant particles with parallel velocities $v_{\|} =(\omega + \Omega_i)/k_{\|} \approx v_{\mathrm{th}i}/k_{\|} 
 \rho_i$. In a plasma with $\beta_i \sim \Delta_i^{-1} \gg 1$, the gyroresonant particles have characteristic velocities $v_{\|} \sim \Delta_i^{-1/2} v_{\mathrm{th}i}$ that are much greater than the ion thermal 
 velocity. This reveals why the growth rates of the unstable modes 
 are very small: due to their long wavelengths, the hydromagnetic waves can only 
 interact resonantly with suprathermal ions, of which there is only a small 
 number compared with the thermal population. The stabilising action of 
 cyclotron damping is weak on such modes, which in turn allows even a small 
 anisotropy to be able to overcome this damping. However, for shorter wavelength modes, cyclotron damping is simply too strong for the instability to operate. { This conclusion is consistent with the findings of~\citet{Matteini_2006}, who presented evidence of distribution functions becoming less distorted by resonant interactions as $\beta_i$ was increased in one-dimensional expanding-box simulations of firehose-unstable plasma with $\beta_i \leq 10$; this finding was attributed to the particles that were resonant with parallel firehose modes being increasingly suprathermal.}

While the resonant parallel firehose instability is generically unimportant in high-$\beta_i$ plasmas with bi-Maxwellian ion distributions, this conclusion does not necessarily hold for plasmas with non-bi-Maxwellian distributions. Indeed, we will show that right-handed circularly-polarised modes can be destabilised by the distribution function that naturally arises during the nonlinear evolution of the oblique firehose instability. These `secondary parallel firehose modes' are characterised and discussed in section~\ref{sec:sims:res:secondparfirehose}. 

 \section{Properties of high-$\beta_i$ plasmas with saturated firehose instability} \label{propfirehosesat_overview}

\subsection{Possible saturated states of the instability}

Once firehose modes are linearly destabilised, they grow until they are able to backreact significantly on the pressure anisotropy that drives their growth. Previous analytical and numerical studies suggest that this backreaction causes a transition from exponential growth of the magnetic energy of the modes to secular, power-law growth~\citep{Schekochihin_2008,Rosin_2011}. 
In turn, the secular-growth phase eventually transitions into saturation, with the magnetic energy no longer growing. Based on both previous studies~\citep[in particular,][]{Melville_2016} and the results of this paper, we claim that there are three qualitatively distinct states -- {\it ultra-high-beta}, {\it Alfv\'en-inhibiting} and {\it Alfv\'en-enabling} -- that can be realised by the saturation of firehose instabilities in high-$\beta$ plasmas. Which of these states is realised depends on the relative magnitude of just two independent parameters: $\beta_i$, and $\tau \Omega_i$, where we formally define the macroscopic evolution time $\tau$ by
\begin{equation}
   \tau \equiv \left|\left(\hat{\bb{b}}\hat{\bb{b}}-\frac{\mathsfbi{I}}{3}\right)\!\bdbldot \bnabla \bb{u}\right|^{-1} = \left| \frac{\mathrm{d}}{\mathrm{d}t} \log{\frac{B}{n^{2/3}}} \right|^{-1} \, .
\end{equation}
We describe each of these states in subsections~\ref{sec:UHBoverview},~\ref{sec:AIoverview}, and~\ref{sec:AEoverview}, respectively. 
To aid comparison between these states, Table~\ref{tab:overviewstates} summarises their key properties. 
\begin{table}
\centering
{\renewcommand{\arraystretch}{1.3}
\renewcommand{\tabcolsep}{0.15cm}
\begin{tabular}[c]{c|c|c|c}
Property & Ultra-high-beta & Alfv\'en-inhibiting & Alfv\'en-enabling \\
\hline
$\Delta_{\rm sat}$ & $ < -2/\beta_{i}$ & $\simeq -2/\beta_{i}$ & $\simeq -1.6/\beta_{i}$ \\
$\nu_{\rm eff}$ & $\approx 0.25 \Omega_i$ & $\simeq 0.5 \beta_{i}/\tau$ & $\simeq 0.4 \beta_{i}/\tau$ \\
$\mu_{\rm B}$ & $\sim 4 p_i/\Omega_i$ & $\simeq \tau B^2/4 \upi$ & $\simeq 0.8 \tau B^2/4 \upi $ \\ 
$\delta B^2/B_0^2$ & $\sim 1$ & $\gtrsim \beta_{i}(\tau \Omega_i)^{-1}$?  & $\sim \beta_i^{1/4} (\tau \Omega_i)^{-1/2}$\\
\renewcommand{\arraystretch}{1} \begin{tabular}{@{}c@{}} Long-wavelength \\ ($k \rho_i \ll 1$) modes? \end{tabular} & Yes & Yes & No
\end{tabular}} 
\caption{Summary of typical values describing the ultra-high-beta, Alfv\'en-inhibiting, and Alfv\'en-enabling states of a firehose-unstable plasma. Properties include the regulated pressure anisotropy in saturation $\Delta_{\rm sat}$, the particle-averaged effective collisionality $\nu_{\rm eff}$, the implied effective Braginskii viscosity $\mu_{\rm B}$, and the characteristic energy $\delta B^2/B_0^2$ of the firehose fluctuations. Whether or not the magnetic-energy spectrum of firehose fluctuations extends to wavelengths much greater than $\rho_i$ is also indicated. We note that $\delta B^2/B_0^2$ in the Alfv\'en-inhibiting state remains uncertain, because our study and that of \citet{Melville_2016} obtain discrepant results (see discussion in section~\ref{sec:AIoverview}).}
\label{tab:overviewstates}
\end{table}

The one commonality between all states is the emergence of an effective collisionality $\nu_{\rm eff}$ associated with the firehose fluctuations, which manifests as an additional isotropisation term in the CGL equations that describes the evolution of parallel and perpendicular pressures in magnetised plasmas~\citep{Chew_1956}:
\begin{subeqnarray}
 \frac{\mathrm{d} p_{\perp}}{\mathrm{d}t} & = & p_{\perp} \frac{\mathrm{d}}{\mathrm{d}t} \log{(n B)} -\nabla \cdot \boldsymbol{q}_{\perp} - q_{\perp} \nabla \cdot \hat{\boldsymbol{b}} - \nu_{\rm eff} \left(p_{\perp}-p_{\|}\right) , \\
 \frac{\mathrm{d} p_{\|}}{\mathrm{d}t}  & = & p_{\|} \frac{\mathrm{d}}{\mathrm{d}t} \log{\frac{n^3}{B^2}}-\nabla \cdot \boldsymbol{q}_{\|} + 2 q_{\perp} \nabla \cdot \hat{\boldsymbol{b}} - 2 \nu_{\rm eff} \left(p_{\|}-p_{\perp}\right) , \label{CGL_withcoll}
\end{subeqnarray}
where $\boldsymbol{q}_{\|}$ and $\boldsymbol{q}_{\perp}$ denote the parallel heat fluxes of parallel and perpendicular temperature, respectively. This effective collisionality in turn gives rise to an anomalous viscous stress tensor $\bb{\mathsf{\rmPi}}$. In a weakly collisional plasma ($\nu_{\rm eff} \ll \Omega_i$), this tensor is approximated well by
\begin{equation}
\bb{\mathsf{\rmPi}} \approx -\mu_{\rm B,eff} \left(\hat{\boldsymbol{b}} \hat{\boldsymbol{b}}-\frac{\msb{I}}{3} \right)\left(\hat{\boldsymbol{b}} \hat{\boldsymbol{b}}-\frac{\msb{I}}{3} \right)\bdbldot\grad \boldsymbol{u} = \frac{\mu_{\rm B,eff}}{\tau} \left(\hat{\boldsymbol{b}} \hat{\boldsymbol{b}}-\frac{\msb{I}}{3} \right) \, , \label{mag_visc_stress}    
\end{equation}
where 
\begin{equation}
\mu_{\rm B,eff} = \frac{p_i}{\nu_{\rm eff}} \simeq -\frac{1}{2} (\Delta_i)_{\rm sat} \beta_{\|i} \frac{B^2}{4 \upi} \,\tau \label{eff_viscosity} 
\end{equation}
is the effective Braginskii viscosity.

\subsubsection{Ultra-high-beta: $\tau \lesssim \beta_i \Omega_i^{-1}$} \label{sec:UHBoverview}

The `ultra-high-beta' state is realised when the effective collisionality required to regulate the pressure anisotropy back to the value required for marginal stability of the firehose instability (viz., $\nu_{\rm eff} \sim \beta_{\|i}/\tau$) becomes larger than $\Omega_i$, and therefore is not realisable. Microphysically, the ultra-high-beta state is characterised by large-amplitude magnetic-field perturbations: after a brief exponential growth phase, the fluctuations grow secularly for a time of order $\tau$ until $\delta B^2/B_0^2 \sim 1$ and a broad spectrum of firehose fluctuations emerge (including wavelengths that are much greater than $\rho_i$). These relatively large-amplitude fluctuations result in an effective collisionality $\nu_{\rm eff} \approx 0.25 \Omega_i$~\citep{Melville_2016}. There is, of course, an effective viscosity associated with this scattering rate, but because its value is ${\sim}p_i/\Omega_i$, the viscous stress tensor may not be sufficiently anisotropic that the form (\ref{mag_visc_stress}) is an adequate description.

\subsubsection{Alfv\'en-inhibiting: $\beta_i \Omega_i^{-1} \ll \tau \lesssim \tau_{\rm cr}(\beta_i)$} \label{sec:AIoverview}

If $\tau \Omega_i/\beta_i$ is much greater than unity, but is not too large (see section~\ref{sec:AEoverview}), then the time $t_{\rm sat} \sim (\beta_i \tau/\Omega_i)^{1/2}$ taken for the firehose instability to saturate, as observed empirically by~\citet{Melville_2016}, becomes much smaller than $\tau$, and a state distinct from the ultra-high-beta one is realised. After a time ${\sim}t_{\rm sat}$ has passed, particle scattering becomes efficient enough to regulate the pressure anisotropy to the marginal value of the long-wavelength firehose instability ($\Delta_{\rm sat} \simeq -2/\beta_{i}$) as well as inhibit further growth of magnetic perturbations. Using shearing-box simulations of collisionless plasmas, \citet{Melville_2016} found that the effective collisionality $\nu_{\rm eff}$ in this state was given approximately by $\nu_{\rm eff} \simeq 0.5 S \beta$, where $S \simeq 1/\tau$ is the rate of shear (i.e., the stretching rate of the magnetic field by the incompressible flow). This effective collisionality gives rise to an effective Braginskii viscosity in the plasma given by $\mu_B \simeq \tau B^2/4 \upi$. 

As for the magnetic-field perturbations themselves, the key difference between the Alfv\'en-inhibiting and ultra-high-beta states is that the characteristic magnitude of the perturbed energy in the former is much smaller than the energy of the background field. However, the precise scaling of $\delta B^2/B_0^2$ with $\beta_i$, $\tau$, and $\Omega_i$ in the Alfv\'en-inhibiting state remains unclear based on relevant studies to date. In their high-$\beta$ shearing-box simulations, both \citet{Kunz2014_b} and \citet{Melville_2016} found empirically that $\delta B^2/B_0^2 \sim (\beta_i/\tau \Omega_i)^{1/2} \ll 1$ over a range of $\beta_i$ and $\tau \Omega_i$, while the HEB simulation study reported in section \ref{sec:sims} of this paper instead obtains $\delta B^2/B_0^2 \sim \beta_i/\tau \Omega_i \ll 1$. One plausible explanation for the discrepancy in these scalings is that our simulation study covers characteristically smaller values of $\beta_i$ and larger values of $\tau$ than considered by \citet{Kunz2014_b} and \citet{Melville_2016}, with only some overlap. The smallest values of $\delta B^2/B_0^2$ in \citet{Kunz_2014a} and \citet{Melville_2016} are comparable to the largest values that we observed in the simulations described in section \ref{sec:sims}, and over this (albeit limited) range, we see evidence of a flatter power-law dependence of $\delta B^2/B_0^2$ on $\beta_i/\tau \Omega_i$ emerging at sufficiently small values of this parameter in our simulations. This would seem to suggest that mechanisms whose efficacy scales strongly with mode amplitude, such as nonlinear mode-coupling or trapping, could start to affect the saturation of the firehose instability at small enough values of $\beta_i/\tau \Omega_i$. Another possibility is the contribution of long-wavelength firehose modes to the total energy budget in the shearing-box simulations; such modes, which are inefficient at causing the pitch-angle scattering of particles, can nonetheless grow significantly if the pressure anisotropy attains a value $\Delta_i \ll -2/\beta_i$ at the time $t_{\rm nl}$ at which secular growth begins, which was generically the case in the prior shearing-box studies and also in a few of our expanding-box simulations. { A third possibility is that, for values of $\tau \Omega_i$ that are only a few orders of magnitude larger than unity, in which saturation occurs on timescales comparable to $\tau$, the type of macroscopic motion that generates pressure anisotropy affects that saturation (see section \ref{sec:applications} for further discussion of this issue). In particular, for the uni-directional expansions we simulate, flux conservation implies that the out-of-plane component of the perturbed magnetic field decreases at the same rate as the macroscopc field, whereas for a two-dimensional shear, the out-of-plane component remains constant. This would give rise to larger values of $\delta B^2/B^2_0$.} Irrespective of the precise scaling of $\delta B^2/B^2_0$ with $\beta/\tau\Omega_i$, in all of the simulations of Alfv\'en-inhibiting states discussed in this paper, the saturated firehose fluctuations satisfy $\delta B^2/B_0^2 \ll 1$ and evidence of a broad spectrum of modes (including long-wavelength modes) is observed.

\subsubsection{Alfv\'en-enabling: $\tau \gtrsim \tau_{\rm cr}(\beta_i)$} \label{sec:AEoverview}

Finally, if $\tau$ exceeds a critical, $\beta_i$-dependent `transition' timescale $\tau_{\rm cr} = \tau_{\rm cr}(\beta_i)$, then another qualitatively distinct state is realised. The key property that
underpins the transition between the Alfv\'en-inhibiting state and this, third, Alfv\'en-enabling state is the wavenumber dependence of the 
firehose instability's threshold: `kinetic' ion-Larmor-scale 
firehose modes are destabilised at smaller characteristic pressure anisotropies than are longer-wavelength firehose modes (see section~\ref{linfiretheory_review}). The instability threshold of the oblique ion-Larmorf-scale firehose modes implies the existence of a timescale $\tau_{\rm cr}$ such that only ion-Larmor-scale firehose modes ever become unstable if $\tau\gtrsim \tau_{\rm cr}$ (see section~\ref{linfiretheory_bitimescaletheory} for a more extended demonstration of this). The condition arises because oblique ion-Larmor-scale firehose modes can undergo significant exponential growth --  and thereby backreact on the plasma -- before a broad spectrum of firehoses modes develops if the pressure anisotropy of the plasma is driven at a slower rate than the  characteristic linear growth rate of the oblique ion-Larmor-scale firehoses. This transition timescale then determines whether the Alfv\'en-enabling or Alfv\'en-inhibiting state is realised. In general, $\tau_{\rm cr}$ is a monotonically increasing function of $\beta_i$. For certain ranges of $\beta_i$, simplified expressions for $\tau_{\rm cr}$ can be determined using analytic approximations for the growth rate of oblique ion-Larmor-scale firehose modes. For plasma with values of $\beta_{i}$ that are not too large ($\beta_{i} \ll 10^5$), we find that $\tau_{\rm cr} \propto \beta_{i}^{1.6} \Omega_i^{-1}$ [cf. (\ref{bifurcationtimescale_modhighbeta})]; for $\beta_i$ very large ($\beta_{i} \gg 10^5$), $\tau_{\rm cr} \propto \beta_{i}^{3/2} \log{\beta_i} \, \Omega_i^{-1}$ [cf. (\ref{bifurcationtimescale_highbeta})]. 

Although there are some commonalities, the Alfv\'en-enabling state differs qualitatively from both the ultra-high-beta and Alfv\'en-inhibiting states in several key regards. Macroscopically, the saturated pressure anisotropy attains a value $\Delta_{\rm sat} \simeq -1.6/\beta_{i}$ that simultaneously marginalises kinetic-scale firehose modes while allowing long-wavelength, linear Alfv\'enic modes to be \emph{stable and propagate} (thus, the moniker `Alfv\'en-enabling'). Microphysically, the firehose-induced effective collisionality $\nu_{\rm eff} \simeq 0.4 \beta_i/\tau$ efficiently regulates the pressure anisotropy, similarly to the Alfv\'en-enabling state, with associated Braginskii viscosity $\mu_{\rm B} \simeq 0.8 \tau B^2/4 \upi$. However, the fundamental nature of the firehose modes themselves differ in the Alfv\'en-enabling state, with the wavelengths of all modes being comparable to the ion-Larmor scale. These modes that are present can be categorised into two populations: oblique firehose modes; and a newly identified population of `secondary' parallel firehose modes, which are initially damped but are then destabilised by the backreaction of the oblique firehose modes on the ion distribution function. The resulting distribution function is notable in not being a bi-Maxwellian, which, in turn, accounts for why $\Delta_{\rm sat} \simeq -1.6/\beta_{i}$ does not attain the marginal stability value for a bi-Maxwellian distribution ($\Delta_i \simeq -1.35/\beta_i$). The presence of the secondary parallel firehose modes -- which generically have a larger amplitude than the oblique modes -- gives rise to the scaling of the perturbed field energy, on account of their distinct saturation mechanism: $\delta B^2/B_0^2 \sim \beta_i^{1/4} (\tau \Omega_i)^{-1/2}$. 

\subsection{Why an Alfv\'en-enabling state is realised when $\tau \gtrsim \tau_{\rm cr}$} \label{linfiretheory_bitimescaletheory}

Although the reduced instability threshold for ion-Larmor-scale firehose modes had been identified previously, and while Alfv\'en-enabling states have been observed in simulations~\citep[see, e.g.,][]{Hellinger_2008,Bott_2021}, we are not aware of any existing theories explaining the physics underpinning the transition in high-$\beta$ plasma between the Alfv\'en-inhibiting and Alfv\'en-enabling states. We therefore outline such a theory here, based on our results from section~\ref{linfiretheory_numres}. 

In a plasma in which $\Delta_i$ is driven increasingly negative at a sufficiently slow rate, resonant oblique firehose modes can grow significantly and regulate the pressure anisotropy before $\Delta_i$ becomes negative enough for fluid firehose modes to be destabilised. More specifically, if the characteristic timescale ${\sim}\gamma_{\rm \perp f}^{-1}$ over which the resonant oblique firehose modes grow linearly is much smaller than the time interval $\Delta t$ over which the pressure anisotropy would be driven by the macroscopic evolution from $\Delta_i \simeq -1.35/\beta_{\|i}$ to $\Delta_i = -2/\beta_{\|i}$ (viz., $\gamma_{\rm \perp f} \Delta t \gg 1$),  
then the growth of these modes will regulate $\Delta_i$ before $\Delta_i$ becomes $\leq -2/\beta_{\|i}$. In this case, an Alfv\'en-enabling state will persist. If, by contrast, $\gamma_{\rm \perp f} \Delta t \ll 1$, then resonant oblique firehose modes will not have had the chance to grow before the plasma attains $\Delta_i\le -2/\beta_{\|i}$ and (linear) Alfv\'en waves  no longer propagate. 

In the case when $\Delta_i$ is driven linearly in time over a timescale $\tau$ (i.e., $\Delta_i \approx -t/\tau$, where $t = 0$ is defined as the time at which the pressure is isotropic), the condition for the Alfv\'en-enabling state to result is $\gamma_{\rm \perp f} \Delta t ={0.65 \gamma_{\rm \perp f} \tau}/{\beta_{i}} \gg 1$. The transition timescale $\tau_{\rm cr}$ is then the characteristic value of $\tau$ at which $\gamma_{\rm \perp f} \Delta t \approx N_{\rm fold}$, where $N_{\rm fold}$ is an order-unity factor equal to the number of e-folding times of the instability required for the resonant oblique firehoses to backreact on the plasma (in our simulations, we find $N_{\rm fold} \approx 5)$. This implies that
\begin{equation}
\tau_{\rm cr}(\beta_{i}) \approx 1.5 N_{\rm fold} \beta_{i} {\gamma}_{\rm \perp f}^{-1} .\label{bifurcationtimescale_general}
\end{equation}
Because ${\gamma}_{\rm \perp f}$ decreases with $\beta_{i}$, we conclude that $\tau_{\rm cr}$ monotonically increases as $\beta_{i}$ does. 
Then using the simplified expressions for the growth rate of resonant oblique firehose modes given in section \ref{linfiretheory_numres}, explicit expressions for $\tau_{\rm cr}$ as a function of $\beta_{i}$ can be found in various different parameter regimes. When $\beta_ i \gg 10^{6}$, substituting (\ref{firehosescaling_fluidthresholdreg_exact_highbeta}) into (\ref{bifurcationtimescale_general}) gives
\begin{equation}
\tau_{\rm cr}(\beta_{i}) \approx 0.9 N_{\rm fold} \beta_{i}^{3/2} \log{\left(\frac{27 \upi \beta_i}{10}\right)}   \Omega_i^{-1}\quad \mathrm{(for \; very \; large \;} \beta_{i} \mathrm{)} . \label{bifurcationtimescale_highbeta}
\end{equation}
By contrast, if $\beta_i$ satisfies $1 \ll \beta_{i} \ll 10^{5}$, then we instead substitute in the empirical scaling (\ref{firehosescaling_fluidthresholdreg_approx}) into \eqref{bifurcationtimescale_general} to obtain
\begin{equation}
\tau_{\rm cr}(\beta_{\|i}) \approx 5 N_{\rm fold} \beta_{i}^{1.6} \Omega_i^{-1} \quad \mathrm{(for \;} 1 \ll \beta_i \ll 10^{5}\mathrm{)} . \label{bifurcationtimescale_modhighbeta}
\end{equation}
In both regimes, the transition timescale is much greater than the ion gyroperiod. 
However, in plasmas where the ratio of the macroscopic evolution timescale $\tau$ to the ion gyroperiod is many orders of magnitude larger than $\beta_i$ -- as is often the cases in astrophysical plasmas of interest (see section~\ref{sec:applications}) -- both (\ref{bifurcationtimescale_highbeta}) and (\ref{bifurcationtimescale_modhighbeta}) imply that $\tau > \tau_{\rm cr}$, with the consequence that such plasmas will always end up in an Alfv\'en-enabling state.

We can also make specific predictions for the relationship between the parameter $\tau$ and the minimum value $(\Delta_i)_{\rm min}$ of the pressure anisotropy attained when $(\Delta_i)_{\rm min}$ is close to $-2/\beta_{\|i}$. For example, assuming that 
the first minimum of $\Delta_i$ is attained when oblique, kinetic-scale firehose fluctuations begin to modify the equilibrium -- {\it viz.}, when $\gamma_{\rm \perp f} \Delta t \approx N_{\rm fold}$ -- it follows that $(\Delta_i)_{\rm min} \approx -1.35/\beta_i - N_{\rm fold}/(\gamma_{\rm \perp f}\tau)$. Then, in the case of moderately large $\beta_i$ ($1 \ll \beta_{i} \ll 10^{5}$), equation~(\ref{firehosescaling_growthratefluidthreshold_approx}) for the peak growth rate of oblique firehose modes when $(\Delta_i)_{\rm min}$ is close to $-2/\beta_{\|i}$ implies that 
\begin{equation}
(\Delta_i)_{\rm min} \approx \Delta_{\rm cr} - 1.8 \left(\frac{N_{\rm fold}}{\tau \Omega_i}\right)^{0.625} . \label{minfirehosethreshold_scalepredicition}
\end{equation}
This states that the difference $(\Delta_i)_{\rm min} - \Delta_{\rm cr}$ does not depend on $\beta_{\|i}$ in this parameter regime, instead being proportional to $(\tau \Omega_i)^{-0.625}$. Another corollary of (\ref{minfirehosethreshold_scalepredicition}) is that $(\Delta_i)_{\rm min} \beta_i$ is not a function of $\tau$, $\Omega_i$ and $\beta_{i}$ independently, but rather only of the specific combination $\tau/\tau_{\rm cr}$.

\subsection{Summary and the rest of this paper}

In this section, we have summarised the possible states that can be realised in firehose-saturated, high-$\beta$ plasmas. These claims obviously require careful justification with recourse to nonlinear analytical studies and/or simulations. While such a study on the transition between ultra-high-beta and Alfv\'en-inhibiting states has already been completed by~\citet{Melville_2016}, no prior study has been done of the analogous transition between the Alfv\'en-inhibiting and Alfv\'en-enabling states. We have carried out such a study and report its results in sections~\ref{sec:sims},~\ref{sec:simtheoryinterp} and~\ref{sec:collisionality}. Readers who are happy to take such results on trust, and are instead keen to consider the relationship of our theory of firehose saturation with previous studies and the implications for astrophysical high-$\beta$ plasmas, are encouraged to skip forward to section~\ref{sec:applications}. 

\section{Kinetic simulations of firehose-susceptible high-$\beta$ plasmas} \label{sec:sims}

\subsection{Overview}

While the existence of Alfv\'en-enabling and Alfv\'en-inhibiting states in firehose-susceptible plasmas can be predicted via the linear theory of the firehose instability, determining the equilibrium properties of these two states necessitates modelling the firehose's nonlinear saturation across a range of different parameters (e.g., $\beta_{\|i}$, $\tau$) as pressure anisotropy is driven by a plasma's macroscopic evolution. This is most effectively done numerically. In this section, we first explain why so-called Hybrid Expanding-Box (HEB) simulations are particularly well suited to this purpose, and describe the method underpinning them. Then, we outline the results of a parameter study of numerous such simulations, characterising the time evolution of quantities such as the pressure anisotropy $\Delta_i$, the effective Alfv\'en speed $v_{\rm A,eff}$, and the magnetic-field strength $\delta B_{\rm f}$ of the firehose fluctuations. This, in turn, allows us to determine the equilibrium thermodynamic and microphysical properties of the Alfv\'en-enabling and Alfv\'en-inhibiting states, respectively.

Our key finding is that, in expanding, high-$\beta_i$ plasmas with an expansion time $\tau_{\rm exp}$ (see section \ref{sec:sims:why_exp}) that satisfies $\tau_{\rm exp} \gg \tau_{\rm cr}(\beta_{\|i})$ ({\it viz.}, the `asymptotic' Alfv\'en-enabling state), the pressure anisotropy is regulated to a value $\Delta_i \simeq -1.6/\beta_{\|i}$ that is above the value $\Delta_i = -2/\beta_{\|i}$ at which Alfv\'en waves cease to propagate. By contrast, if $\beta_i \Omega_i^{-1} \ll \tau_{\rm exp} \ll \tau_{\rm cr}(\beta_{\|i})$ ({\it viz.}, an Alfv\'en-inhibiting state), then $\Delta_i \simeq -2/\beta_{\|i}$. We also show that the firehose fluctuations are qualitatively distinct in the two regimes. In the Alfv\'en-inhibiting state, a broad spectrum of magnetic fluctuations (including long-wavelength modes) is excited; in the Alfv\'en-enabling state, magnetic energy is primarily concentrated in fluctuations at ion-Larmor scales. In the latter case, there are two types of modes: oblique  modes, and parallel ion-Larmor-scale modes. The latter are not, in fact, resonant parallel firehose modes of the conventional type, but are instead a secondary instability associated with the (non-bi-Maxwellian) ion distribution that is created by resonant scattering of suprathermal ions by the oblique firehose modes. That the saturated value of $\Delta_i$ is somewhat more negative (${\simeq}-1.6/\beta_{\|i}$) than the linear threshold of the resonant oblique firehose instability in a bi-Maxwellian plasma (${\simeq}-1.35/\beta_{\|i}$) can also be attributed to the non-bi-Maxwellian form of the ion distribution function in saturation. 

Finally, we characterise the velocity-space-averaged effective collisionality $\nu_{\rm eff}$ for firehose-susceptible plasmas in both the Alfv\'en-enabling and the Alfv\'en-inhibiting states (see section \ref{sec:boxavcollisionality}). We confirm that, for all of our expanding-box simulations, $\nu_{\rm eff} \sim \beta_{i}/\tau_{\rm exp}$, in agreement with previous shearing-box simulations of the firehose instability that are not in the ultra-high-beta regime~\citep{Kunz2014_b,Melville_2016}. We then provide quantitative estimates of the plasma's effective parallel Braginskii viscosity $\mu_{\rm B}$ in our simulations. 

\subsection{Simulation set-up}

\subsubsection{Why simulate an expanding plasma?} \label{sec:sims:why_exp}

As was mentioned in the Introduction, a range of different macroscopic bulk-flow fluid motions -- including shearing and expanding motions -- can give rise to negative ion pressure anisotropy ($\Delta_i < 0$) in collisionless, magnetised plasma. To see this in more detail, let us assume that the parallel and perpendicular pressures evolve according to the double-adiabatic equations ({\it viz.}, the CGL equations (\ref{CGL_withcoll}) after dropping the heat fluxes and effective collisionality):
\begin{equation}\label{eqn:cgl}
\frac{\mathrm{d}}{\mathrm{d}t} \log{\frac{p_{\perp}}{nB}} = 0 \, , \quad \frac{\mathrm{d}}{\mathrm{d}t} \log{\frac{p_{\|} B^2}{n^3}} = 0 \, .
\end{equation}
In any collisionless plasma governed by these equations whose initial temperature is isotropic ({\it viz.}, $T_{\|} = T_{\perp }$ at some time $t = 0$), the pressure anisotropy satisfies
\begin{equation}
\Delta_i = \frac{B^3/B_0^3}{n^2/n_0^2} - 1 \, ,
\end{equation}
where the subscript `$0$' is from here on used to denote the value of a  quantity at $t = 0$. Thus, the ion pressure anisotropy will decrease in any double-adiabatic plasma in which $B^3/n^2$ decreases due to the plasma's macroscopic evolution. 

Of the motions that cause $B^3/n^2$ to decrease, a particularly advantageous one to choose for our purposes is that of spatially uniform expansion at a constant rate in one direction transverse to the mean magnetic field. { There are a few different physical situations in which this type of expansion could arise: during the motion of a macroscopic, linearly polarised magnetosonic wave travelling perpendicularly to the  background magnetic field; in certain regions of compressive turbulence; and, finally, the expansion of cylindrical, magnetised plasma -- for example, generated by an exploding wire array in a laboratory astrophysics experiment.} Assuming that expansion occurs at a rate $1/\tau_{\rm exp}$, where $\tau_{\rm exp}$ is the expansion time, it follows that $B \propto n = n_0/(1+t/\tau_{\rm exp})$, and so
\begin{equation}
\Delta_i = \frac{n}{n_0} - 1 = -\frac{t}{t+\tau_{\rm exp}} \, . \label{presanisop_DA}
\end{equation}
{ Beyond studying specific physical systems,} there are three pragmatic reasons for this choice of motion in order to study firehose instabilities. First of these is the possibility of simulating such an expansion exactly via a coordinate transform method, allowing for a simulation domain that is both fixed and homogeneous to be used. Using a coordinate transform method (of which a shearing box is another example) maximises the effective separation between macro- and micro-scales for a fixed simulation domain size; it also allows for simulation-domain-averaged properties of the plasma (including the ion distribution function) to be used as a reasonable analogue for that plasma's `equilibrium' properties, minimising the uncertainty that could be introduced by macroscopic spatial variation of the plasma. The second reason is that, in contrast to a shearing-box simulation~\citep[e.g.,][]{Kunz2014_b}, the direction of the macroscopic magnetic field does not change as the motion proceeds, which simplifies comparing different times in the simulation. 
Finally, compared to other `simple' motions, an expansion in a direction transverse to the background magnetic field gives rise to a comparably slow evolution of the pressure anisotropy over a fixed period of time. For example, a two-dimensional incompressible motion in which there is simultaneously expansion in one direction transverse to the background magnetic field and and contraction in the parallel direction, causing the background magnetic-field strength to vary as $B = B_0/(1+t/\tau_{\rm exp})$, would give rise to a value of $|{\mathrm{d} \Delta_i}/{\mathrm{d}t}|$ that is initially three times larger than the analogous one-dimensional transverse expansion. As we show in section~\ref{sec:sims:res:pressanisop}, accessing the Alfv\'en-enabling regime when $\beta_{i} \gg 1$ requires macroscopic evolution rates that are at least several orders of magnitude smaller than the ion-Larmor frequency $\Omega_i$; because such simulations are expensive, choosing a motion that minimises the rate of change of $\Delta_i$ at a fixed time period is desirable. 
Motivated by these considerations, we choose in this paper to simulate plasmas expanding in a single transverse direction. { In addition to its application to the specific physical systems mentioned earlier in this paragraph, we anticipate that the evolution and saturation of the firehose instability becomes insensitive to the specifics of the macroscopic motion driving it provided there is sufficient separation of relevant timescales; we revisit this assumption after describing our simulation results in section \ref{sec:applications}.}

\subsubsection{Hybrid Expanding Box (HEB) simulations with {\tt Pegasus++}}

To capture all ion firehose instabilities correctly, the plasma's ions (but not necessarily the electrons) must be modelled kinetically. We therefore choose to conduct HEB simulations. Although this approach and its implementation have been described elsewhere~\citep[e.g.,][]{Hellinger_2005,Bott_2021}, we explain the method here for the completeness. All HEB simulations reported in this paper were carried out using the PIC code {\tt Pegasus++}~\citep{Kunz_2014a,Bott_2021}.

In the HEB approach, one transforms from the locally co-moving frame of the expanding plasma to a co-moving frame in which the metric extends as the plasma expands, and then performs all subsequent calculations in this expanding frame. Denoting position in the co-moving, non-expanding frame by $\bb{r}$, and in the co-moving, expanding frame by $\bb{r}'$, the frame transformation is characterised by a matrix $\metric\equiv \partial \bb{r}/\partial \bb{r}'$, with determinant $\lambda = \mathrm{det} \, \metric$. For HEB PIC simulations using {\tt Pegasus++}, we evolve two sets of equations: those describing the motion of ion macroparticles, and those describing the evolution of electromagnetic fields. 
The former, which constitute evolution equations for the primed-frame positions $\bb{r}'_p=\metric^{-1}\bb{r}_p$ and velocities $\bb{v}'_p = \metric^{-1} \bb{v}_p$ of macroparticles, are given by
\begin{subeqnarray}
    \frac{\mathrm{d} \bb{r}'_p}{\mathrm{d}t'} &=& \bb{v}'_p , \label{eqn:drdt} \\
    \frac{\mathrm{d}\bb{v}'_p}{\mathrm{d} t'} &= &  \frac{e}{m_i} \, \metric^{-2} \left[ \bb{E}'(t',\bb{r}'_p) + \frac{\bb{v}'_p}{c}\btimes\bb{B}'(t',\bb{r}'_p)\right] - 2 \metric^{-1}\frac{\mathrm{d}\metric}{\mathrm{d}  t'}\,\bb{v}'_p , \label{eqn:dvdt}
\end{subeqnarray}
where the fields $\bb{E}'$ and $\bb{B}'$ are related to the physical electric field $\bb{E}$ and magnetic field $\bb{B}$ in the unprimed frame via\footnote{The fields $\bb{E}'$ and $\bb{B}'$ are not the physical transformations of $\bb{E}$ and $\bb{B}$ into the primed frame, but are instead convenient proxy fields to evolve.}
\begin{equation}
\bb{B}' = \lambda \metric^{-1} \bb{B} \quad{\rm and}\quad \bb{E}' = \metric
\bb{E} .
\end{equation}
To solve (\ref{eqn:dvdt}), {\tt Pegasus++} employs a straightforward modification of the Boris push that groups the velocity-dependent non-inertial force with the $\bb{v}'_p \btimes\bb{B}$ rotation.
The fields $\bb{B}'$ and $\bb{E}'$ in turn satisfy modified versions of Faraday's law and a generalised Ohm's law, respectively:\footnote{The (transformed) Hall term, $[\bnabla'\btimes (\metric^2\bb{B}')]\btimes{\bb{B}'}/{4\upi e n'\lambda}$, in the Ohm's law was incorrectly reported as $(\bnabla'\btimes \bb{B}')\btimes{\metric^2 \bb{B}'}/{4\upi e n'\lambda}$ in both~\citet{Hellinger_2005} and~\citet{Bott_2021}. This error was not replicated in {\tt Pegasus++} itself, neither here nor for \citet{Bott_2021}.}
\begin{subeqnarray}
\frac{\partial \bb{B}'}{\partial t'} & = & -c\bnabla'\btimes\bb{E}' , \\
\bb{E}' & = & -\frac{\bb{u}'}{c}\btimes\bb{B}' - \frac{T_e}{en'}\nabla' n' + [\bnabla'\btimes (\metric^2\bb{B}')]\btimes\frac{\bb{B}'}{4\upi e n'\lambda} .
\end{subeqnarray} 
Fluid quantities in the primed frame are calculated in the usual way by taking moments of the primed-frame ion distribution function. This is done by summing up the (weighted) phase-space contributions from each ion macroparticle of shape $S$ centered on the phase-space position $(\bb{r}'_p,\bb{v}'_p)$ to the phase-space position ($\bb{r}',\bb{v}'$). For example, the primed-frame ion density $n'$ and primed-frame ion-flow velocity $\bb{u}'$, which are related to their unprimed analogues via $n' = \lambda n$ and $\bb{u}' = \metric^{-1} \bb{u}$, are computed via
\begin{equation}
n'(\bb{r}') = \sum\nolimits_p \, S[\bb{r}'-\bb{r}'_p(t')] \quad{\rm and}\quad \bb{u}'(\bb{r}') = \frac{1}{n'} \sum\nolimits_p \, \bb{v}'_p \, S[\bb{r}'-\bb{r}'_p(t')] .
\end{equation}
At any given time, physical variables can be computed directly from the primed-frame variables using the appropriate inverse coordinate transform. 

For our simulations, we set
\begin{equation}
\metric(t) = \left(1+\frac{t}{\tau_{\rm exp}}\right)\hat{\bb{x}} \hat{\bb{x}} + \hat{\bb{y}} \hat{\bb{y}} + \hat{\bb{z}} \hat{\bb{z}} , \label{Lambda_sims}
\end{equation}
where $\{\hat{\bb{x}},\hat{\bb{y}},\hat{\bb{z}}\}$ is a set of basis vectors of an orthogonal coordinate system in which $\hat{\bb{z}}$ is parallel to $\bb{B}(t = 0)$ (and remains parallel to the mean `guide' field in the simulation domain throughout). In terms of the evolution of the side lengths $[L_x(t),L_z(t)]$ of {a two-dimensional spatial} domain, equation~\eqref{Lambda_sims} gives
\begin{equation}
L_x(t) = L_{x0}  \left(1+\frac{t}{\tau_{\rm exp}}\right) , \quad L_z(t) = L_{z0} .
\end{equation}
We define the effective expansion time $\tau_{\rm exp,eff}$ via
\begin{equation}
\tau_{\rm exp,eff} \equiv \left(\D{t}{}\log{\frac{B^3}{n^2}}\right)^{-1}= \left(\D{t}{}\log{B}\right)^{-1} = t + \tau_{\rm exp} \, .
\end{equation}
We choose this definition for three reasons. First, in the limit of small pressure anisotropy, it is a simple matter to show that $\mathrm{d} \Delta_i/{\mathrm{d} t} \approx \mathrm{d} (\Delta_i \beta_{\|i})/{\mathrm{d} t} \approx -1/\tau_{\rm exp,eff}$ as the plasma expands. Secondly, in the saturated phase of the firehose instability, it can be shown that the box-averaged effective collisionality associated with firehose fluctuations is inversely proportional to $\tau_{\rm exp,eff}$ (see section \ref{sec:boxavcollisionality}).
Finally, although $\tau_{\rm exp,eff}$ increases in time, the ion cyclotron frequency $\Omega_i \propto B = B_0 \tau_{\rm exp}/\tau_{\rm exp,eff}$ decreases in such a way that their product is constant: $\tau_{\rm exp,eff} \Omega_i =  \tau_{\rm exp} \Omega_{i0}$. 

We ran numerous HEB simulations of this type with different values of $\tau_{\rm exp}$ and $\beta_{i0}$. { We chose to perform these simulations in a 2.5D geometry -- that is, particles move in three dimensions and the electromagnetic fields are three-dimensional, but spatial gradients are restricted to the two-dimensional $(x,z)$ plane -- because such simulations can capture the relevant physics at significantly reduced computational costs. Periodic boundary conditions were applied in all spatial directions.}
Table~\ref{tab:simulations} outlines the key parameters of all of the simulations reported in the paper. 
\begin{table}
  \begin{center}
\def~{\hphantom{0}}
  \begin{tabular}{lcccccccc}
     Run  & $\beta_{i0}$  & $\tau_{\rm exp} \Omega_{i0}$   & $\tilde{\tau}_0$ & $\beta_{i}({t_{\rm min}})$ & $\tilde{\tau}_{\rm eff}$ & $N_{\rm ppc}$ & { $t_{\rm end}/\tau_{\rm exp}$} & { $t_{\rm end} \Omega_{i0}$}  \\[3pt]
       A   & ~~6~ & ~~2$ \times 10^3$ & 4.2~ & ~~~7.77 & 2.8~ & $5 \times 10^{3}$ & { 2.1~ } & { $4.2 \times 10^3$} \\
       BI   & ~12 & ~~5$ \times 10^2$ & 0.35 & ~15.1 & 0.24 & $5 \times 10^{3}$ & {5.2~ } & { $2.6 \times 10^3$} \\
       BII  & ~12 & ~~2$ \times 10^3$ & 1.4~ & ~14.1 & 1.1~  & $5 \times 10^{3}$ & { 3.0~ } & { $6.0 \times 10^3$}  \\
       BIII   & ~12 & ~~8$ \times 10^3$ & 5.6~ & ~13.7 & 4.5~ & $1 \times 10^{4}$ & { 1.9~ } & { $1.5 \times 10^4$} \\
       BIV   & ~12 & ~~2$ \times 10^4$ & 14~~~ & ~13.6 & 11~~~ & $1 \times 10^{4}$ & { 1.0~ } & { $2.0 \times 10^4$}\\
       CI & ~25 & ~~2$ \times 10^3$ & 0.43 & ~27.6 & 0.37 & $1 \times 10^{4}$ & { 0.72 }  & { $1.4 \times 10^3$} \\
       CII & ~25 & ~~5$ \times 10^3$ & 1.1~ & ~27.0 & 0.95 & $1 \times 10^{4}$ & { 1.0~ } & { $5.0 \times 10^3$} \\
       CIII & ~25 & ~~8$ \times 10^3$ & 1.7~ & ~26.9 & 1.5~ & $1 \times 10^{4}$ & { 0.72 }  & { $5.8 \times 10^3$} \\
       CIV & ~25 & ~~2$ \times 10^4$ &  4.3~ & ~26.7 &  3.9~ & $1 \times 10^{4}$ &{ 0.66 } & { $1.3 \times 10^4$} \\
       CV & ~25 & ~~5$ \times 10^4$ & 11~~~ & ~26.6 & 9.7~ & $2 \times 10^{4}$ & { 0.49 }  & { $2.5 \times 10^4$} \\
       DI & ~50 & ~~2$ \times 10^2$ & 0.01 & ~65.4 & 0.01 & $1 \times 10^{4}$ & { 3.1~ }  & { $6.3 \times 10^2$} \\
       DII & ~50 & ~~5$ \times 10^2$ & 0.04 & ~58.2 & 0.03 & $1 \times 10^{4}$ & { 2.0~ } & { $1.0 \times 10^3$} \\
       DIII & ~50 & ~~2$ \times 10^3$ & 0.14 & ~53.8 & 0.13 & $1 \times 10^{4}$ & { 2.0~ } & { $4.0 \times 10^3$} \\
       DIV & ~50 & ~~5$ \times 10^3$ & 0.35 & ~52.5 & 0.33 & $1 \times 10^{4}$ & { 0.82 } & { $4.1 \times 10^3$} \\
       DV & ~50 & 1.25$ \times 10^4$ & 0.89 & ~51.8 & 0.84 & $1 \times 10^{4}$ & { 0.75 } & { $9.4 \times 10^3$} \\
       DVI & ~50 & ~~2$ \times 10^4$ & 1.4~ & ~51.7 & 1.3~ & $1 \times 10^{4}$ & { 0.38 } & { $7.6 \times 10^3$} \\
       DVII & ~50 & ~~5$ \times 10^4$ & 3.5~ & ~51.3 & 3.4~ & $1 \times 10^{4}$ & { 0.37 } & { $1.9 \times 10^4$} \\
       EI & 100 & ~~2$ \times 10^3$ & 0.05 & 106~ & 0.04 & $1 \times 10^{4}$ & { 3.0~ } & { $6.0 \times 10^3$}\\
       EII & 100 & ~~5$ \times 10^3$ & 0.12 & 104~ & 0.11 & $1 \times 10^{4}$ & { 1.2~ } & { $6.0 \times 10^3$} \\
       EIII & 100 & 1.25$ \times 10^4$ & 0.29 & 102~ & 0.28 & $1 \times 10^{4}$ & { 0.78 } & { $9.8 \times 10^3$} \\
       EIV & 100 & ~~3$ \times 10^4$ & 0.70 & 102~ & 0.68 & $1 \times 10^{4}$ & { 0.50 } & { $1.5 \times 10^4$} \\
       FI & 200 & ~~5$ \times 10^2$ & ~0.004 & 227~ & ~0.003 & $1 \times 10^{4}$ & { 3.3~ } & { $1.7 \times 10^3$}\\
       FII & 200 & ~~1$ \times 10^3$ & 0.01 & 216~ & ~0.007 & $1 \times 10^{4}$ & { 2.7~ } & { $2.7 \times 10^3$} \\
       FIII & 200 & ~~2$ \times 10^3$ & 0.02 & 210~ & ~0.014 & $1 \times 10^{4}$ & { 2.7~ } & { $5.4 \times 10^3$} \\
       FIV & 200 & ~~5$ \times 10^3$ & 0.04 & 205~ & ~0.037 & $1 \times 10^{4}$ & { 2.0~ } & { $1.0 \times 10^4$} \\
       FV & 200 & 1.25$ \times 10^4$ & 0.10 & 203~ & ~0.094 & $1 \times 10^{4}$ & { 0.64 } & { $8.0 \times 10^3$} \\
  \end{tabular}
  \caption{Parameters of all HEB simulations performed in this study. Here, $t_{\rm min}$ is the time at which the (first) minimum of the pressure anisotropy is attained, { $t_{\rm end}$ is the time at which the simulation run was ended}, and $\tilde{\tau}_0$ and $\tilde{\tau}_{\rm eff}$ are defined by $\tilde{\tau}_0 \equiv \tau_{\rm exp} \Omega_{i0}\beta_{i0}^{-1.6}/27$, and $\tilde{\tau}_{\rm eff} \equiv \tau_{\rm exp,eff} \Omega_i({t_{\rm min}})\beta_{i}({t_{\rm min}})^{-1.6}/27$, respectively. The empirical factor of $27$ is introduced so that runs with $\tilde{\tau}_{\rm eff} \gtrsim 1$ are at all times in an Alfv\'en-enabling state (see section \ref{sec:sims:res:pressanisop}). { In units of $\rho_i$, all simulations were run with the same numerical resolution ($\Delta x = \Delta z = 0.26 \rho_i$), and with the same initial side lengths of the simulation domain ($L_{z0} = 1.5 L_{x0} = 300 \rho_i$).}}
  \label{tab:simulations}
  \end{center}
\end{table}
{ All simulations were initialised with equal parallel and perpendicular temperatures. The numerical resolution of the simulations was chosen ($\Delta x = \Delta z = 0.26 \rho_i$) such that the characteristic wavenumbers of firehose modes would be sufficiently well resolved: the maximum wavenumber $k_{\rm max}$ of modes that could be resolved in our simulations was $k_{\rm max} = \upi/\Delta z \simeq 12.1 \rho_i^{-1}$. We note that, as a result of this choice, for those of our simulations with $\beta_{i0} = 200$, the wavenumber $k = d_i^{-1}$ at which fluctuations on the scale of ion skin depth $d_i$ exist was not resolved; however, we believe that this is an acceptable limitation, because in high-$\beta$ plasmas, the characteristic scale of firehose fluctuations is $\rho_i$, not $d_i$~\citep[see, e.g.,][]{Bott_2023}. Simulations were run until what appeared to be a saturated state was obtained.}
The large number $N_{\rm ppc}$ of particles per cell used in these simulations is necessary in order to suppress the effect of numerical collisions (arising from the Poisson noise due to finite sampling of the ion distribution function) on both the evolution of the pressure anisotropy and the firehose instability itself. Even with such large values of $N_{\rm ppc}$, we find that numerical collisionality has a quantitative (but not a qualitative) effect on some of our results (see Appendix~\ref{append:collisionality}). 

\subsection{Results} \label{sec:sims:res}

\subsubsection{Box-averaged pressure anisotropy and effective Alfv\'en speed} \label{sec:sims:res:pressanisop}

\begin{figure}
  \centering
  \includegraphics[width=\linewidth]{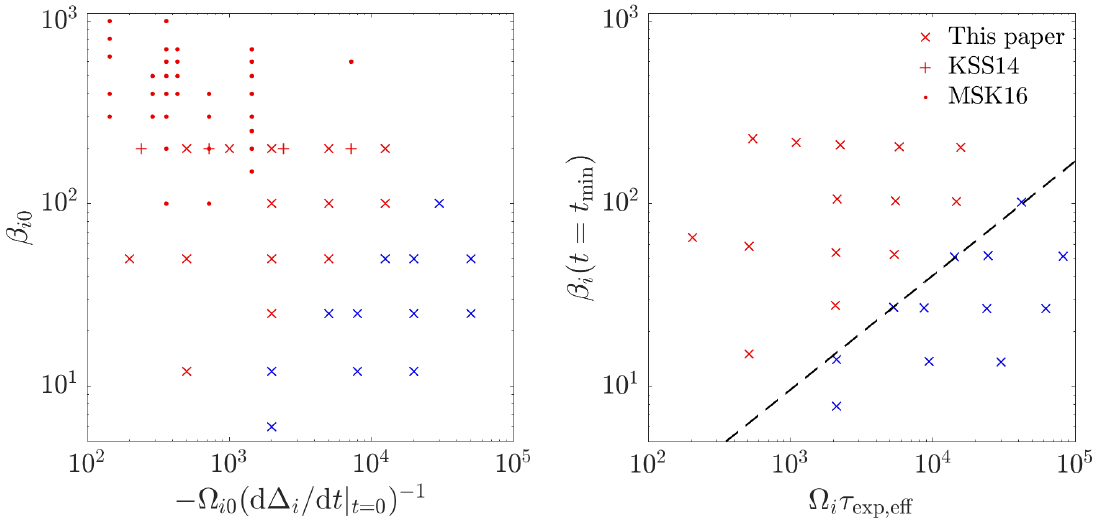}
  \caption{Phase-space maps of various simulations of high-$\beta_{i}$ firehose-susceptible plasmas, which indicate whether an Alfv\'en-enabling state ($\Delta_i > -2/\beta_{\|i}$) is maintained { at all times} (blue points) or not (red points). In the left map, we include the HEB simulations completed for this paper (denoted by `$\times$'), as well as the shearing-box hybrid-kinetic simulations reported in~\citet{Kunz2014_b} (`$+$') and~\citet{Melville_2016} (`$\bcdot$'). The dashed line that provides an accurate delineation of { consistently Alfv\'en-enabling states and other states} is given by the equation $\tau_{\rm exp} \Omega_i = 27 \beta_i^{1.6}$ [cf. figure~\ref{tab:overviewstates}, and eqn.  (\ref{bifurcationtimescale_modhighbeta})].}
\label{fig:sims_phasespace}
\end{figure}

The existence of both Alfv\'en-enabling and Alfv\'en-inhibiting states among the expanding plasmas we have simulated is illustrated in figure~\ref{fig:sims_phasespace}($a$), which positions each simulation in a $[-\Omega_i (\mathrm{d}\Delta_i/\mathrm{d}t)|_{t = 0})^{-1},\beta_{i0}] = [\tau_{\rm exp}\Omega_{i0},\beta_{i0}]$ phase space, and indicates whether the value of the ion pressure anisotropy $(\Delta_i)_{\rm min}$ attained at the first minimum (at some time $t = t_{\rm min}$) is more (red) or less (blue) negative than $\Delta_i = -2/\beta_{\|i}$. 
Qualitatively, at fixed $\beta_{i0}$ the simulated plasmas transition from being in an { Alfv\'en-enabling state (for which $(\Delta_i)_{\rm min} > -2/\beta_{\|i}$) to an Alfv\'en-inhibiting state (for which $(\Delta_i)_{\rm min} \ll -2/\beta_{\|i}$) as the expansion time is decreased.} Furthermore, the characteristic expansion time at which this transition occurs is (for the suite of simulations we have conducted) a monotonically increasing function of $\beta_{\|i}$. More quantitatively, figure~\ref{fig:sims_phasespace}($b$) shows that the scaling (\ref{bifurcationtimescale_modhighbeta}) of $\tau_{\rm cr}$ with $\beta_{\|i}$ derived in Section \ref{sec:theory} (dashed line) is an excellent fit to the measured (effective) expansion time $\tau_{\rm exp,eff}$ at which the instantaneous value of $\beta_i$ and $(\Delta_i)_{\rm min}$ satisfy $(\Delta_i)_{\rm min} \approx -2/\beta_{\|i}$. For reference, on figure~\ref{fig:sims_phasespace}($a$) we plot also the positions in the same $[-\Omega_i (\mathrm{d}\Delta_i/\mathrm{d}t)|_{t = 0})^{-1},\beta_{i0}]$ phase space of previously published high-$\beta_i$ shearing-box simulations of the firehose instability~\citep{Kunz2014_b,Melville_2016}; these simulations all realised the Alfv\'en-inhibiting state in saturation, a finding consistent with their initialised parameters. 

\begin{figure}
  \centering
  \includegraphics[width=\linewidth]{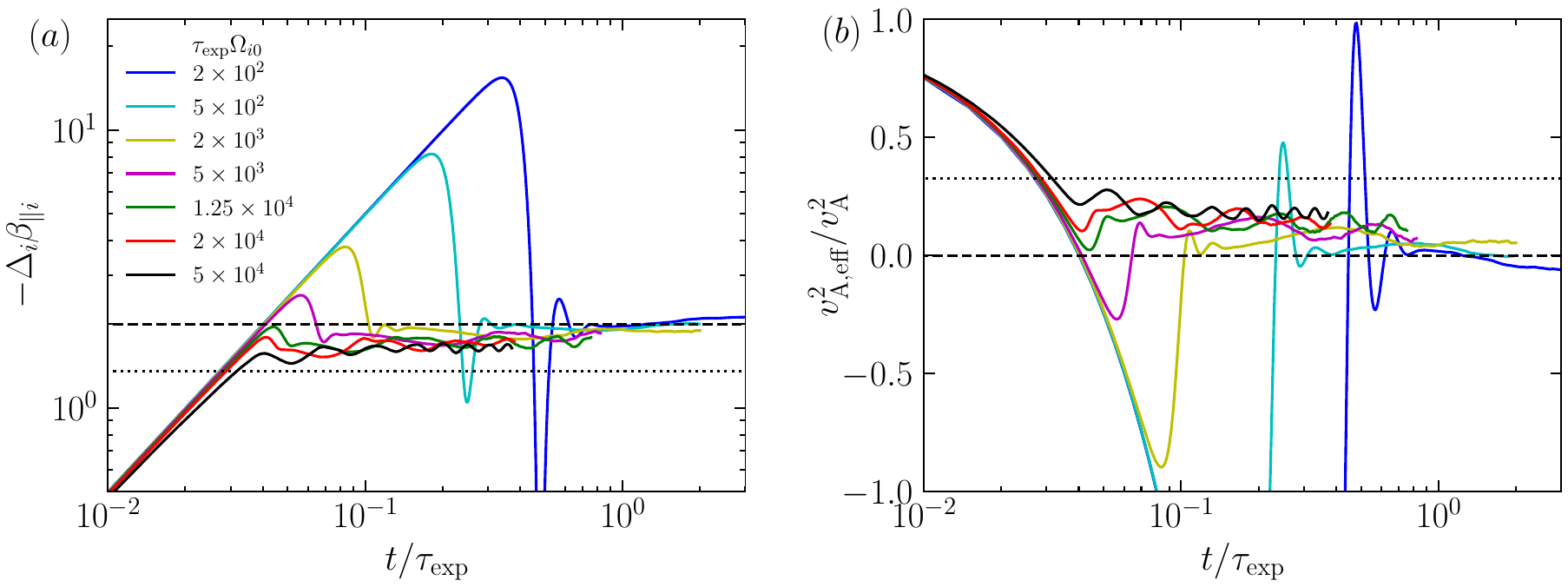}
  \caption{Time evolution of $(a)$ the firehose-instability parameter $-\Delta_i \beta_{\|i}$ and $(b)$ the squared and normalised effective Alfv\'en speed $v_{\rm A,eff}^2/v_{\rm A}^2$ for all of the D runs ($\beta_{i0} = 50$). On panel $(a)$, the dotted black line denotes the threshold $\Delta_i \beta_{\|i} = -1.35$ of the oblique firehose instability when $\beta_i \gg 1$, while the dashed black line shows the fluid firehose threshold $\Delta_i \beta_{\|i} = -2$. On panel $(b)$, the dotted (dashed) black line denotes the corresponding value $v_{\rm A,eff}^2 = 0.32 v_{\rm A}^2$ ($v_{\rm A,eff}^2 = 0 v_{\rm A}^2$) of the squared effective Alfv\'en speed at the threshold of the oblique firehose instability (fluid firehose threshold).}
\label{fig:sims_constantbetaexample}
\end{figure}

A simple way to illustrate how the evolution of the pressure anisotropy and the effective Alfv\'en speed changes as the expanding plasma transitions from being Alfv\'en-inhibiting to Alfv\'en-enabling is to fix $\beta_{i0}$, and compare the evolution of $\Delta_i \beta_{\|i}$ and $v_{\rm A,eff}^2/v_{\rm A}^2$ over time for a selection of increasing values of $\tau_{\rm exp}$. This comparison is made in figure~\ref{fig:sims_constantbetaexample}. 
It is clear from figure~\ref{fig:sims_constantbetaexample}$(a)$ that, for $\Delta_i > \Delta_{\rm cr} \simeq -1.35/\beta_{\|i}$, the initial evolution of $\Delta_i \beta_{\|i}$ is independent of $\tau_{\rm exp}$, as predicted by (\ref{presanisop_DA}).\footnote{The one exception to this is run DVII (with $\tau_{\rm exp}\Omega_{i0} = 5 \times 10^{4}$, $\beta_{i0} = 50$; black line), in which $-\Delta_i \beta_{\|i}$ seems to increase slightly less quickly than in the other runs. This is due to the cumulative effect of numerical collisionality over such a long run time (see Appendix~\ref{append:collisionality}).} However, once the oblique firehose instability is triggered, we see that for comparatively larger expansion times (e.g., black and red lines in figure~\ref{fig:sims_constantbetaexample}$(a)$), $-\Delta_i \beta_{\|i}$ stops increasing at smaller characteristic values of $t/\tau_{\rm exp}$, and attains a less positive value  $-\Delta_i \beta_{\|i}$ at the time $t_{\rm min}$ at which the first minimum of $\Delta_i$, $(\Delta_i)_{\rm min}$, is attained. At times $t > t_{\rm min}$, $-\Delta_i \beta_{\|i}$ is regulated, eventually converging to an order-unity value in all of our simulations. For the largest values of $\tau_{\rm exp}$, we find that the pressure anisotropy is ultimately regulated to values $\Delta_i \approx -1.6/\beta_{\|i}$. By contrast, for the comparatively smaller expansion times (e.g., cyan and blue lines), $\Delta_i \approx -2/\beta_{\|i}$, characteristic of an Alfv\'en-inhibiting state. These saturated values of $\Delta_i$ imply that, for the simulations with comparatively smaller expansion times that we have run, the plasma attains an Alfv\'en-inhibiting state with $v_{\rm A,eff}^2/v_{\rm A}^2 \approx 0$, while for the larger expansion times, $v_{\rm A,eff}^2/v_{\rm A}^2 \approx 0.2$ (see figure~\ref{fig:sims_constantbetaexample}$(b)$). For intermediate values of $\tau_{\rm exp}$, $(\Delta_i)_{\rm min}$ drops below $\Delta_i = -2/\beta_{\|i}$ by an $\textit{O}(1/\beta_{\|i})$ value, but the `steady-state' values of $\Delta_i$ that are subsequently attained imply that the state in these runs is, in saturation, Alfv\'en-enabling (albeit with a reduced value of $v_{\rm A,eff}^2/v_{\rm A}^2$ compared to runs in which $(\Delta_i)_{\rm min} > -2/\beta_{\|i}$).   

\begin{figure}
  \centering
  \includegraphics[width=\linewidth]{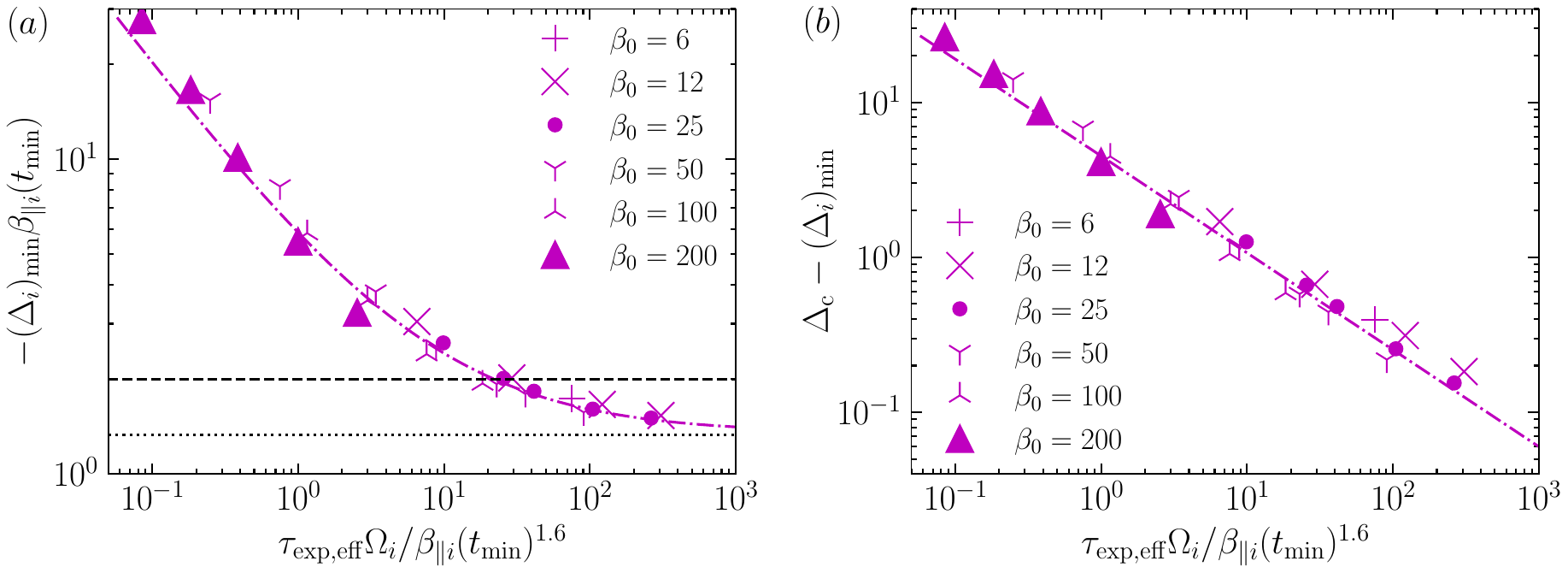}
  \caption{Panel $(a)$: values of the firehose-instability parameter $-\Delta_i \beta_{\|i}$ at the time $t_{\rm min}$ at which the pressure anisotropy attains its first minimum, $(\Delta_i)_{\rm min}$, for all runs, as a function of $\tau_{\rm exp,eff} \Omega_i/\beta_{\|i}(t_{\rm min})^{1.6}$. The dotted (dashed) black line denotes the threshold $\Delta_i \beta_{\|i} = -1.35$ of the oblique firehose instability (the fluid firehose instability threshold, $\Delta_i \beta_{\|i} = -2$) when $\beta_i \gg 1$; the dotted-dashed purple line denotes equation (\ref{minfirehosethresholdpredicition}). Panel $(b)$: values of the difference between $(\Delta_i)_{\rm min}$ and the value $\Delta_{\rm cr}$ at which the oblique firehose becomes unstable as a function of $\tau_{\rm exp,eff} \Omega_i/\beta_{\|i}(t_{\rm min})^{1.6}$, for all runs. The dotted-dashed purple line denotes equation (\ref{minfirehosethresholdpredicition}).}
\label{fig:sims_params_fig1}
\end{figure}

A key prediction of the theory outlined in section~\ref{sec:theory} is that the transition between Alfv\'en-enabling and Alfv\'en-inhibiting states in firehose-susceptible high-$\beta_{i}$ plasmas is a function of the parameter $\tau_{\rm exp,eff} \Omega_i/\beta_i^{1.6}$ (in the limit where $1 \ll \beta_i \ll 10^{5}$). We test this prediction in figure~\ref{fig:sims_params_fig1}$(a)$ by plotting for each of our simulations the relationship between $\tau_{\rm exp,eff} \Omega_i/\beta_i^{1.6}$ and $(\Delta_i)_{\rm min} \beta_{\|i}(t_{\rm min})$. 
{We see that the value of $\tau_{\rm exp,eff} \Omega_i/\beta_i^{1.6}$ is predictive of $(\Delta_i)_{\rm min} \beta_{\|i}(t_{\rm min})$ for all of our simulations, with the decreasing nonlinear relationship 
\begin{equation}
(\Delta_i)_{\rm min} \beta_{\|i}(t_{\rm min}) \approx -1.35 - 5.1 \frac{ \beta_{\|i}(t_{\rm min})}{\left(\tau_{\rm exp,eff} \Omega_i\right)^{0.625}} \label{minfirehosethresholdpredicition}
\end{equation}
between the two parameters being a good fit to our data. } This relationship is consistent with the prediction (\ref{minfirehosethreshold_scalepredicition}) that was based on the linear theory of the firehose instability (with $N_{\rm fold} \simeq 5.4$). It follows that $(\Delta_i)_{\rm min} \beta_{\|i}(t_{\rm min}) \approx -2$ when $\tau_{\rm exp,eff} \Omega_i \approx 27 \beta_i^{1.6}$. 
Furthermore, figure~\ref{fig:sims_params_fig1}$(b)$ shows that the power-law dependence of $(\Delta_i)_{\rm min} - \Delta_{\rm cr}$ on $(\tau_{\rm exp,eff} \Omega_i)^{-0.625}$ that was predicted by (\ref{minfirehosethreshold_scalepredicition}) is well satisfied.

\begin{figure}
  \centering
  \includegraphics[width=\linewidth]{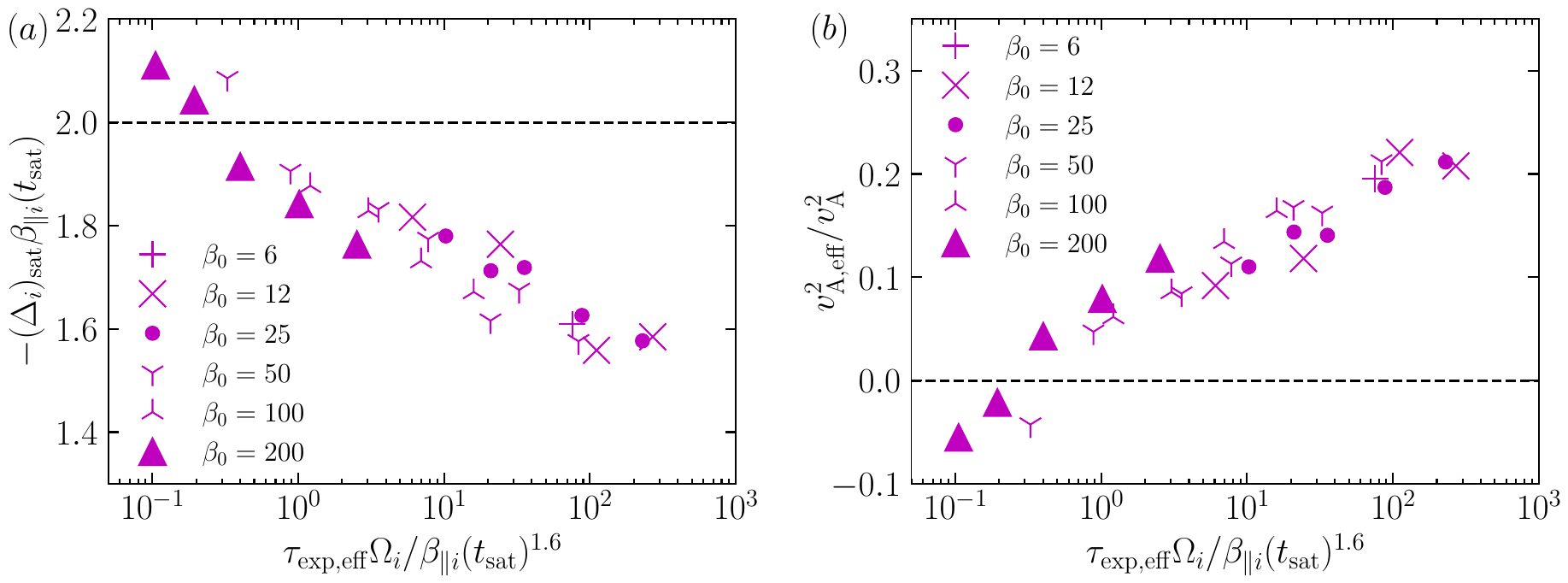}
  \caption{Panel $(a)$: values of the firehose-instability parameter $-\Delta_i \beta_{\|i}$ at the time $t_{\rm sat}$ at which the square of the perturbed magnetic-field strength $\delta B_{\rm f}^2/B_0^2$ associated with the firehose fluctuations attains its maximum value, $(\delta B_{\rm f}^2/B_0^2)_{\rm max}$, for all runs as a function of $\tau_{\rm exp,eff} \Omega_i/\beta_{\|i}(t_{\rm sat})^{1.6}$. The dashed black line denotes the fluid firehose instability threshold, $\Delta_i \beta_{\|i} = -2$, when $\beta_i \gg 1$. Panel $(b)$: values of the square of the effective Alfv\'en speed, $v_{\rm A,eff}^2/v_{\rm A}^2$ at $t = t_{\rm sat}$, for all runs as a function of $\tau_{\rm exp,eff} \Omega_i/\beta_{\|i}(t_{\rm min})^{1.6}$. }
\label{fig:sims_params_fig2}
\end{figure}

The parameter $\tau_{\rm exp,eff} \Omega_i/\beta_i^{1.6}$ also has a quasi-deterministic relationship with the values of $\Delta_i \beta_{\|i}$ and $v_{\rm A,eff}^2/v_{\rm A}^2$ in our simulations \emph{once the firehose instability has saturated}. We illustrate this in figure~\ref{fig:sims_params_fig2} by plotting 
$-\Delta_i \beta_{\|i}$ and $v_{\rm A,eff}^2/v_{\rm A}^2$ at the time $t_{\rm sat}$ at which the firehose fluctuations attain their peak magnetic-field strength; we denote the value of $\Delta_i$ attained at this time as $(\Delta_i)_{\rm sat}$.  
Figure~\ref{fig:sims_params_fig2}$(a)$ shows that, as $\tau_{\rm exp,eff} \Omega_i/\beta_{\|i}(t_{\rm sat})^{1.6}$ increases from below unity to much greater values, $(\Delta_i)_{\rm sat}\beta_{\|i}(t_{\rm sat})$ increases monotonically from a value close to $-2$ to a less negative value of {$\simeq$}$-1.6$; equivalently, $v_{\rm A,eff}^2/v_{\rm A}^2$ increases from being close to zero to {$\simeq$}$0.2$. For the simulations we have performed, we find that for $\tau_{\rm exp,eff} \Omega_i/\beta_{\|i}(t_{\rm sat})^{1.6} \gtrsim 80$, $(\Delta_i)_{\rm sat}\beta_{\|i}(t_{\rm sat})$ does not become less negative if $\tau_{\rm exp,eff} \Omega_i/\beta_{\|i}(t_{\rm sat})^{1.6}$ is increased further still (and $v_{\rm A,eff}^2/v_{\rm A}^2$ does not increase). We infer from this that such a state is the `asymptotic' Alfv\'en-enabling state for asymptotically large values of $\tau_{\rm exp,eff} \Omega_i/\beta_{\|i}(t_{\rm sat})^{1.6}$. 
Given that the relevance of the parameter $\tau_{\rm exp,eff} \Omega_i/\beta_i^{1.6}$ is derived entirely from the linear theory of the firehose instability, it is perhaps unsurprising that the correlation between $\tau_{\rm exp,eff} \Omega_i/\beta_{\|i}(t_{\rm sat})^{1.6}$ and $(\Delta_i)_{\rm sat}\beta_{\|i}(t_{\rm sat})$ is indeed less strong than that between $\tau_{\rm exp,eff} \Omega_i/\beta_{\|i}(t_{\rm min})^{1.6}$ and $(\Delta_i)_{\rm min}\beta_{\|i}(t_{\rm min})$; however, the existence of any correlation at all suggests that the initial evolution of the firehose instability has a qualitative effect on the subsequent dynamics. 
Furthermore, the spread in values of $(\Delta_i)_{\rm sat}\beta_{\|i}(t_{\rm sat})$ at particular values of $\tau_{\rm exp,eff} \Omega_i/\beta_{\|i}(t_{\rm sat})^{1.6}$ is partially explained by the fact that, for plasmas in an Alfv\'en-enabling state, $\Delta_i\beta_{\|i}$ periodically fluctuates once the firehose instability has saturated; our chosen measure of $\Delta_i\beta_{\|i}$ in saturation is pointwise in time, and so does not account for this effect. Comparison with time-dependent phase-space plots of $[\tau_{\rm exp,eff} \Omega_i/\beta_{\|i}(t_{\rm sat})^{1.6},(\Delta_i)_{\rm sat}\beta_{\|i}(t_{\rm sat})]$ (not shown) supports this explanation, and also recovers the same general trend that is observed in figure~\ref{fig:sims_params_fig2}$(a)$. 

In summary, our simulation results confirm that the parameter $\tau_{\rm exp,eff} \Omega_i/\beta_i^{1.6}$ is indeed a key metric for determining whether a firehose-susceptible high-$\beta_i$ plasma attains an Alfv\'en-inhibiting or Alfv\'en-enabling state once the firehose instability has saturated. 

\subsubsection{Magnetic-field fluctuations} \label{sec:sims:res:magfluc}

\begin{figure}
  \centering
  \includegraphics[width=\linewidth]{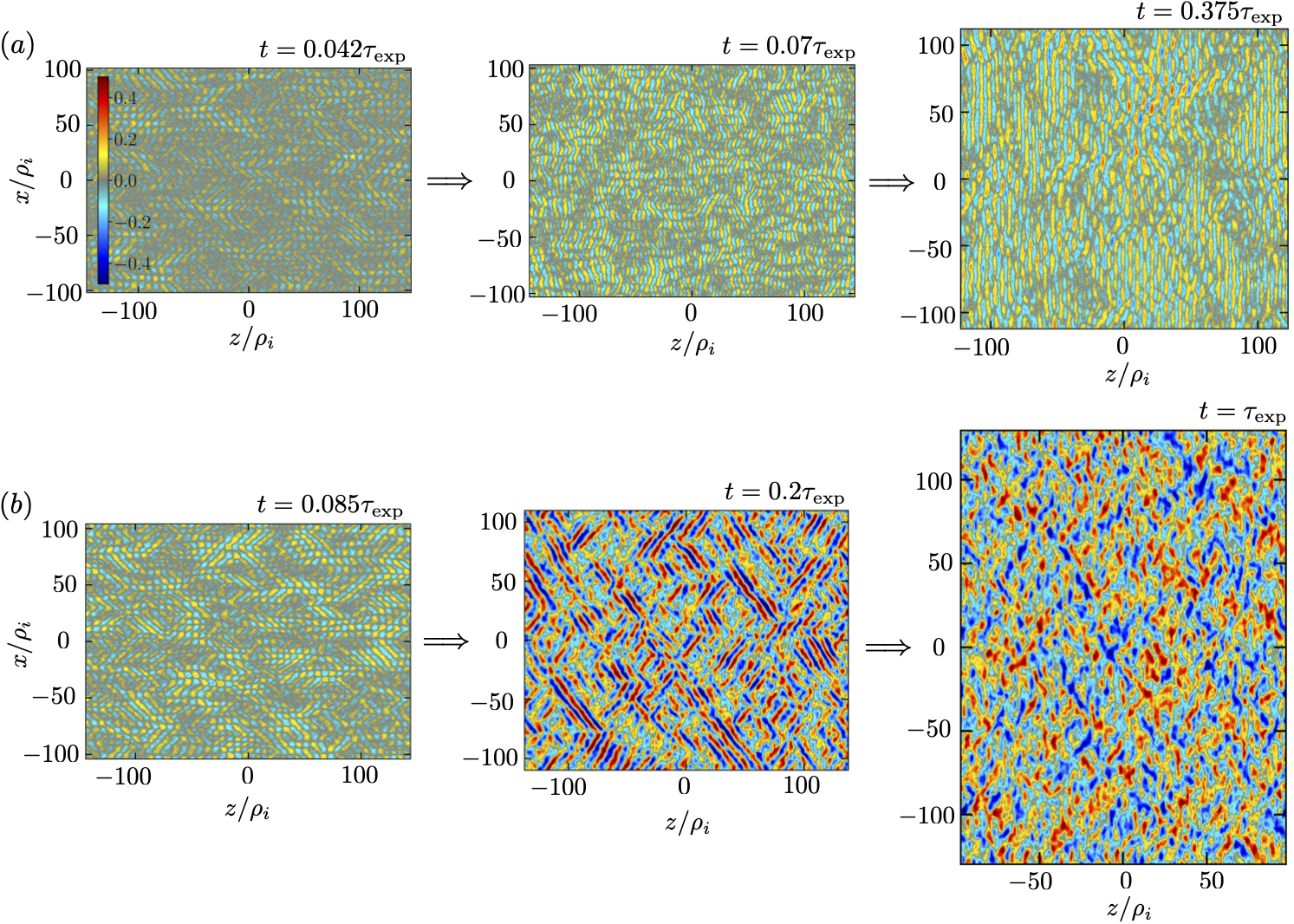}
  \caption{2D visualisations of the out-of-plane component of the perturbed magnetic field in two simulations at $\beta_{i0} = 50$ representing $(a)$ an Alfv\'en-enabling state (run DVI) and $(b)$ an Alfv\'en-inhibiting state (run DIII). The perturbed field appears at three different times: in the linear, nonlinear, and saturated phases of the firehose instability, respectively. We note that, as a fraction of the expansion time, the characteristic times at which the linear, nonlinear and saturated states are realised is longer in the Alfv\'en-inhibiting than Alfv\'en-enabling regime; this is because the firehose instability develops at a comparatively slower rate in this case when compared to the expansion rate.}
\label{fig:sims_vis}
\end{figure}

In addition to having distinct macroscopic properties -- specifically, different equilibrium pressure anisotropies and effective Alfv\'en speeds -- the Alfv\'en-enabling and Alfv\'en-inhibiting states are different microphysically. One manifestation of this is the nature of the firehose fluctuations that arise.  Figure~\ref{fig:sims_vis} visualises the out-of-plane (dominant) component of the perturbed magnetic field for two simulations having $\beta_{i0}=50$ but differing $\tau_{\rm exp}\Omega_{i0}$, such that one realises an Alfv\'en-enabling state (panel ($a$); run DVI) while the other realises an {Alfv\'en-inhibiting} state (panel ($b$); run DIII). 
Initially, in both simulations oblique firehose fluctuations with characteristic wavenumber $k_{\|} \rho_i \sim k_{\perp} \rho_i \approx 0.45$ are destabilised. However, the magnitude of the magnetic-field perturbations in both the nonlinear regime and the saturated states is larger in the simulation that realizes {an Alfv\'en-inhibiting state (relative to an Alfv\'en-enabling state)}. In addition, oblique fluctuations occurring over a range of scales are much more prominent in the Alfv\'en-inhibiting state.

\begin{figure}
  \centering
  \includegraphics[width=\linewidth]{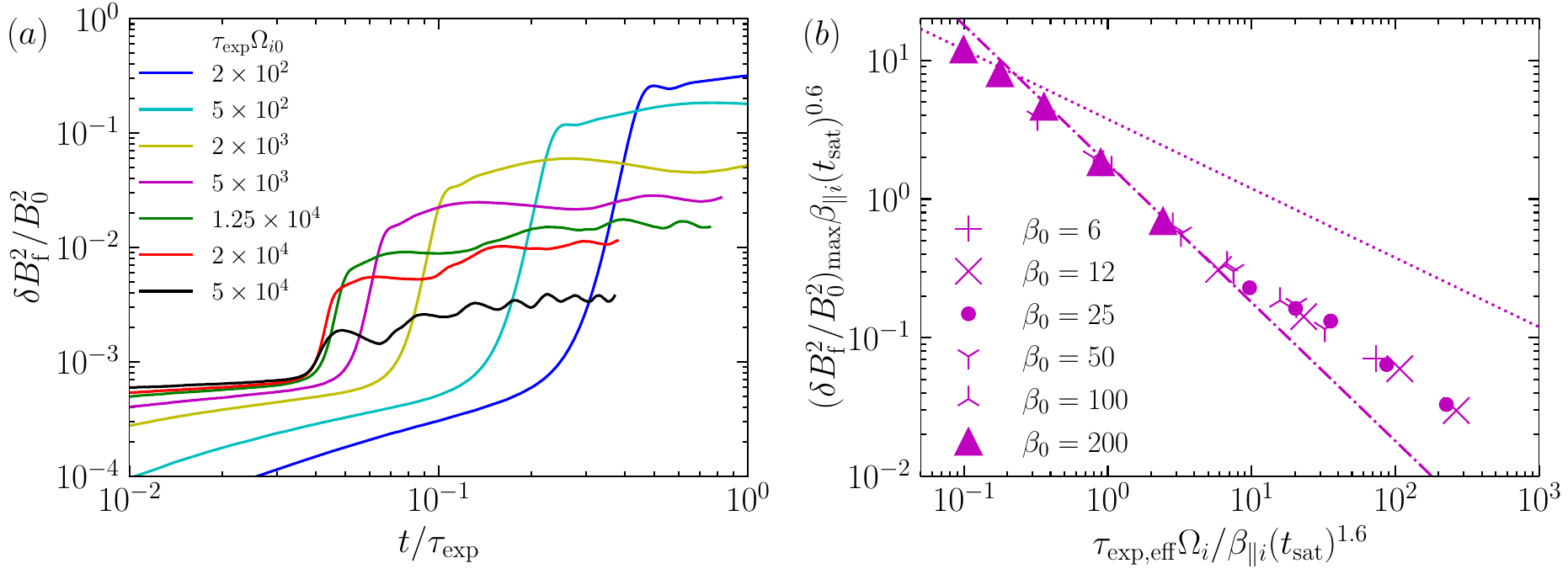}
  \caption{$(a)$ Time evolution of the square of the perturbed magnetic-field strength $\delta B_{\rm f}^2/B_0^2$ associated with the firehose fluctuations for all of the D runs ($\beta_{i0} = 50$). $(b)$ Maximum value of $\delta B_{\rm f}^2/B_0^2$, $(\delta B_{\rm f}^2/B_0^2)_{\rm max}$, as a function of $\tau_{\rm exp,eff} \Omega_i/\beta_{\|i}(t_{\rm sat})^{1.6}$, for all runs. The dashed [dotted] line shows the relationship $\delta B_{\rm f}^2/B_0^2 \simeq 1.6\beta_{\|i}(t_{\rm sat})/\tau_{\rm exp,eff} \Omega_i$ [$\delta B_{\rm f}^2/B_0^2 \simeq 0.77 (200)^{0.3} \beta_{\|i}(t_{\rm sat})^{0.2}/(\tau_{\rm exp,eff} \Omega_i)^{0.5}$]. }
\label{fig:sims_dB_fig1}
\end{figure}

How the key parameters of the expanding plasma affect the characteristic amplitude of magnetic fluctuations can be most simply explored by considering the evolution of the box-averaged perturbed magnetic energy, $\delta B_{\rm f}^2/B_0^2$. Figure~\ref{fig:sims_dB_fig1}$(a)$ shows the evolution of $\delta B_{\rm f}^2/B_0^2$ in time at fixed $\beta_{i0}$. 
The evolution of $\delta B_{\rm f}^2/B_0^2$ in all of our simulations proceeds through four phases. First, there is a pre-firehose phase, in which the box-averaged magnetic-field strength of the fluctuations is simply that associated with random grid-scale fluctuations; next, a linear growth stage, during which the amplitude of firehose fluctuations grows exponentially; third, a nonlinear phase, in which the amplitude of fluctuations continue to grow, but no longer exponentially; finally, saturation. How $\delta B_{\rm f}^2/B_0^2$ evolves qualitatively as a function of time during the nonlinear and saturated phase of the firehose depends on $\beta_{\|i}$ and $\tau_{\rm exp} \Omega_i$. At sufficiently large values of $\tau_{\rm exp}$ (at fixed $\beta_{\|i}$), $\delta B_{\rm f}^2/B_0^2$ does not grow monotonically during the nonlinear phase, nor is it constant in the `saturated' state (see especially the black line in figure~\ref{fig:sims_dB_fig1}($a$)). Instead, the magnetic energy oscillates around a mean value with a characteristic period that is much smaller than $\tau_{\rm exp,eff}$. { These oscillations correlate with those seen in the pressure anisotropy in section \ref{sec:sims:res:pressanisop}, implying a direct link between the amplitude of the firehose fluctuations in saturation, and the regulation of the pressure anisotropy.} For smaller values of $\tau_{\rm exp}$ (again at fixed $\beta_{\|i}$), $\delta B_{\rm f}^2/B_0^2$ does not oscillate in saturation. We also find that, as the expansion time $\tau_{\rm exp}$ is decreased, the characteristic magnitude at which $\delta B_{\rm f}^2/B_0^2$ attains its maximum, $(\delta B_{\rm f}^2/B_0^2)_{\rm max}$, {increases}.

Similarly to the pressure anisotropy and effective Alfv\'en speed, the specific value of $(\delta B_{\rm f}^2/B_0^2)_{\rm max}$, when renormalised by $\beta_{\|i}^{0.6}$, can be predicted with a high degree of confidence by the parameter $\tau_{\rm exp,eff} \Omega_i/\beta_{\|i}^{1.6}$ for any given $\beta_{i}$ and effective expansion time $\tau_{\rm exp,eff} \Omega_i$. This is demonstrated in figure~\ref{fig:sims_dB_fig1}$(b)$. However, the exact relationship between $(\delta B_{\rm f}^2/B_0^2)_{\rm max} \beta_{\|i}^{0.6}$ and $\tau_{\rm exp,eff} \Omega_i/\beta_{\|i}^{1.6}$ is not simply a power law.  For values of $\tau_{\rm exp,eff} \Omega_i/\beta_{\|i}^{1.6}$ of order unity, $(\delta B_{\rm f}^2/B_0^2)_{\rm max} \propto \beta_{\|i}/ \tau_{\rm exp,eff} \Omega_i$, a prediction that arises from a na\"{i}ve quasilinear scattering model (see section \ref{sec:sims:res:satfirehose}). However, a shallower power law dependence arises for either sufficiently small or sufficiently large values of $\tau_{\rm exp,eff} \Omega_i/\beta_{\|i}^{1.6}$. That $(\delta B_{\rm f}^2/B_0^2)_{\rm max}$ is not inversely proportional to $\tau_{\rm exp,eff} \Omega_i/\beta_{\|i}$ at sufficiently small values of the latter parameter is consistent with previous shearing-box simulations of firehose-susceptible high-$\beta$ plasma~\citep{Kunz2014_b,Melville_2016}; for example, \citet{Melville_2016} found that $(\delta B_{\rm f}^2/B_0^2)_{\rm max} \approx 0.77 (\beta_i/\tau_{\rm exp,eff} \Omega_i)^{0.5}$. Computing this formula for our $\beta_{i0} = 200$ runs, we find reasonable agreement for those of our runs with the smallest values of $\tau_{\rm exp,eff} \Omega_i/\beta_{\|i}^{1.6}$ (figure ~\ref{fig:sims_dB_fig1}$(b)$, dotted line).   
That the same also holds at sufficiently large values of $\tau_{\rm exp,eff} \Omega_i/\beta_{\|i}^{1.6}$ is a new finding, suggesting that the nature of the firehose modes present in this {scenario } is distinct. 

\begin{figure}
  \centering
  \includegraphics[width=\linewidth]{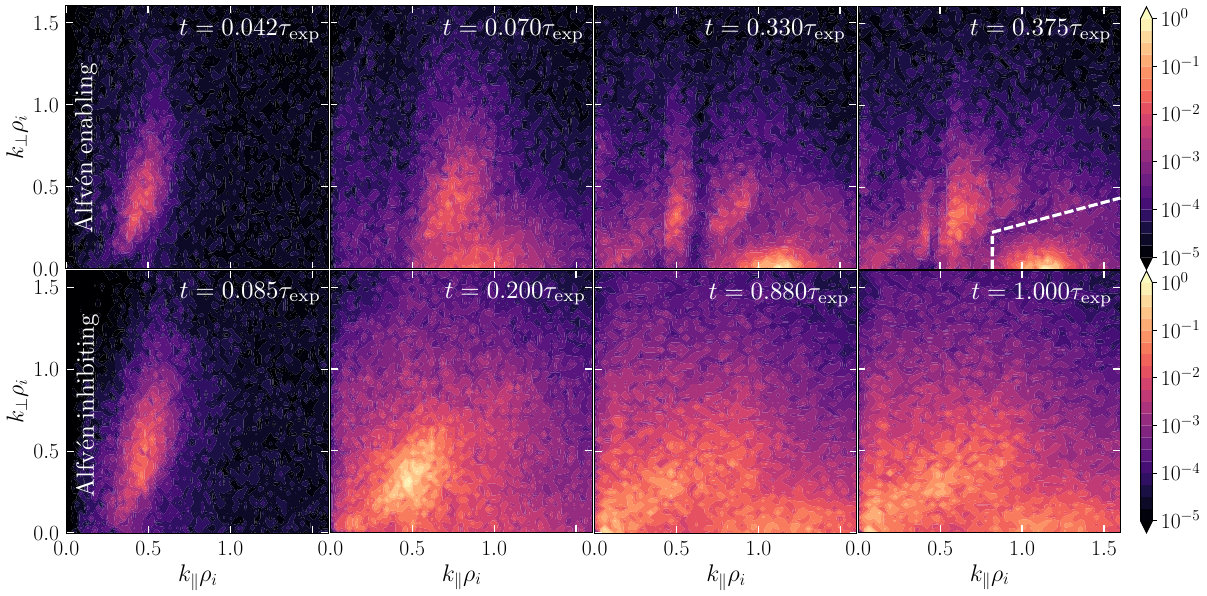}
  \caption{Two-dimensional magnetic-energy spectra $E_B(k_{\|},k_{\perp})$ of the firehose fluctuations at a selection of different times during the firehose instability's evolution: linear phase (far left), nonlinear phase (near left), and two times during the saturated state (near and far right). The top row corresponds to an Alfv\'en-enabling state (run DVI), while the bottom row corresponds to an Alfv\'en-inhibiting state (run DIII). The region circumscribed by the dashed line indicates the region of wavenumber space ($k_{\|} \rho_i \geq 0.8, \; k_{\|} > k_{\perp}/\tan{(15^{\circ})} \approx 3.7 k_{\perp}$) that is used when calculating the magnetic energy of quasi-parallel firehose modes for figure~\ref{fig:sims_dB_fig3}.}
\label{fig:sims_dB_fig2}
\end{figure}

To explore this possibility, Figure~\ref{fig:sims_dB_fig2} displays the evolution of the magnetic-energy spectrum, $E_B(k_{\|},k_{\perp})$, corresponding to the fluctuations visualised in figure~\ref{fig:sims_vis}, with the top (bottom) row pertaining to the Alfv\'en-enabling (-inhibiting) state. 
As expected, the magnetic-energy spectra are initially very similar, indicating oblique modes with $k_\parallel\rho_i\approx k_\perp\rho_i\approx 0.5$. However, in the nonlinear phases of the instability, clear differences emerge. In the saturated Alfv\'en-inhibiting state (bottom row), a wide range of wavenumbers is excited (including fluctuations with characteristic wavelengths that are much larger than the ion Larmor radius), and $E_B(k_{\|},k_{\perp})$ attains a quasi-steady state. By contrast, in the saturated Alfv\'en-enabling state (top row), the magnetic energy is primarily concentrated in two distinct populations of fluctuations whose scales are comparable to the ion-Larmor radius: oblique firehose modes and quasi-parallel modes (the latter circumscribed in the top-right panel by the dashed line). As was also clear from figure~\ref{fig:sims_dB_fig1}$(a)$, the `saturated' Alfv\'en-enabling state is not quasi-steady, but instead is quasi-periodic: while the spectrum of quasi-parallel modes does not change significantly, the spectrum of oblique firehose modes evolves periodically. In section~\ref{sec:sims:res:secondparfirehose}, we argue that the quasi-parallel modes are associated with a secondary parallel firehose instability 

\begin{figure}
  \centering
  \includegraphics[width=\linewidth]{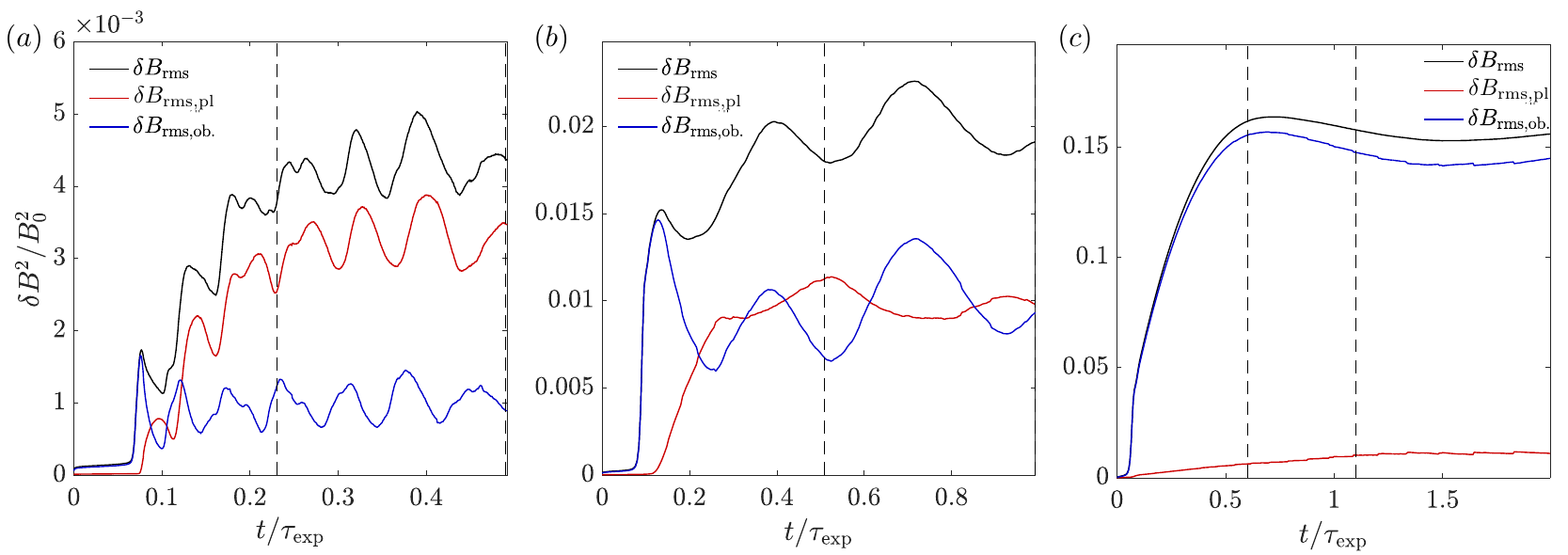}
  \caption{Time evolution of the square of the perturbed magnetic-field strength $\delta B_{\rm f}^2/B_0^2$ (solid black line) associated with the firehose fluctuations, along with the analogous quantity $\delta B_{\rm f,pl}^2/B_0^2$ for quasi-parallel fluctuations (solid red line) and $\delta B_{\rm f,ob}^2/B_0^2$ for oblique fluctuations (solid blue line), for three different simulations: $(a)$ run CV ($\tau_{\rm exp} \Omega_{i0} = 5 \times 10^4$, $\beta_{i0} = 25$), $(b)$ run CII ($\tau_{\rm exp} \Omega_{i0} = 5 \times 10^3$, $\beta_{i0} = 25$), and $(c)$ run FIII ($\tau_{\rm exp} \Omega_{i0} = 2 \times 10^3$, $\beta_{i0} = 200$).}
\label{fig:sims_dB_fig3}
\end{figure}

A simple way to illustrate the quasi-periodic behaviour of firehose-instability saturation in the Alfv\'en-enabling state is to examine the individual components of the perturbed magnetic energy, $\delta B_{\rm f}^2/B_0^2$, {\it viz.}, the component associated with the quasi-parallel modes, $\delta B_{\rm f,pl}^2/B_0^2$, and the component associated with the oblique modes, $\delta B_{\rm f,ob}^2/B_0^2$. These components are obtained by dividing the $(k_{\|},k_{\perp})$ plane into a quasi-parallel region and a non-quasi-parallel region (see figure~\ref{fig:sims_dB_fig2}), and then calculating the total magnetic energies residing within these two separate regions. Figure~\ref{fig:sims_dB_fig3} shows the evolution of $\delta B_{\rm f}^2/B_0^2$, $\delta B_{\rm f,pl}^2/B_0^2$, and $\delta B_{\rm f,ob}^2/B_0^2$ for a selection of different simulations: specifically, a simulation of an asymptotic Alfv\'en-enabling state (figure~\ref{fig:sims_dB_fig3}$(a)$), a marginal Alfv\'en-enabling state (figure~\ref{fig:sims_dB_fig3}$(b)$), and an Alfv\'en-inhibiting state (figure~\ref{fig:sims_dB_fig3}$(c)$).
In the Alfv\'en-enabling states, we observe that the $\delta B_{\rm f,ob}^2/B_0^2$ oscillates quasi-periodically, with the magnitude of that oscillation being comparable to its mean value; $\delta B_{\rm f,pl}^2/B_0^2$ also oscillates with a similar period, but {with a comparatively smaller amplitude relative to its mean}. As the parameter $\tau_{\rm exp,eff} \Omega_i/\beta_{\|i}^{1.6}$ decreases from large to small (left to right in figure~\ref{fig:sims_dB_fig3}), both the absolute and relative amplitudes of quasi-parallel and non-quasi-parallel modes changes. This can be attributed to the distinct saturation mechanisms of the quasi-parallel and oblique firehose modes (see section~\ref{sec:sims:res:satfirehose}). {In Alfv\'en-inhibiting states, the saturated value of $\delta B_{\rm f,ob}^2/B_0^2$ does not change on a period smaller than the expansion time}. Furthermore, deviations from the maximum value of $\delta B_{\rm f,ob}^2/B_0^2$ are much smaller than the maximum value itself, in contrast to the Alfv\'en-enabling states. 

{ Computing $\delta B_{\rm f}^2/B_0^2$, $\delta B_{\rm f,pl}^2/B_0^2$, and $\delta B_{\rm f,ob}^2/B_0^2$ for all of our simulations in the Alfv\'en-enabling state gives a simple way to quantify -- and thereby interpret -- the oscillation period of the perturbed magnetic energy. In particular, for each of these simulations, we identify a period of the simulation in which firehose instabilities have saturated, and then directly calculate the period $\tau_{\rm osc}$ between the maximum value of the perturbed magnetic energy, and the next minimum value. The results of this analysis are shown in figure~\ref{fig:sims_osc}. 
\begin{figure}
  \centering
  \includegraphics[width=0.9\linewidth]{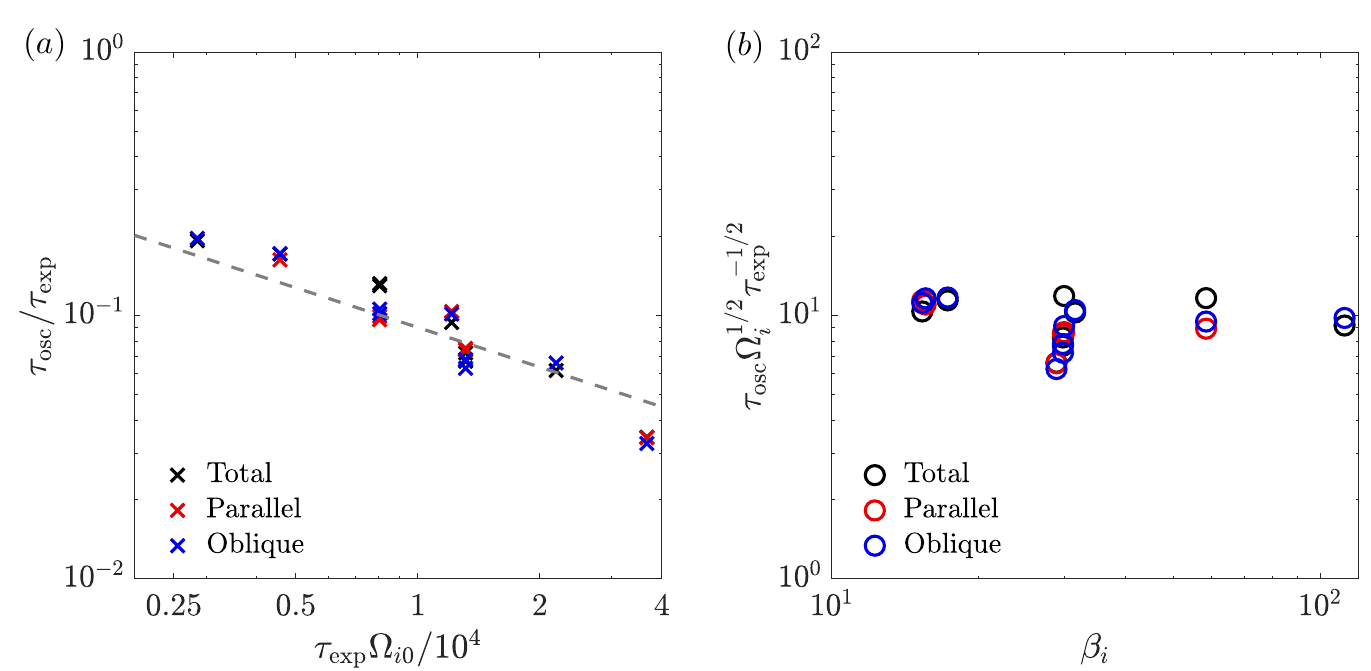}
  \caption{{$(a)$ Numerically determined (half)-period $\tau_{\rm osc}$ of oscillation of the perturbed magnetic energy $\delta B_{\rm f}^2/B_0^2$ associated with all firehoses modes (black), of the magnetic energy $\delta B_{\rm f,pl}^2/B_0^2$ associated with parallel modes (red), and the magnetic energy $\delta B_{\rm f,ob}^2/B_0^2$ associated with oblique modes for all of our Alfv\'en-enabling simulations as a function of the expansion time. The dashed grey line shows the theoretical prediction $\tau_{\rm osc} \propto \tau_{\rm exp}^{1/2} \Omega_i^{-1/2}$. $(b)$ Same as in panel $(a)$, but as a function of $\beta_i$.}}
\label{fig:sims_osc}
\end{figure}
We find that $\tau_{\rm osc}$ is, indeed, much smaller than $\tau_{\rm exp}$ for all of our simulations that attain Alfv\'en-enabling states. Furthermore, $\tau_{\rm osc}/\tau_{\rm exp}$ is, to a reasonable degree of approximation, inversely proportional to the square root of $\tau_{\rm exp} \Omega_{i0}$ (see figure~\ref{fig:sims_osc}a), whilst being approximately independent of $\beta_i$ (see figure~\ref{fig:sims_osc}b). This finding is consistent with the oscillation period being comparable in magnitude to the scattering rate of particles by the quasi-parallel modes which, in the Alfv\'en-enabling state, have the largest amplitude of all firehose-unstable modes (see section \ref{sec:sims:res:satfirehose_parfirehose}). This conclusion does not seem to depend on whether $\tau_{\rm osc}$ is computed from $\delta B_{\rm f}^2/B_0^2$, $\delta B_{\rm f,pl}^2/B_0^2$, or $\delta B_{\rm f,ob}^2/B_0^2$ (see figure~\ref{fig:sims_osc}).}

\subsubsection{Ion distribution functions} \label{sec:sims:res:distfunc}

\begin{figure}
  \centering
  \includegraphics[width=\linewidth]{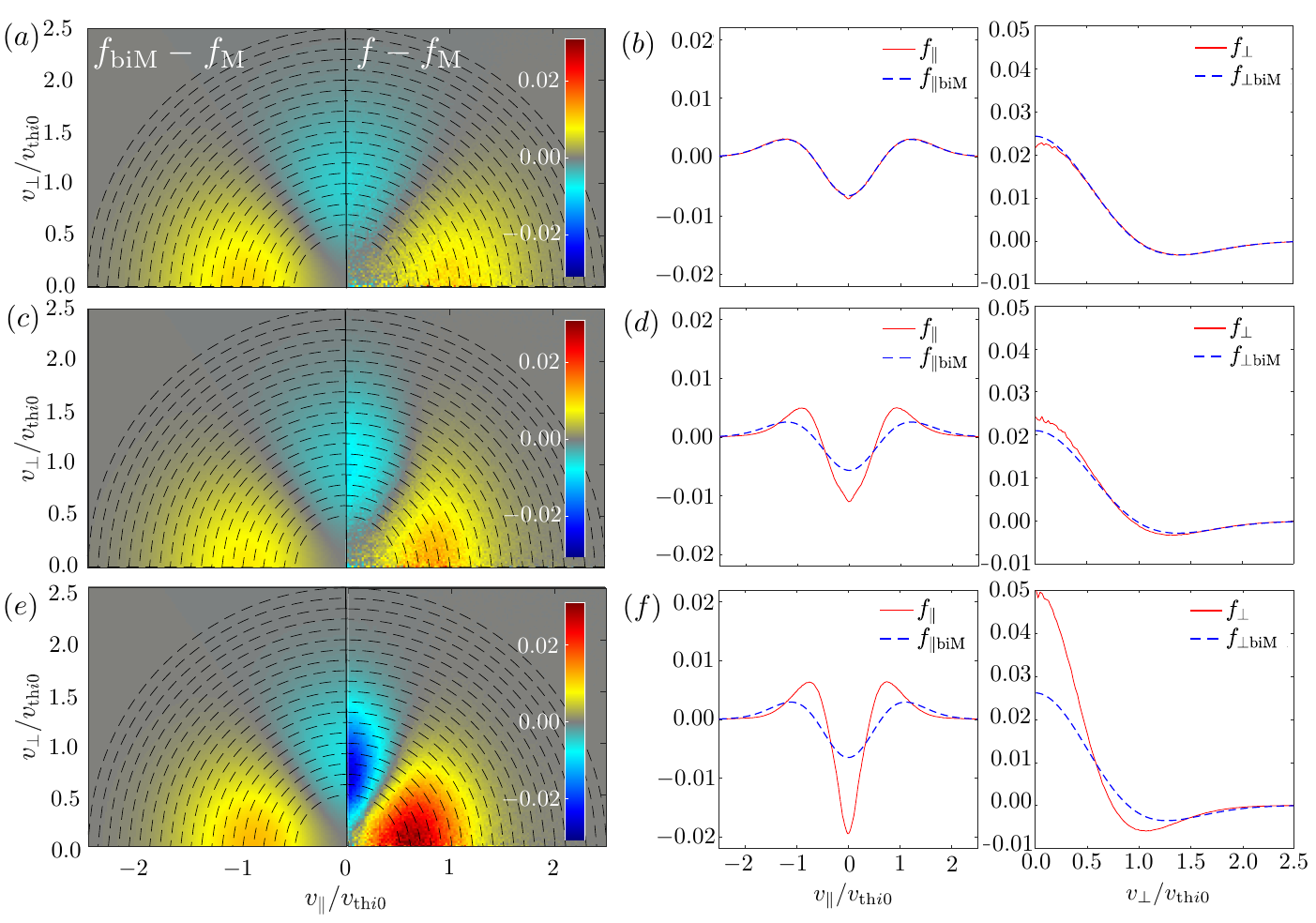}
  \caption{Domain-averaged ion-distribution function $f(v_{\|},v_{\perp})$ in a simulation representative of an Alfv\'en-enabling state (run DVI) during the ($a$) linear ($t = 0.042 \tau_{\rm exp}$), ($c$) nonlinear ($t = 0.07 \tau_{\rm exp}$) and ($e$) saturated ($t = 0.375 \tau_{\rm exp}$) stages of the firehose instability. The right half of each panel shows $f-f_{\rm M}$, where $f_{\rm M}$ is a Maxwellian distribution with the same temperature as $f$; the left half of each panel shows $f_{\rm biM}-f_{\rm M}$, where $f_{\rm biM}$ is a bi-Maxwellian with the same parallel and perpendicular temperatures as $f$. Panels ($b$), ($d$) and ($f$) show the non-Maxwellian component of the parallel ($f(v_{\|})-f_{\rm M}(v_{\|})$, left panel) and perpendicular ($f(v_{\perp})-f_{\rm M}(v_{\perp})$, right panel) distribution functions at the same times, respectively. Dashed lines denote the corresponding $f_{\rm biM}$. }
\label{fig:sims_dist_AE}
\end{figure}
\begin{figure}
  \centering
  \includegraphics[width=\linewidth]{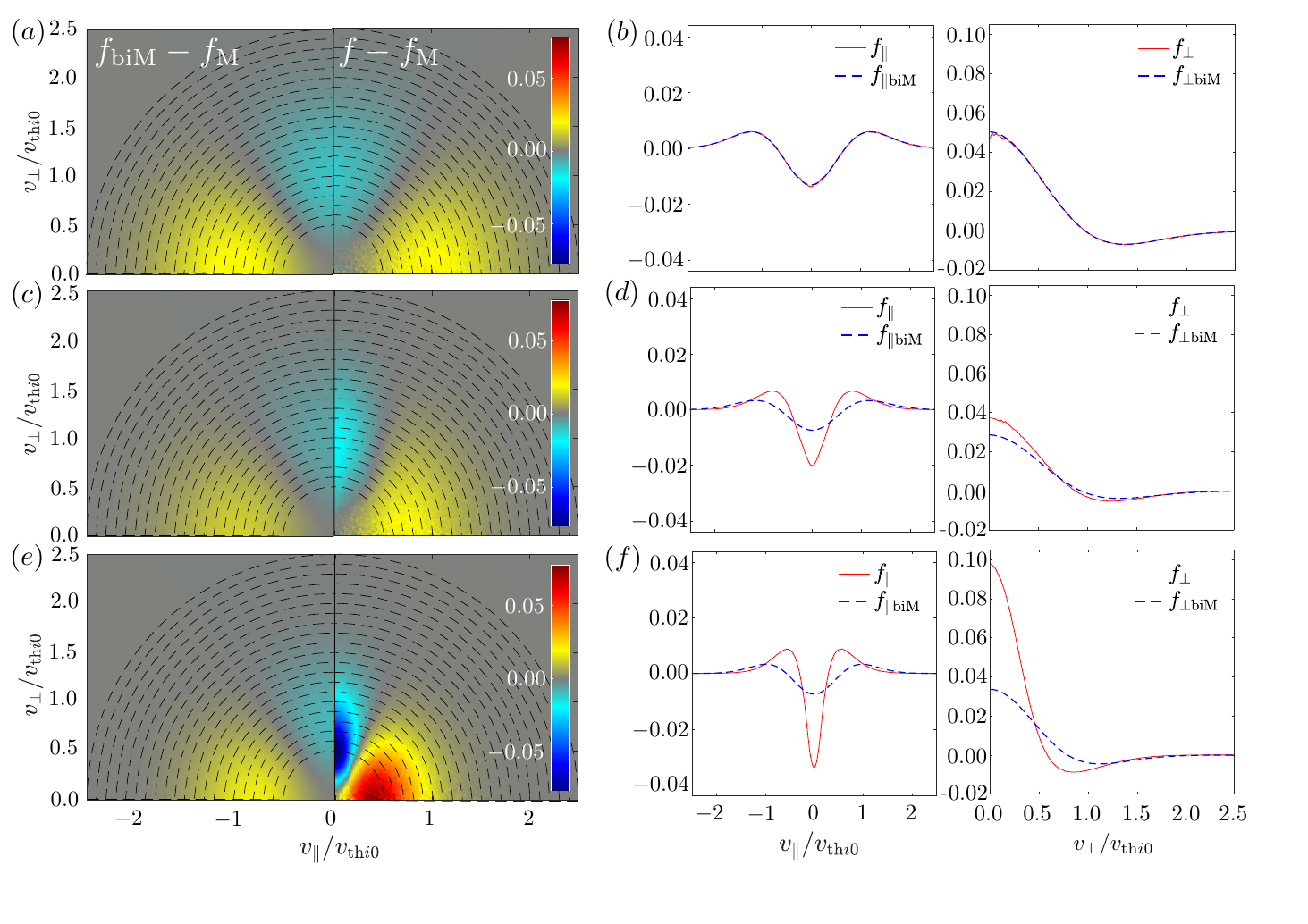}
  \caption{Domain-averaged ion-distribution function $f(v_{\|},v_{\perp})$ in a simulation representative of an Alfv\'en-inhibiting state (run DIII) during the ($a$) linear ($t = 0.075 \tau_{\rm exp}$), ($c$) nonlinear ($t = 0.2 \tau_{\rm exp}$) and ($e$) saturated ($t = \tau_{\rm exp}$) stages of the firehose instability. As in figure~\ref{fig:sims_dist_AE}, the right half of each panel shows $f-f_{\rm M}$, and the left half of each panel shows $f_{\rm biM}-f_{\rm M}$. Panels ($b$), ($d$) and ($f$) show the non-Maxwellian component of the parallel ($f(v_{\|})-f_{\rm\| M}$, left panel) and perpendicular ($f(v_{\perp})-f_{\rm\perp M}$, right panel) distribution functions at the same times, respectively. Dashed lines denote the corresponding $f_{\rm biM}$.}
\label{fig:sims_dist_AI}
\end{figure}

Another, more subtle manifestation of the distinct microphysics of Alfv\'en-enabling and Alfv\'en-inhibiting states can be seen by comparing the domain-averaged ion distribution functions $f(v_{\|},v_{\perp})$ arising in the two states. The time-dependent evolution of $f(v_{\|},v_{\perp})$ in representative Alfv\'en-enabling and Alfv\'en-inhibiting states during the linear, nonlinear and saturated stages of the firehose instability is shown in figures~\ref{fig:sims_dist_AE} and~\ref{fig:sims_dist_AI}, respectively. As follows directly from double-adiabatic conservation laws \eqref{eqn:cgl}, the ion distribution functions in all runs initially evolve to become bi-Maxwellian, with $T_{\|i} \approx T_{\|i0}$ and $T_{\perp i} \approx T_{\perp i0}/(1+t/\tau_{\rm exp})$; indeed, figures~\ref{fig:sims_dist_AE}($a$) and~\ref{fig:sims_dist_AI}($a$) indicate little difference between $f_{\rm biM}-f_{\rm M}$ (left halves of these plots) and $f-f_{\rm M}$ (right halves), {where $f_{\rm biM}$ is the bi-Maxwellian distribution with the parallel and perpendicular temperatures computed from $f(v_{\|},v_{\perp})$, and $f_{\rm M}$ the Maxwellian distribution function with its isotropic temperature computed from $f(v_{\|},v_{\perp})$}.  However, once the firehose fluctuations acquire a sufficient magnitude to backreact on the ions, the distribution functions are no longer described well as bi-Maxwellians (see figures~\ref{fig:sims_dist_AE}$(c)$ and~\ref{fig:sims_dist_AI}$(c)$). In saturation (figures~\ref{fig:sims_dist_AE}$(e)$ and~\ref{fig:sims_dist_AI}$(e)$), the difference becomes even more pronounced. 

\begin{figure}
  \centering
  \includegraphics[width=\linewidth]{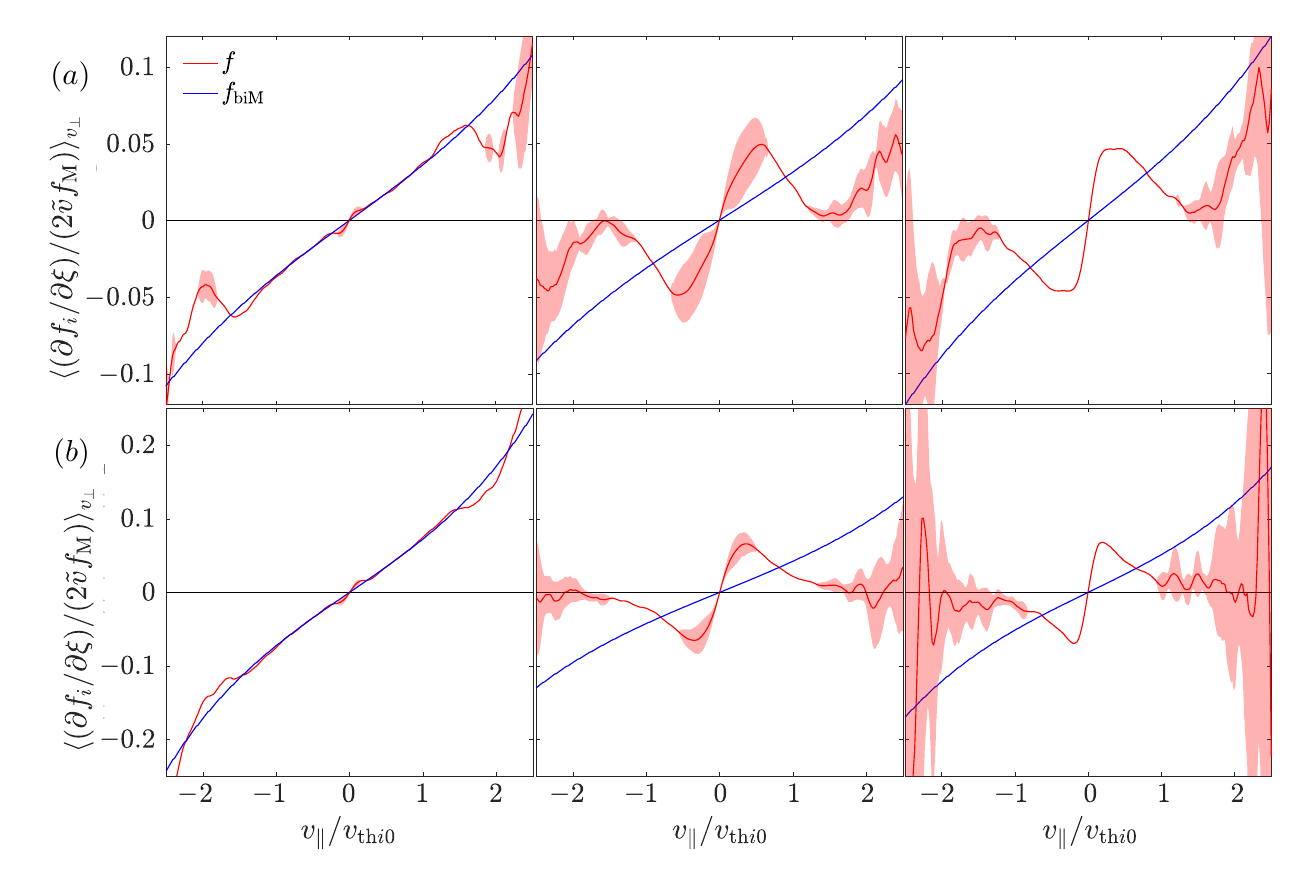}
  \caption{Pitch-angle gradient of the ion distribution function $f$ divided by $2 \tilde{v} f_{\rm M}$ (solid red line), where $\tilde{v} = v/v_{\mathrm{th}i}$, and $f_{\rm M}$ is a Maxwellian distribution with the same temperature as $f$, averaged over $v_{\perp}$. The solid blue line is the analogous quantity, but calculated using $f_{\rm biM}$, the bi-Maxwellian distribution function with the same parallel and perpendicular temperatures as $f$. The red-pink bar denotes the standard deviation of $(\partial f_i/\partial \xi)/(2 \tilde{v} f_{\rm M})$, determined from the range of $v_{\perp}$ over which the average is computed. Row~$(a)$: Alfv\'en-enabling state (run DV) in the linear phase (left panel), nonlinear phase (middle panel) and saturation (right panel). Row~$(b)$: Alfv\'en-inhibiting state (run DIII).}
\label{fig:sims_dist_pitchangle_grad}
\end{figure}

To characterise the departures from bi-Maxwellian distribution functions more carefully -- and thereby identify the subtle differences between the Alfv\'en-enabling and Alfv\'en-inhibiting states -- it is helpful to define one-dimensional distribution functions: the distribution function integrated over perpendicular and parallel velocities, $f(v_\parallel) \equiv \int_{0}^{\infty} \rmd v_{\perp} v_{\perp}\, f$ and $f(v_\perp) \equiv \int_{-\infty}^{\infty} \rmd v_{\|}\, f$, respectively. The clearest non-bi-Maxwellian feature in the nonlinear phase of both states (figures~\ref{fig:sims_dist_AE}$(d)$ and~\ref{fig:sims_dist_AI}$(d)$) and in saturation (figures~\ref{fig:sims_dist_AE}$(f)$ and \ref{fig:sims_dist_AI}$(f)$) is the comparatively more pronounced anisotropy of the distribution function at subthermal velocities. But the main difference between the two states is the distribution function of ions with suprathermal velocities: in the Alfv\'en-inhibiting state (figure~\ref{fig:sims_dist_AI}$(f)$), the distribution function is quasi-isotropic for all velocities $|v_{\|}| \gtrsim 1.25 v_{\mathrm{th}i}$, whereas in the Alfv\'en-enabling state (figure \ref{fig:sims_dist_AE}$(f)$), a significant anisotropy is retained at specific velocities that evolve periodically as a function of time\footnote{{ In $\beta_i \gtrsim 1$ firehose-unstable plasma,~\citet{Matteini_2006} observed the development of power-law tails at suprathermal velocities. We do not observe the development of such tails in our (comparatively much larger $\beta_i$) simulations; the distribution function remains quasi-Maxwellian.}}. The difference is challenging to discern from the distribution functions themselves, but can be more clearly seen by comparing the pitch-angle gradient of the distribution function (see figure~\ref{fig:sims_dist_pitchangle_grad}). 
Row~$(a)$ of figure~\ref{fig:sims_dist_pitchangle_grad} demonstrates that in the nonlinear and saturation phases of the Alfv\'en-enabling state, the pitch-angle gradient of the ion distribution is not close to zero for $|v_{\|}| \gtrsim 1.75 v_{\mathrm{th}i}$, whereas the opposite is true for the Alfv\'en-inhibiting state. 
These features of the distribution function are directly related to properties of the effective collision operator associated with the firehose fluctuations (see section~\ref{sec:collisionality}).

\subsection{Velocity-averaged collisionality and effective viscosity} \label{sec:boxavcollisionality}

\begin{figure}
  \centering
  \includegraphics[width=\linewidth]{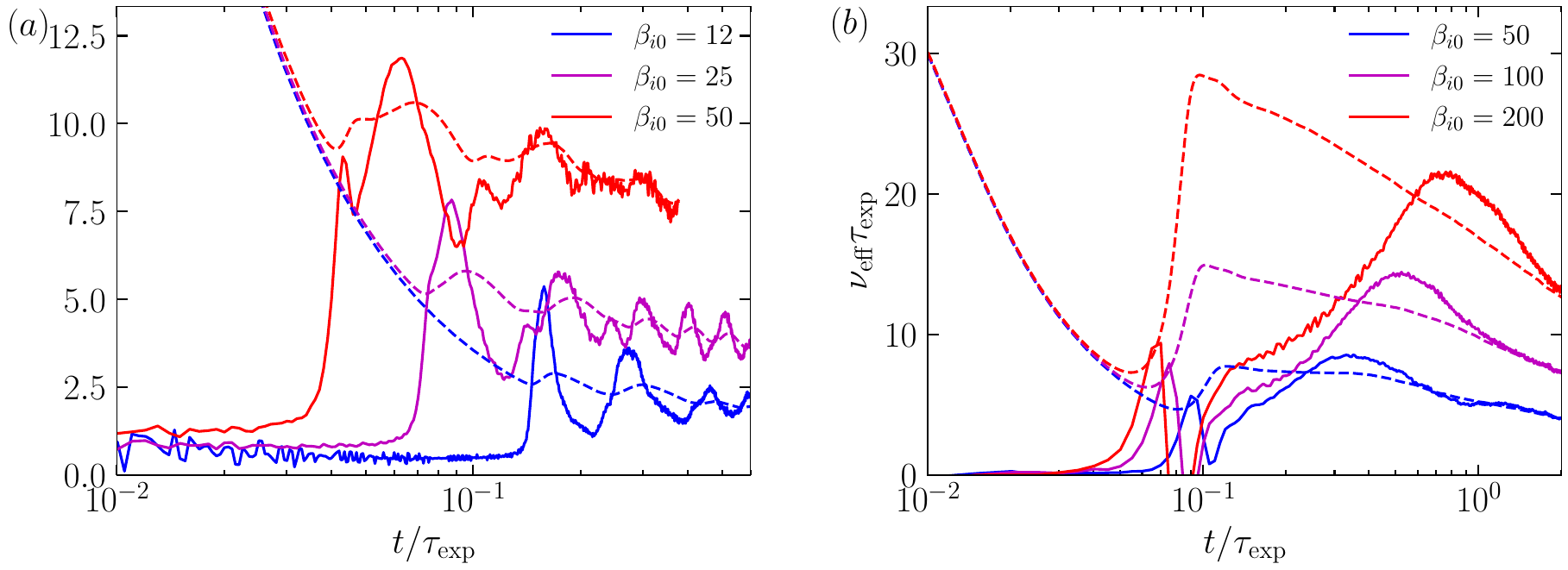}
  \caption{$(a)$ Values of the effective collisionality $\nu_{\rm eff}$ measured directly in the simulations (solid lines) with Alfv\'en-enabling states (runs BIV, CIV, and DVI). The expansion time in these simulations is $\tau_{\rm exp} \Omega_{i0} = 2 \times 10^{4}$. The effective collisionalities predicted by the simple model (\ref{effective_coll}) for each simulation are shown by the dashed lines, to which the curves asymptote at late times. $(b)$ Same as in panel $(a)$, but for three simulations (runs DIII, EI, and FIII) with $\tau_{\rm exp} \Omega_{i0} = 2 \times 10^{3}$ and therefore Alfv\'en-inhibiting states.}
\label{fig:coll_boxavtimeplot}
\end{figure}

Finally, we characterise the average collisionality $\nu_{\rm eff}$ of all particles in our simulation. There are various approaches for measuring $\nu_{\rm eff}$ in PIC simulations; we adopt that taken in~\citet{Riquelme_2015} and~\citet{Bott_2021}, and calculate $\nu_{\rm eff}$ via the rate of change of the simulation-domain-averaged first adiabatic invariant $\mu$: $\nu_{\rm eff} = \dot{\bar{\mu}}/\overline{(T_{\|i}-T_{\perp i})/B}$. We adopt this measure because, in a plasma without collisionality, $\mu$ is well conserved, so its non-conservation is a clear signature of collisionality. More practically, this measure allows for a time-resolved estimate of the effective collisionality to be computed. Figure~\ref{fig:coll_boxavtimeplot} shows $\nu_{\rm eff}$ as a function of time for two representative sets of simulations, each at fixed $\tau_{\rm exp}$: panel~$(a)$ shows three simulations in the Alfv\'en-enabling regime with $\tau_{\rm exp} \Omega_{i0} = 2 \times 10^{4}$, while panel~$(b)$ shows three Alfv\'en-inhibiting simulations with $\tau_{\rm exp} \Omega_{i0} = 2 \times 10^{3}$.
Qualitatively, it is clear that $\nu_{\rm eff}$ increases with increasing $\beta_{i0}$ (blue to red) and decreasing $\tau_{\rm exp}$ (left to right). 

Similarly to~\citet{Bott_2021}, we can derive a theoretical estimate for $\nu_{\rm eff}$ by using the firehose-collisionality-modified CGL equations (\ref{CGL_withcoll}). 
To derive an estimate for $\nu_{\rm eff}$, we make three simplifying assumptions: first, that heat fluxes are negligible [and so all terms proportional to $q_{\|}$ or $q_{\perp}$ in~(\ref{CGL_withcoll}) can be ignored]; secondly, that the dimensionless pressure anisotropy is small; and thirdly, that the expansion rate is much smaller than the effective collision rate. It follows from these three assumptions that (cf.~\citealt{Braginskii_1965})
\begin{equation}
 \nu_{\rm eff} \Delta_i \simeq \D{t}{} \log{\frac{B}{n^{2/3}}} \, . \label{press_coll_relation}  
\end{equation}
Finally, noting that for a transversely expanding plasma, $B \propto n$, we deduce from (\ref{press_coll_relation}) that
\begin{equation}
 \nu_{\rm eff}^{\rm CGL} \simeq \frac{1}{3 \Delta_i} \frac{\mathrm{d}}{\mathrm{d}t} \log{B} = -\frac{1}{3 \Delta_i \tau_{\rm exp,eff}} \, . \label{effective_coll}  
\end{equation}
This prediction is plotted in figure~\ref{fig:coll_boxavtimeplot} (dashed lines). In the Alfv\'en-inhibiting regime, (\ref{effective_coll}) compares very favourably to our numerical estimates of $\nu_{\rm eff}$ in the saturated states of the simulations we show (panel~$(a)$). In the Alfv\'en-enabling regime, (\ref{effective_coll}) agrees well with the numerical collisionality averaged over the saturated state, but does not capture significant time-dependent fluctuations (panel~$(b)$). Because $-\Delta_i \beta_{\|i} \sim 1$ in the saturated states of our simulations, it follows that $\nu_{\rm eff} \sim \beta_{\|i}/\tau_{\rm exp,eff}$, as expected. 

\begin{figure}
  \centering
  \includegraphics[width=\linewidth]{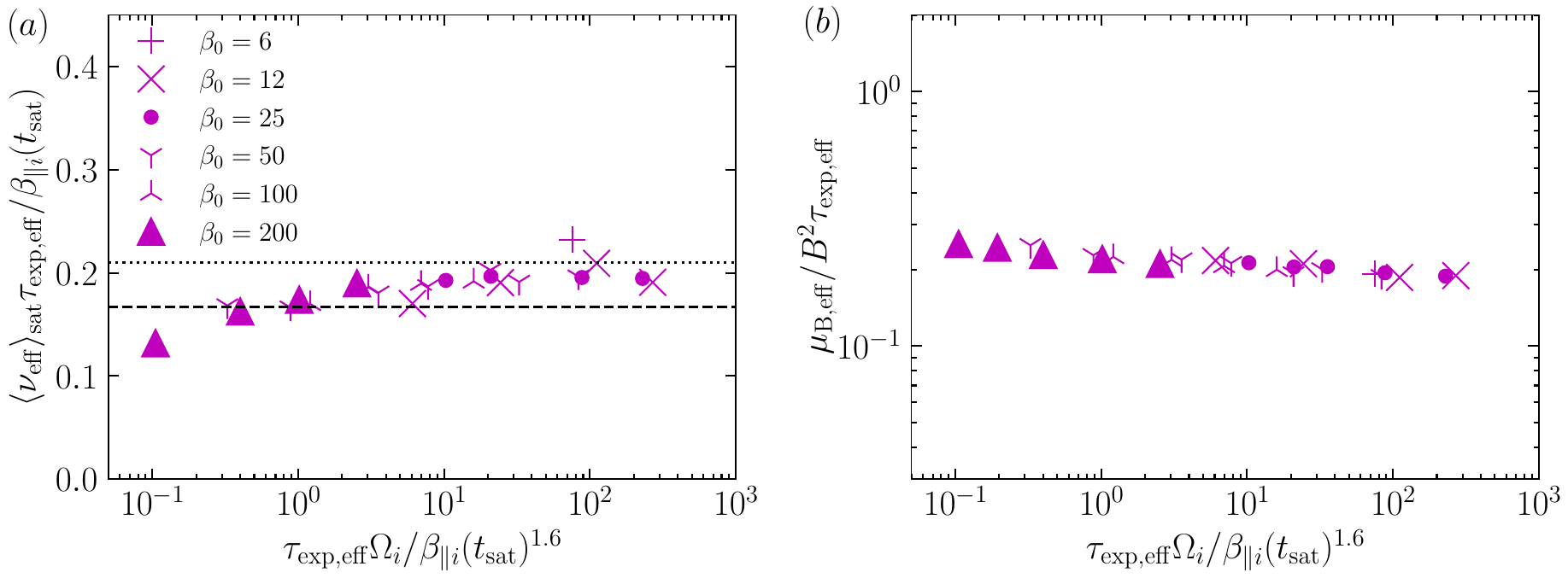}
  \caption{$(a)$ Values of the effective collisionality $\langle \nu_{\rm eff} \rangle_{\rm sat}$ measured directly in all simulations, averaged over the time interval between the time at which the firehose fluctuations attain their peaks strength and the time at which the next local minimum is obtained. The dashed line indicates the effective (time-averaged) value $\nu_{\rm eff} = \beta_{\|i}/6 \tau_{\rm exp,eff}$ of the collisionality predicted in asymptotic Alfv\'en-inhibiting states, while the dotted line shows the value $\nu_{\rm eff} \simeq 0.21 \beta_{\|i}/ \tau_{\rm exp,eff}$ appropriate for asymptotic Alfv\'en-enabling states. $(b)$ Effective parallel Braginskii viscosity $\mu_{\rm B,eff}$ associated with the collisionality measured directly in all simulations.}
\label{fig:coll_boxav_allplots}
\end{figure}

Turning to our complete set of runs, figure~\ref{fig:coll_boxav_allplots}$(a)$ shows the numerical estimates of the characteristic collisionality in the saturated state of all of our simulations. 
To account for the time variation of the collisionality in Alfv\'en-enabling states, we average it over a time interval in which the saturated state is realised.  
The effective collisionality is consistent across all of our simulations, but $\nu_{\rm eff}$ does increase slightly as the critical parameter $\tau_{\rm eff,exp} \Omega_i/\beta_{\|i}^{1.6}$ increases. This trend follows directly from our prior result that, in saturation, $\Delta_i$ increases from $\Delta_i \simeq -2/\beta_{\|i}$ to 
$\Delta_i \simeq -1.6/\beta_{\|i}$ as $\tau_{\rm eff,exp} \Omega_i/\beta_{\|i}^{1.6}$ increases from small ({\it viz.}, plasma in an Alfv\'en-inhibiting state) to large ({\it viz.}, plasma in an Alfv\'en-enabling state). Based on these values and (\ref{effective_coll}), it follows that we would expect $\nu_{\rm eff} \simeq \beta_{\|i}/6 \tau_{\rm exp,eff}$ in Alfv\'en-inhibiting states (dashed line in figure~\ref{fig:coll_boxav_allplots}), while $\nu_{\rm eff} \simeq 0.21 \beta_{\|i}/ \tau_{\rm exp,eff}$ in asymptotic Alfv\'en-enabling states (dotted line). The prediction is realised in our simulations. 

Having computed the domain-averaged collisionality, we can then determine the plasma's effective parallel Braginskii viscosity $\mu_{\rm B,eff}$. By comparison with (\ref{eff_viscosity}), it follows that in our HEB simulations, 
\begin{equation}
\mu_{\rm B,eff} \approx -\frac{3}{2} (\Delta_i \beta_{\|i})_{\rm sat} \frac{B^2}{4 \upi} \tau_{\rm exp,eff} \, . \label{eff_viscosity_expbox} 
\end{equation}
This estimate agrees well with the value of $\mu_{\rm B,eff}$ that is directly computed from our simulations (figure~\ref{fig:coll_boxav_allplots}). That $\mu_{\rm B,eff}$ is given by (\ref{eff_viscosity}) is striking for two reasons: (i) in stark contrast to classical, strongly collisional plasmas, the plasma's viscosity is dependent upon the magnetic-field strength; and (ii) the viscosity coefficient decreases as the expansion rate increases, \emph{viz.}, weakly collisional plasmas behave like non-Newtonian fluids. 

\section{Theoretical interpretation of results} \label{sec:simtheoryinterp}

\subsection{Overview}

While some of the results from our HEB simulations -- for example, the particle-averaged effective collisionality, or the regulation of pressure anisotropy in Alfv\'en-inhibiting states -- are consistent with the results of previous simulations of firehose instabilities~\citep[e.g.,][]{Hellinger_2008,Kunz2014_b,Riquelme_2015,Melville_2016}, other results are not, and so require further analysis and interpretation. Three findings about the Alfv\'en-enabling state in particular are unexpected, and so warrant additional investigation. First of these is the emergence of ion-Larmor-scale parallel firehose modes, which are specifically predicted not to be present by the linear theory of the firehose instability in a bi-Maxwellian plasma that was outlined in section~\ref{linfiretheory_resparfire}. Secondly, the regulated pressure anisotropy $(\Delta_i)_{\rm sat} \simeq -1.6/\beta_{\|i}$ in the Alfv\'en-enabling state does not correspond to the linear threshold $\Delta_{\rm cr} \simeq -1.35/\beta_{\|i}$ for the oblique firehose instability in a bi-Maxwellian plasma (see section~\ref{linfiretheory_numres}). The third unexpected finding is  that the box-averaged perturbed magnetic energy, $\delta B_{\rm f}^{2}/B_0^2$, does not scale as $\delta B_{\rm f}^{2}/B_0^2 \propto \beta_i/\tau_{\rm exp,eff} \Omega_i$ as might be na\"{i}vely anticipated, but instead has a weaker dependence. These findings are discussed in sections~\ref{sec:sims:res:secondparfirehose},~\ref{sec:sims:res:stapresaniso}, and~\ref{sec:sims:res:satfirehose} respectively.  

\subsection{Secondary parallel firehose instability in the Alfv\'en-enabling regime} \label{sec:sims:res:secondparfirehose} 

A notable result from our simulations is the presence of ion-Larmor-scale parallel firehose modes in the Alfv\'en-enabling regime. The presence of such modes is, at first glance, inconsistent with the linear theory of the firehose instability in a bi-Maxwellian plasma with a negative pressure anisotropy (\S \ref{linfiretheory_resparfire}), which predicts that the resonant parallel firehose should be subdominant to oblique firehose modes in high-$\beta$ plasma. 
However, it can, in fact, be shown that these modes are not whistler/fast magnetosonic modes destabilised by the resonant parallel firehose instability (as would occur in the plasma with $\beta_i \sim 1$), but are instead a lower-frequency mode excited by a (newly identified) secondary instability associated with the non-bi-Maxwellian form of the distribution function. This form, presented in section~\ref{sec:sims:res:distfunc}, is caused by the backreaction of the oblique firehose modes on the otherwise bi-Maxwellian distribution function that is driven by the plasma's expansion. 

To understand this secondary parallel firehose instability better, it is helpful to describe qualitatively the types of parallel modes (and their growth) that the (high-$\beta$) plasma can support linearly as the expansion proceeds. Initially, at the start of the simulation, when the ion distribution function is Maxwellian, there are two types of forward-propagating parallel modes with $\beta_i^{-1/2} \ll k_{\|} \rho_i \lesssim 1$: right-handed whistler/magnetosonic modes, which have characteristic real frequencies $\varpi \sim k_{\|}^2 \rho_i^2 \Omega_i$, and left-handed, ion-cyclotron modes which, {in high-$\beta_i$ plasma, have $\varpi \sim \Omega_i/\beta_i$~\citep{Foote_1979}\footnote{{We note that the frequency $\varpi$ of these modes is not described by the cold-plasma dispersion relation because  $\varpi \ll k_{\|} v_{\mathrm{th}i}$.}}. }  These two types of modes have very different characteristic frequencies because of their distinct physical mechanisms; the characteristic oscillation of the higher-frequency whistler/magnetosonic modes is supported by inertial and gyroviscous forces acting out of phase (with the action of the Alfv\'en restoring force being negligible), while for the lower-frequency ion-cyclotron modes, it is the out-of-phase action of the Alfv\'en restoring force and the gyroviscous force that gives rise to oscillatory dynamics. Despite their distinct mechanisms, both of these modes are damped ($\gamma < 0$). As the plasma expands, the pressure anisotropy becomes increasingly negative, which changes the character of the ion cyclotron mode; specifically, the real frequency of this mode becomes negative for $k_{\|} \rho_i \sim 1$ (though the mode remains damped). Because $k_{\|} > 0$, this change of sign corresponds to initially forward-propagating ion-cyclotron modes becoming backward propagating (and vice versa); in short, the initially left-handed forward-propagating ion-cyclotron mode becomes a type of right-handed (forward-propagating) mode that is qualitatively distinct to the whistler/magnetosonic mode. Physically, this change of handedness can be attributed to the Alfv\'enic restoring force being weakened by increasingly strong parallel pressure forces associated with the negative pressure anisotropy. The damping of these ion-cyclotron modes finally becomes growth once the oblique firehose fluctuations begin to backreact on the ion distribution function. These fluctuations, which have a characteristic parallel wavenumber $k_{\|} \approx 0.5 \rho_i^{-1}$ efficiently scatter particles with a characteristic velocity $v_{\|} \approx v_{\mathrm{th}i}/(k_{\|} \rho_i)_{\rm ob} \approx 2 v_{\mathrm{th}i}$, and isotropise the distribution in a narrow $v_{\|}$ interval. This, in turn, enables the right-handed ion-cyclotron modes to extract energy from these same particles, and thereby grow.  

With some effort, we can characterise the growth of the secondary parallel firehose modes (and their analogous damped modes in the initial stage of the simulation) analytically. For arbitrary background distribution functions ${f}_{s0}$ of species $s$, the linear dispersion relation of parallel modes in a hot plasma is, neglecting the displacement current, 
\begin{eqnarray}
D^{\pm} & = & k_{\|}^2 c^2 + \sum_{s} \frac{\omega_{\mathrm{p}s}^2}{n_{s0}} \Bigg\{\upi \int_{C_{\rm L}} \mathrm{d}v_{\|} \int_0^{\infty} \mathrm{d}v_{\perp} \frac{v_{\perp}^2}{k_{\|}v_{\|}-\omega \mp \Omega_s} \nonumber \\*
&& \qquad \qquad \times \left[k_{\|} \left(v_{\perp} \frac{\partial {f}_{s0}}{\partial v_{\|}}-v_{\|} \frac{\partial {f}_{s0}}{\partial v_{\perp}}\right)+\omega \frac{\partial {f}_{s0}}{\partial v_{\perp}}\right]\Bigg\} = 0 ,
\end{eqnarray}
where $n_{s0}$ is the equilibrium number density of species $s$, $\omega_{\mathrm{p}s}$ is the plasma frequency, $C_{\rm L}$ is the usual Landau (`L') contour, and we have assumed that $k_{\|} > 0$. In a Maxwellian plasma, the `$+$' and `$-$' roots with $\omega > 0$ correspond to the whistler/magnetosonic modes and ion-cyclotron modes, respectively. Motivated by our simulation results, we
further specialise to the `low-frequency' ion-cyclotron modes with $k_{\|} \rho_i \sim 1$, which satisfy $\omega \sim k_{\|} v_{\mathrm{th}i}/\beta_{i} \ll k_{\|} v_{\mathrm{th}i}$. We also assume a Maxwellian electron population (as in the hybrid-kinetic simulations), and that the ion distribution function's anisotropy is small compared to its characteristic magnitude:  
\begin{equation}
\frac{v_{\mathrm{th}i}}{v}\frac{\partial {f}_{i0}}{\partial \xi} = v_{\mathrm{th}i} \left(\frac{\partial {f}_{i0}}{\partial v_{\|}}-\frac{v_{\|}}{v_{\perp}} \frac{\partial {f}_{i0}}{\partial v_{\perp}} \right) \sim \frac{\omega}{k_{\|} v_{\mathrm{th}i} } f_{\mathrm{M}} \ll f_{\mathrm{M}} \, ,   
\end{equation}
where $\xi \equiv v_{\|}/v$ is the pitch angle.
Under these assumptions, simplified expressions can be derived for the real frequency $\varpi$ and growth rate $\gamma$ of these modes:
\begin{subeqnarray}\label{eqn:prlmodes}
\frac{\varpi}{\Omega_i}  &  \approx & \pm \Bigg\{ \mathcal{G}(k_{\|} \rho_i)\left[k_{\|}^2 d_i^{2} + \left(\frac{\upi}{n_{i0}} \mathcal{P}\int_{-\infty}^{\infty} \mathrm{d} {v}_{\|} \int_0^{\infty} \mathrm{d}{v}_{\perp} \frac{k_{\|} {v}_{\perp}^3}{k_{\|}v_{\|} \mp \Omega_i} \frac{1}{v} \frac{\partial {f}_{i0}}{\partial \xi}\right) \right] \nonumber \\*
&&  \quad - \left[\frac{\upi^2}{n_{i0}} \int_0^{\infty} \mathrm{d}v_{\perp} v_{\perp}^{3} \left(\frac{1}{v} \frac{\partial {f}_{i0}}{\partial \xi}\right)\Bigg|_{v_{\|}  = v_{\|\mathrm{res}}^{\pm}}\right] \left[\frac{\sqrt{\upi}}{k_{\|} \rho_i} \exp{\left(-\frac{1}{k_{\|}^2 \rho_i^2}\right)} \right] \nonumber \Bigg\} \\
&& \times \left\{\left[\mathcal{G}(k_{\|} \rho_i)\right]^{2} + \frac{\upi}{k_{\|}^2 \rho_i^2} \exp{\left(-\frac{2}{k_{\|}^2 \rho_i^2}\right)} \right\}^{-1}, \\
 \frac{\gamma}{\Omega_i} & \approx & \pm \left[\mathcal{G}(k_{\|} \rho_i)\right]^{-1} \left[\frac{\upi^2}{n_{i0}} \int_0^{\infty} \mathrm{d}v_{\perp} v_{\perp}^{3} \left(\frac{1}{v} \frac{\partial {f}_{i0}}{\partial \xi}-\frac{2 v_{\rm wv}}{v_{\mathrm{th}i}^2} {f}_{\rm M}\right)\Bigg|_{v_{\|} = v_{\|\mathrm{res}}^{\pm}}\right] \, ,  \label{gamma_prlmodes}
\end{subeqnarray}
where $d_i = \beta_i^{-1/2} \rho_i$ is the ion inertial length, 
\begin{equation}
\mathcal{G}(k_{\|} \rho_i) = 1 +  \frac{1}{k_{\|} \rho_i} \mathrm{Re} \, Z\left(\frac{1}{k_{\|} \rho_i}\right)  
\end{equation}
is a special function related to the plasma dispersion function $Z(x)$ whose only root occurs at $k_{\|} \rho_i \simeq 1.08$, 
$v_{\mathrm{wv}} \equiv \varpi/k_{\|}$ is the parallel phase velocity of the wave, and $v_{\|\mathrm{res}}^{\pm} \equiv (\varpi \pm \Omega_i)/k_{\|} \approx \pm \Omega_i/k_{\|}$ is the parallel velocity of particles that are resonant with that mode. We note that, due to our assumed ordering, we have removed the whistler/magnetosonic branch, and so (\ref{gamma_prlmodes}) describes just the real frequency and growth rate of (both forward- and backward-propagating) modes of the ion-cyclotron type. It follows that the damping or growth of such parallel modes depends upon the sign of the quantity 
\begin{equation}
\mathcal{I}_{\pm} \equiv \frac{1}{v} \frac{\partial {f}_{i0}}{\partial \xi}-\frac{2 v_{\rm wv}}{v_{\mathrm{th}i}^2} {f}_{\rm M}
\end{equation}
evaluated near the resonant velocity $v_{\|\mathrm{res}}^{\pm}$. For $k_{\|} \rho_i < 1.08$, growth occurs whenever $\pm \mathcal{I}_{\pm} > 0$, and vice versa for $k_{\|} \rho_i > 1.08$.\footnote{The apparent singularity in the expression (\ref{gamma_prlmodes}$b$) for $\gamma$ at $k_{\|} \rho_i \approx 1.08$ -- that is, the value of $k_{\|} \rho_i$ at which $\mathrm{Re} \, Z\left({1}/{k_{\|} \rho_i}\right) \approx -k_{\|} \rho_i$ -- is an artifact, because the numerator also vanishes at this value.}

We can use \eqref{eqn:prlmodes} to evaluate $\varpi$ and $\gamma$ as the ion distribution function evolves from a Maxwellian via a bi-Maxwellian distribution to the non-bi-Maxwellian state associated with scattering by oblique firehoses. In a plasma with a bi-Maxwellian ion distribution, equations (\ref{gamma_prlmodes}) simplify considerably, because
we have 
 \begin{equation}
 \frac{v_{\mathrm{th}i}}{v}\frac{\partial {f}_{i0}}{\partial \xi} = -2 \Delta_i \frac{v_{\|}}{v_{\mathrm{th}i}} {f}_{\rm M} , \quad \mathcal{I}_{\pm}|_{v_{\|} = v_{\|\mathrm{res}}^{\pm}} = \frac{2}{v_{\mathrm{th}i}}\left(\mp \frac{\Delta_i}{k_{\|} \rho_i} - \frac{v_{\rm wv}}{v_{\mathrm{th}i}}\right) {f}_{\rm M} , \label{pitchangle_gradient_biMax}
 \end{equation}
so that 
\begin{subeqnarray}
\varpi  &  \approx & \pm \Delta_i \pm \Omega_i \frac{k_{\|}^2 \rho_i^{2}}{\beta_i} \mathcal{G}(k_{\|} \rho_i) \left\{\left[\mathcal{G}(k_{\|} \rho_i)\right]^{2} + \frac{\upi}{k_{\|}^2 \rho_i^2} \exp{\left(-\frac{2}{k_{\|}^2 \rho_i^2}\right)}\right\}^{-1} , \\
 \gamma & \approx & - \Omega_i \frac{\sqrt{\upi} k_{\|} \rho_i}{\beta_i} \exp{\left(-\frac{1}{k_{\|}^2 \rho_i^2}\right)} \left\{\left[\mathcal{G}(k_{\|} \rho_i)\right]^{2} + \frac{\upi}{k_{\|}^2 \rho_i^2} \exp{\left(-\frac{2}{k_{\|}^2 \rho_i^2}\right)} \right\}^{-1} . \quad   \label{gamma_prlmodes_bimax}
\end{subeqnarray} 
In the plasma's initial state, in which $\Delta_i = 0$, the forward-propagating modes are indeed those associated with the `$-$' root, as expected, and so are left-handed, because the numerator of (\ref{gamma_prlmodes_bimax}$a$) is negative for $k_{\|} \rho_i < 1.08$. These equations further imply that $\gamma < 0$ initially. In the bi-Maxwellian stage, equation (\ref{gamma_prlmodes_bimax}$a$) indicates that, for $\Delta_i < 0$, the `$-$' mode transitions from being forward-propagating to backward-propagating at a smaller value of $k_{\|} \rho_i$ than for a Maxwellian distribution (and vice versa for the `$+$' mode). When $\Delta_i < -2/\beta_i$, $\varpi < 0$ at all wavenumbers $k_{\|} \sim \rho_i^{-1}$ for the `$-$' mode, and $\varpi > 0$ for the `$+$' mode. 
  However, both the `$+$' and `$-$' mode are still damped at this stage by ions with $v_{\|} \approx \pm \Omega_i/k_{\|} $. Finally, in the state with the non-bi-Maxwellian distribution, scattering by the oblique firehoses causes ${\partial {f}_{i0}}/{\partial \xi}|_{v_{\|} = v_{\|\mathrm{res}}}$ to decrease in magnitude near $v_{\|} \approx \pm 2 v_{\mathrm{th}i}$, and $\varpi$ does not change its sign when these resonant particles start to be isotropised, because (\ref{gamma_prlmodes}$a$) implies that $\varpi$ is less sensitive than $\gamma$ to the value of $f_i$ at specific $v_{\|}$. Once $\mathcal{I}_{\pm}$ reverse their sign for forward- and backward-propagating resonant parallel modes, respectively, it then follows that their growth rate becomes positive.

\begin{figure}
  \centering
  \includegraphics[width=\linewidth]{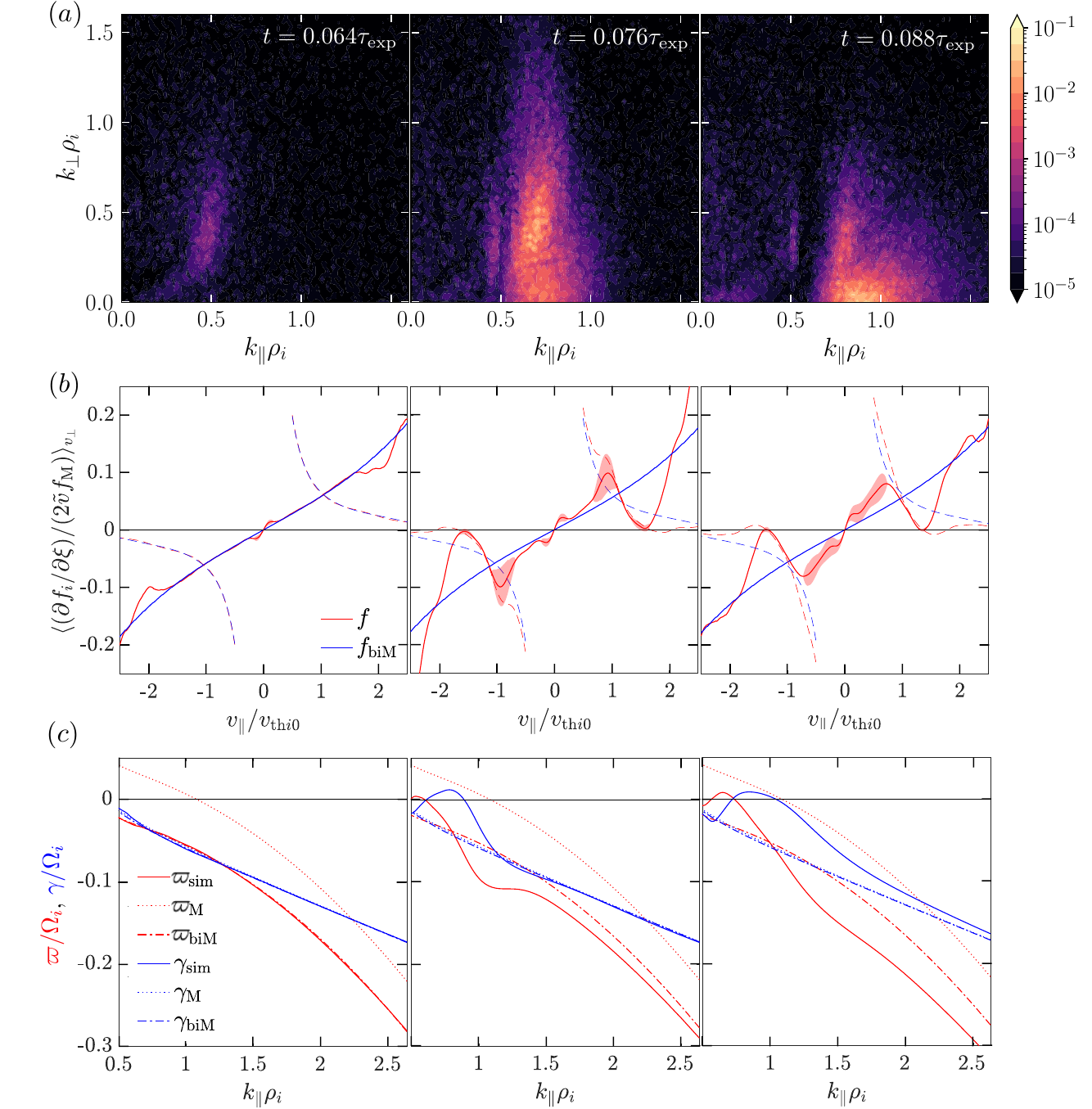}
  \caption{$(a)$ Two-dimensional magnetic-energy spectra of the firehose fluctuations in run~CV at a selection of different times around the emergence of the parallel secondary firehose instability. $(b)$ Pitch-angle gradient of the ion distribution function $f$ divided by $2 \tilde{v} f_{\rm M}$ (solid red line), where $\tilde{v} = v/v_{\mathrm{th}i}$, and $f_{\rm M}$ is a Maxwellian distribution with the same temperature as $f$, averaged over $v_{\perp}$, at the same times shown in panel $(a)$. The solid blue line is the analogous quantity, but calculated using $f_{\rm biM}$, the bi-Maxwellian distribution function with the same parallel and perpendicular temperatures as $f$. The dotted red and blue lines show $\tilde{v}_{\rm wv} = {v}_{\rm wv}/v_{\mathrm{th}i}$ calculated using a linear dispersion relation solver that finds the complex frequency of low-frequency modes with an given input numerical distribution function. $(c)$  Approximate real frequencies (red) and damping rates (blue) [cf.~(\ref{gamma_prlmodes})] of the `$-$' root, for $f$ (solid lines), $f_{\rm M}$ (dotted lines), and $f_{\rm biM}$ (dot-dashed lines) at the same times indicated in panel $(a)$.}
\label{fig:sims_secondparfire_AE}
\end{figure}

This evolution is illustrated using one of our simulations (run CV, an `asymptotic' Alfv\'en-enabling simulation) in figure~\ref{fig:sims_secondparfire_AE}. 
Panel $(a)$ shows the two-dimensional magnetic-energy spectrum at various times in the simulation around the time at which the parallel modes are observed; panel $(b)$ shows the pitch-angle gradient of the ion distribution function $f_{i}$ at those same times; and panel $(c)$ shows $\varpi$ and $\gamma$ of the `$-$' modes, which we calculate using the approximate expressions (\ref{gamma_prlmodes}). The integrals in these expressions for $\varpi$ and $\gamma$ are evaluated numerically, taking as their input the numerical distribution function. We see that, in the initial stages of the growth of oblique firehose modes (Figure~\ref{fig:sims_secondparfire_AE}$a$, left panel), when  $f_{i}$ is still approximately bi-Maxwellian (and so the pitch-angle gradient of $f_i$ is well described by (\ref{pitchangle_gradient_biMax}) -- see figure~\ref{fig:sims_secondparfire_AE}$b$, left panel), parallel `$-$' modes with $k_{\|} \rho_i > 0.5$ have a negative sign, but are damped (Figure~\ref{fig:sims_secondparfire_AE}$c$, left panel). However, concurrently with the emergence of parallel modes (Figure~\ref{fig:sims_secondparfire_AE}$a$, middle panel), the ion distribution function becomes non-bi-Maxwellian (Figure~\ref{fig:sims_secondparfire_AE}$b$, middle panel), and parallel ion-cyclotron modes become linearly unstable (Figure~\ref{fig:sims_secondparfire_AE}$c$, middle panel), albeit over quite a narrow range of wavenumbers. For the fastest growing modes, 
\begin{equation}
v_{\mathrm{th}i} \mathcal{I}_{-} = \frac{v_{\mathrm{th}i}}{v} \frac{\partial {f}_i}{\partial \xi}-\frac{2}{k_{\|} v_{\mathrm{th}i}} \varpi\left(-\frac{1}{k_{\|}\rho_i}\right) {f}_{\rm M} \ll \left|\frac{\varpi}{k_{\|} v_{\mathrm{th}i}}\right| {f}_{\rm M} \sim \frac{1}{\beta_i} {f}_{\rm M} , 
\end{equation}
and so, in contrast to the bi-Maxwellian, $f_i$ has the property that its pitch-angle gradient is approximately equal to twice the (normalised) phase velocity $v_{\rm wv} \sim v_{\mathrm{th}i}/\beta_{i}$ of the linear modes that $f_i$ supports. In other words, the ion distribution function's anisotropy is constrained by the parallel modes' phase velocity. As the simulation progresses further, the unstable parallel modes tend to acquire slightly larger wavenumbers (Figure~\ref{fig:sims_secondparfire_AE}$a$, right panel), with those that were initially unstable becoming forward-propagating, stable modes again (Figure~\ref{fig:sims_secondparfire_AE}$c$, right panel). 

In summary, scattering by the ion-Larmor-scale oblique firehose modes that initially arise due to the negative pressure anisotropy is responsible for the development of a non-bi-Maxwellian distribution function, which in turn is subject to an instability of right-handed parallel modes that would not be present if the distribution function were to have remained bi-Maxwellian. This secondary firehose instability could also explain the persistent parallel modes with $k_{\|} \rho_i \sim 1$ seen in regions of negative pressure anisotropy within the hybrid-kinetic simulation of a long-wavelength, large-amplitude Alfv\'en wave reported by~\citet{Squire_2017}.

\subsection{Why $(\Delta_i)_{\rm sat} \simeq -1.6/\beta_{\|i}$ in high-$\beta_i$ Alfv\'en-enabling states} \label{sec:sims:res:stapresaniso}

One finding of our simulation results that was not anticipated from the linear theory of the firehose instability outlined in section~\ref{sec:theory} is that, in saturation, $\Delta_i \simeq -1.6/\beta_{\|i}$. If, as we argued in section~\ref{sec:sims:res:satfirehose}, the saturation of the oblique firehose instability can be described by quasi-linear theory, then it must be the case that the plasma attains a saturated state that is close to marginality with respect to the oblique firehose instability. However, we showed in section~\ref{linfiretheory_numres} that, in a bi-Maxwellian plasma,   the oblique firehose instability's threshold is given by $\Delta_{\rm cr} \simeq -1.35/\beta_{\|i} > (\Delta_i)_{\rm sat}$. Na\"{i}vely, it might therefore be expected that oblique firehose modes should still grow, and, adopting the estimate (\ref{firehosescaling_thresholdreg}) for these modes' growth rate, will do so at a rate that far exceeds the rate of the plasma's expansion, $\gamma_{\perp \mathrm{f}} \approx 0.7 \Omega_i/\beta_{i} \gtrsim 19 \beta_{i}^{0.6}/\tau_{\rm exp,eff}$ . 

This seeming contradiction is resolved by the fact that the plasma's ion distribution $f_i$ is not well modelled as a bi-Maxwellian distribution, but instead has a distinct form of anisotropy. More specifically, as was illustrated in section \ref{sec:sims:res:distfunc}, the anisotropy of $f_i$ is concentrated at smaller characteristic values of $v$ compared with those of a bi-Maxwellian. This has the consequence of bringing the threshold of ion-Larmor-scale oblique firehose modes closer to the fluid firehose threshold. That the modified form of the anisotropy alters the oblique firehose instability's threshold can be demonstrated mathematically by considering the leading-order FLR corrections to the fluid firehose threshold, which are computed for a general ion distribution function in Appendix~\ref{append_firehose_threshold_condition} [cf. (\ref{threshold_nonbimax_Append})]: 
\begin{equation}
\frac{2}{\beta_{\|i}} + \Delta_i +k_{\|}^2 {\rho}_i^2 \mathcal{A}_{4i} - \frac{3}{16} k_{\perp}^2 {\rho}_i^2 \mathcal{B}_{4i} = \mathcal{O}(\Delta_i k^4 
{\rho}_i^4), \label{threshold_nonbimax}
\end{equation}
where $\mathcal{A}_{4i}$ and $\mathcal{B}_{4i}$ are given by [cf. (\ref{fourthordermoments})]
\begin{subeqnarray}
\mathcal{A}_{4i} & = & -8 \upi \frac{T_{\|i}}{T_{i}} \frac{1}{n_{i} v_{\mathrm{th}i}^4} \int_{-\infty}^{\infty} \mathrm{d} {v}_{\|} \int_0^{\infty} \mathrm{d} {v}_{\bot} \, {v}_{\bot} {v}_{\|}^2 \left({v}_{\|}^2-\frac{1}{4} {v}_{\perp}^2 \right) {f}_{i} , \\
 \mathcal{B}_{4i} & = & -4 \upi \frac{T_{\|i}}{T_{i}} \frac{1}{n_{i} v_{\mathrm{th}i}^4} \int_{-\infty}^{\infty} \mathrm{d} {v}_{\|} \int_0^{\infty} \mathrm{d} {v}_{\bot} \, {v}_{\bot}^3 \left({v}_{\|}^2-\frac{1}{2} {v}_{\perp}^2 \right) {f}_{i} . \label{fourthordermoments_coeffs}
\end{subeqnarray}
Inspecting the velocity-space integrands in (\ref{fourthordermoments_coeffs}) and comparing them with the analogous integrand for the pressure anisotropy, 
\begin{equation}
\Delta_i = - 4 \upi \frac{T_{\|i}}{T_{i}} \frac{1}{n_{i} v_{\mathrm{th}i}^2} \int_{-\infty}^{\infty} \mathrm{d} {v}_{\|} \int_0^{\infty} \mathrm{d} {v}_{\bot} \, {v}_{\bot} \left({v}_{\|}^2-\frac{1}{2} {v}_{\perp}^2 \right) {f}_{i} , \\
\end{equation}
it is clear that concentrating the anisotropy of a distribution function at smaller characteristic velocities will in general reduce the values of the ratios $\mathcal{A}_{4i}/\Delta_i$ and $\mathcal{B}_{4i}/\Delta_i$. Thus, the distribution functions attained in the saturated state of the firehose instability simultaneously maintain comparatively larger values of $(\Delta_i)_{\rm sat}$ than a bi-Maxwellian distribution and smaller values of $\mathcal{A}_{4i}$ and $\mathcal{B}_{4i}$. { Computing $\mathcal{A}_{4i}$ and $\mathcal{B}_{4i}$ directly for our `asymptotic' Alfv\'en-enabling regime simulation (run CV), we find $\mathcal{A}_{4i} \simeq -1.6/\beta_{\|i}$, and $\mathcal{B}_{4i} \simeq -0.8/\beta_{\|i}$; setting $k_{\|} \rho_i \approx k_{\perp} \rho_i \simeq 0.5$ to match those of the dominant oblique firehose mode, (\ref{threshold_nonbimax}) predicts that $(\Delta_i)_{\rm sat} \approx -1.6/\beta_{\|i}$. This agrees very well with its actual value in the simulation.}

An outstanding question that follows naturally from our result is why, in prior $\beta_{\|i} \gtrsim 1$ simulations of firehose-susceptible plasmas~\citep[see, e.g.,][]{Hellinger_2008,Hellinger_2019,Bott_2021}, it was found that $(\Delta_i)_{\rm sat} \simeq -1.4/\beta_{\|i}$, in closer agreement with the bi-Maxwellian threshold of the oblique firehose instability. The most plausible explanation of this (small) discrepancy pertains to the different linear characteristics of firehose instabilities at $\beta_{\|i} \gg 1$ vs. $\beta_{\|i} \gtrsim 1$. Specifically, as we demonstrated in section \ref{linfiretheory_resparfire}, when $\beta_{\|i} \gtrsim 1$, the growth rate of resonant parallel firehose modes tends to be comparable to those of oblique modes. The presence of a saturated population of such modes, which would not be present in high-$\beta_i$ plasma, would be expected to affect the specific value of $\Delta_i$ attained in the saturated state. We note that, though the specific values of $(\Delta_i)_{\rm sat}$ are distinct, both are such that the plasma still attains an Alfv\'en-enabling state.

\subsection{The perturbed magnetic energy of firehose fluctuations in saturation: part I} \label{sec:sims:res:satfirehose}

It was shown in section \ref{sec:sims:res:magfluc} that the relationship between the perturbed magnetic energy associated with the firehose fluctuations in saturation and macroscopic plasma parameters is not simply a power law across all values of the key parameter $\tau_{\rm exp} \Omega_i/\beta_{\|i}^{1.6}$, with a change occurring near the transition between the Alfv\'en-enabling and  Alfv\'en-inhibiting states. This implies that the saturation physics in these two states must be distinct. 

Such a conclusion is, at first glance, counter-intuitive. For the ion-Larmor-scale firehose modes that we observe in both the Alfv\'en-enabling and Alfv\'en-inhibiting states, which to a good approximation consist of perturbations to the direction of the magnetic field, a saturated state is most plausibly maintained via pitch-angle scattering at a rate sufficient to maintain near-marginality with respect to the firehose instability's threshold. Assuming that the rate $\nu_{\rm eff} \sim \beta_i/\tau_{\rm exp}$ of pitch-angle scattering by (ion-Larmor-scale) fluctuations is related to their amplitude $\delta B_{\rm f}/B_0$ by $\nu_{\rm eff} \sim \Omega_i \delta B_{\rm f}^2/B_0^2$ -- in effect, adopting a quasilinear scattering model based on the assumption that $\delta B_{\rm f} \ll B_0$ -- we deduce that
\begin{equation}
\frac{\delta B_{\rm f}^2}{B_0^2} \sim \frac{\beta_i}{\tau_{\rm exp} \Omega_i} . \label{obliquefirehose_dB_scale}
\end{equation}
We note that such a quasilinear model should be self-consistent for any firehose-susceptible plasma in an Alfv\'en-enabling state, because ${\delta B_{\rm f}^2}/{B_0^2} \sim {\beta_i}/{\tau_{\rm exp} \Omega_i} \lesssim \beta_i^{-0.6} \ll 1$. 

This argument, which provides testable predictions for the dependence of ${\delta B_{\rm f}^2}/{B_0^2}$ on $\beta_i$, $\tau_{\rm exp}$, and $\Omega_i$, only partially accounts for the results of our numerical study. The scaling $\nu_{\rm eff} \sim \beta_i/\tau_{\rm exp}$ for the effective collisionality is indeed the same as reported in section \ref{sec:boxavcollisionality}. However, the scaling (\ref{obliquefirehose_dB_scale}) only agrees for our simulations in Alfv\'en-inhibiting states, not Alfv\'en-enabling ones. We conclude that the argument must overlook aspects of firehose-instability saturation that affect the scaling of the perturbed magnetic energy. 

In order to resolve this discrepancy, a more nuanced understanding of scattering of particles by both oblique firehose and secondary parallel firehose modes in Alfv\'en-enabling states -- and how this leads to the saturation of both types of firehose instability -- is required. We therefore characterise an effective `firehose collision operator' in the next section.   

\section{Effective collisionality for the firehose instability} \label{sec:collisionality}

\subsection{Overview}

One key property of the firehose instability in its saturated state is that it provides the plasma
with an effective collisionality, $\nu_{\rm eff}$. 
Particles in the plasma experience this collisionality predominantly as pitch-angle scattering. In this section, 
we move beyond previous velocity-averaged estimates of this collisionality, and propose a model in the Alfv\'en-enabling state for the velocity-dependent pitch-angle scattering rate of particles with speeds of order the thermal speed. This allows us to construct a simple `effective firehose collision operator', given by
\begin{equation}
    \mathfrak{C}_{\rm f}[f] = \frac{1}{2} \frac{\partial}{\partial \xi}\Bigg\{(1-\xi^2)\nu_{\rm eff,pl}(v \xi)\left[\frac{\partial f_{1}}{\partial \xi} - 2 \tilde{w} \tilde{v}_{\rm wv,pl}(v \xi) f_{\rm M}\right] + (1-\xi^2)\nu_{\rm eff,ob}(v \xi) \frac{\partial f_{1}}{\partial \xi} \Bigg\} \, , \label{quasilinearscatteringrate_par_intro}
\end{equation}
where
\begin{subeqnarray}
     \nu_{\rm eff,pl}(v_{\|}) & = & \frac{0.15 v_{\mathrm{th}i}}{|{v}_{\|}|} \frac{\beta_i^{1/4} \Omega_i^{3/4}}{\tau_{\rm exp}^{1/4}} \exp{\left[-0.31(\tau_{\rm exp}\Omega_i)^{1/2} \left(\frac{v_{\mathrm{th}i}}{|{v}_{\|}|}-1.2\right)^2\right]} \nonumber \\
     && \mbox{} + 0.09 H(k_{\|} \rho_i - 1.2) ({\tau \Omega_i})^{-1/2} (k_{\|} \rho_i)^{-2.7} , \\
     v_{\rm wv,pl}(v_{\|}) & = & \mathrm{sgn}(v_{\|})\frac{v_{\mathrm{th}i}}{\beta_i}\left(4.9-2.9 \frac{|v_{\|}|}{v_{\mathrm{th}i}} \right) ,  \\
     \nu_{\rm eff,ob}(v_{\|})
     & = & \frac{1.4 v_{\mathrm{th}i}}{|{v}_{\|}|} \frac{\beta_i}{\tau_{\rm exp}} \exp{\left[-13 \left(\frac{v_{\mathrm{th}i}}{|{v}_{\|}|}-0.75\right)^2\right]}  ,
    \label{colloperator_scatter_intro}
\end{subeqnarray}
where $H(x)$ denotes the Heaviside step function. 
We then compare the predicted properties of this collision operator with two different numerical diagnostics applied to our simulations, and confirm that the model collision operator accounts for both the characteristic anisotropy of the ion distribution function and the root mean square of the firehose fluctuations' magnetic-field strength. In turn, this collision operator allows us to advance our qualitative understanding of the anomalous scaling of the perturbed magnetic energy in Alfv\'en-enabling states discussed in section \ref{sec:sims:res:satfirehose}.

\subsection{An effective firehose collision operator} \label{sec:firehosecollopp}

 Beyond accounting for the saturated amplitude of firehose-unstable modes, there are two other motivations for investigating the velocity-dependent collisionality associated with firehose fluctuations. First, it is the velocity dependence of effective collisions that determines the ion distribution function's anisotropy, and thereby the specific saturation value of the pressure anisotropy at which further growth of firehose-unstable modes is inhibited. As discussed in section~\ref{sec:theory}, long-wavelength firehose modes are insensitive to the form of the ion distribution function's anisotropy, but kinetic-scale firehose modes are sensitive to it. Because it is these kinetic-scale modes that have the least stringent threshold for instability, the specific form of anisotropy is pertinent. Secondly, for certain other problems in astrophysical plasmas such as modelling cosmic-ray transport, understanding the effective collisionality of particles with specific velocities due to firehose fluctuations is a crucial component of the problem's solution.

In general, characterising the effective collision operator associated with arbitrary electromagnetic fluctuations, which could cause slowing, parallel diffusion, and/or perpendicular diffusion of particles, is quite challenging. However, in the specific case of the effective collision operator associated with firehose fluctuations, various simplifying assumptions can be reasonably adopted. Based on the small amplitude of firehose fluctuations realised in the Alfv\'en-enabling state (figure~\ref{fig:sims_dB_fig1}$(b)$ implies that the total magnetic energy of fluctuations satisfies $\delta B_{\rm f}^2/B_0^2 \ll \beta_{\|i}^{-0.6} \ll 1$) and their broad spectra (see figure~\ref{fig:sims_dB_fig2}), we assume that the collision operator can be described by quasilinear theory. Furthermore, we neglect the electric  
fields associated with the firehose fluctuations on the grounds that the electric contribution to the total Lorentz force is subdominant to the magnetic force; it follows from Faraday's law that, for firehose fluctuations, $c \delta \bb{E}/|\bb{v} \btimes \delta\bb{B}| \sim \omega/k v_{\mathrm{th}i} \sim 1/\beta_i \ll 1$. Finally, we assume (based on our simulation results) that the magnetic-field perturbations caused by the firehose instability satisfy $\delta \boldsymbol{B} \approx \delta \boldsymbol{B}_{\perp}$. Taking these assumptions together, the quasilinear collision operator arising from magnetic fluctuations is simply a resonant pitch-angle-scattering operator that isotropises the distribution function in the frame moving at the (parallel) phase velocity $v_{\rm wv} = v_{\mathrm{th}i} \tilde{v}_{\rm wv}$ of the firehose modes (the \emph{wave frame}) at a velocity-dependent scattering rate $\nu_{\rm eff}(v_{\|}',v_{\perp})$ given by~\citep[e.g.,][]{Kulsrud_1969}
\begin{equation}
\nu_{\rm eff}(v_{\|}',v_{\perp}) = \upi \frac{\Omega_i^2}{v_{\|}'}\frac{ \tilde{\mathcal{E}}_{B}\bigl({\Omega_i}/{v_{\|}'}\bigr)}{B_0^2/8\upi} , \quad \tilde{\mathcal{E}}_{B}(k_{\|}) \equiv \sum_{n \neq 0} n^2 \int \mathrm{d}^2 \boldsymbol{k}_{\perp} \, E_{B}(n k_{\|},k_{\perp}) \frac{[\besselJ_n(k_{\perp} v_{\perp}/\Omega_i)]^2}{k_{\perp}^2 v_{\perp}^2/\Omega_i^2} \, , \label{quasilinearscatteringrate_full}
\end{equation}
where the primes denote parallel velocities evaluated in the wave frame, and $\besselJ_{n}(x)$ is the $n$th order Bessel function of the first kind. 

If we also assume that both the anisotropy of the distribution function and $v_{\rm wv}$ are small -- more precisely, that $({\partial f_i}/{\partial \xi})/f_{\rm M} \sim \tilde{v}_{\rm wv} \sim 1/\beta_i \ll 1$ -- it can be shown~\citep[see, e.g., ][]{Yerger_2024} that the quasilinear pitch-angle operator in the plasma's rest frame has the following form:
\begin{equation}
\mathfrak{C}[f] = \frac{1}{2} \frac{\partial}{\partial \xi}\left\{(1-\xi^2)\nu_{\rm eff}(v,\xi)\left[\frac{\partial f_{i}}{\partial \xi} - 2 \tilde{w} \tilde{v}_{\rm wv}(v,\xi) f_{\rm M}\right] \right\} \, ,
\end{equation}
where we remind the reader that $\xi = v_{\|}/v$ is the pitch angle, $v \equiv \sqrt{v_{\|}^2 + v_{\perp}^2}$ is the particle speed, and $\tilde{v}_{\rm wv}(v,\xi)$ is the parallel phase velocity of the firehose modes with which specific particles having peculiar velocity ($v,\xi$) are resonant. Note that if there are separate populations of modes with different characteristics that are responsible for scattering -- as is the case in firehose-infested plasma in an Alfv\'en-enabling state, in which there are both oblique firehose and secondary parallel firehose modes -- a collision operator associated with both populations is required. 

Finally, to be able to write down simple expressions for $\nu_{\rm eff}(v,\xi)$ and $\tilde{v}_{\rm wv}(v,\xi)$, we make one final assumption: that the fluctuations can be treated as being quasi-parallel in the sense that, for most particles, $v_{\perp}^2 \ll \Omega_i^2/k_{\perp}^2$. The assumption simplifies the sum in  (\ref{quasilinearscatteringrate_full}): all terms with $|n| > 1$ are then negligible, and the Bessel functions in the $n = \pm 1$ terms can be simplified using the identity $\besselJ_{\pm 1}(x) \approx \pm (x/2)(1-x^2/8+\dots)$ for $x \ll 1$. Under this final assumption, the effective pitch-angle scattering operator $\mathfrak{C}[f]$ associated with firehose modes in the Alfv\'en-enabling regime simplifies to
\begin{equation}
    \mathfrak{C}[f] = \frac{1}{2} \frac{\partial}{\partial \xi}\Bigg\{(1-\xi^2)\nu_{\rm eff,pl}(v \xi)\left[\frac{\partial f_{1}}{\partial \xi} - 2 \tilde{v} \tilde{v}_{\rm wv,pl}(v \xi) f_{\rm M}\right] + (1-\xi^2)\nu_{\rm eff,ob}(v \xi) \frac{\partial f_{1}}{\partial \xi} \Bigg\} \, , \label{quasilinearscatteringrate_par}
\end{equation}
where the velocity-dependent pitch-angle scattering rates $\nu_{\rm eff,pl}$ and $\nu_{\rm eff,ob}$ associated with the secondary parallel firehose modes and the oblique firehose modes, respectively, are now only functions of the parallel particle velocity $v_{\|} = v \xi$; they are directly related to the magnetic-energy spectra of the two firehose populations by
\begin{subeqnarray}
     \nu_{\rm eff,pl}(v_{\|}) & \simeq & \frac{\upi}{2} \frac{\Omega_i^2}{v_{\|}}\frac{ E_{B,\mathrm{pl}}\!\left({\Omega_i}/{v_{\|}}\right)}{B_0^2/8\upi} , \\
     \nu_{\rm eff,ob}(v_{\|}) & \simeq & \frac{\upi}{2} \frac{\Omega_i^2}{v_{\|}}\frac{ E_{B,\mathrm{ob}}\!\left({\Omega_i}/{v_{\|}}\right)}{B_0^2/8\upi}
     . \qquad \label{colloperator_scatter}
\end{subeqnarray}
Here, $E_{B,\mathrm{pl}}(k_{\|})$ and $E_{B,\mathrm{ob}}(k_{\|})$ are the 1D parallel magnetic energy spectra of the secondary parallel and oblique firehose fluctuations, respectively, while
\begin{equation}
\tilde{v}_{\rm wv,pl}(v_{\|}) = \frac{{v}_{\rm wv,pl}(v_{\|})}{v_{\mathrm{th}i}} \simeq \varpi\bigl({\Omega_i}/{v_{\|}}\bigr) \frac{v_{\|}}{v_{\mathrm{th}i}}
\end{equation}
is an approximation (to leading order in the small parameter $1/\beta_{i}$) of the parallel phase velocity of the modes with which particles having parallel velocity $v_{\|}$ are resonant. 
Because the oblique firehose modes do not have a parallel phase velocity, the pitch-angle scattering operator associated with them is already in the plasma rest frame. The quasi-parallel assumption is  reasonable for the secondary parallel firehose modes, but is less clearly appropriate for the oblique firehose modes. For the latter case, we estimate the error introduced in this approximation by using the numerical result that, in the saturated state of the firehose instability, $k_{\perp} \lesssim 0.5 \rho_i^{-1}$. It follows that the magnitude of the first-order term in the Bessel function expansion is $k^2 v_{\perp}^2/8 \Omega_i^2 \approx v_{\perp}^2/16 v_{\mathrm{th}i}^2$. For particles with $v_{\perp} \lesssim 2 v_{\mathrm{th}i}$ (the majority of thermal particles), the error introduced by the approximation is therefore 25\% or less. 

Thus we have constructed a simple model for the effective firehose collision operator that takes as its inputs two velocity-dependent scattering rates ($\nu_{\rm eff,pl}(v_{\|})$ and $\nu_{\rm eff,ob}(v_{\|})$) and the parallel phase velocity $v_{\rm wv,pl}(v_{\|})$ of the secondary firehose modes. The scattering rates are given directly by the 1D parallel magnetic-energy spectra of oblique and secondary parallel firehose modes $E_{B,\mathrm{ob}}(k_{\|})$ and $E_{B,\mathrm{pl}}(k_{\|})$, respectively, while $v_{\rm wv,pl}(v_{\|})$ depends on the real frequency $\varpi(k_{\|})$ of the secondary firehose modes. Therefore, to compute the effective firehose collision operator, all that remains is to determine $E_{B,\mathrm{ob}}(k_{\|})$, $E_{B,\mathrm{pl}}(k_{\|})$, and $\varpi(k_{\|})$. We compute these functions numerically for all of the expanding-box simulations that we have conducted that attain Alfv\'en-enabling states. In order to obtain a time-averaged collision operator, for each simulation we choose a time interval during which the firehose instability has saturated, and then calculate averaged values of the oblique and parallel magnetic-energy spectra and the secondary firehose mode frequencies\footnote{A time-averaged collision operator is arguably of most relevance for astrophysical applications, because the time-dependent evolution of the collisionality -- first, the progression to a saturated state, then fluctuations around the average saturated state -- occurs over timescales that are much shorter than the timescale over which the collisionality affects the plasma's evolution.}. 

\begin{figure}
  \centering
  \includegraphics[width=\linewidth]{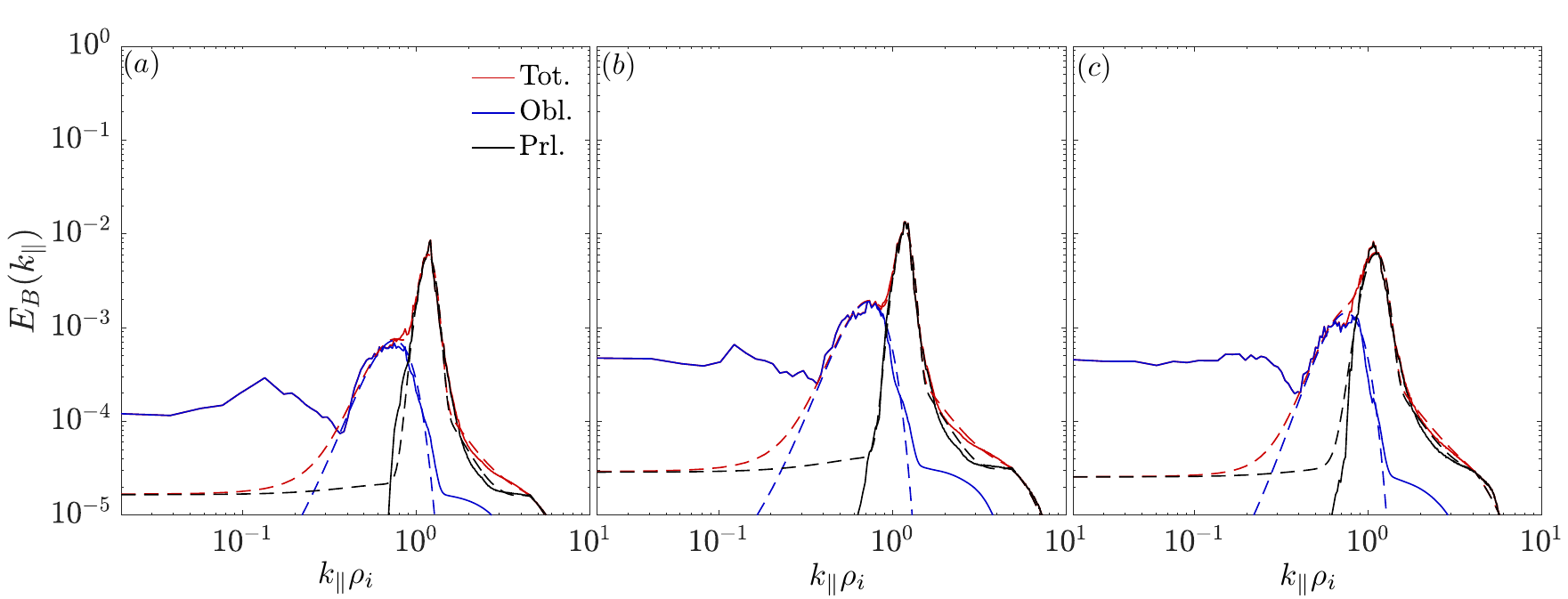}
  \caption{$(a)$ 1D (parallel) magnetic-energy spectrum $E_B(k_{\|})$ of all firehose fluctuations in the saturated, Alfv\'en-enabling state of run CV (solid red line). Also plotted are the magnetic-energy spectra of non-quasi-parallel fluctuations (blue solid line) and quasi-parallel ones (black solid line), as well as fits for these spectra discussed in the main text (dashed lines; equations~\eqref{magnengspec}). $(b)$ Same as in panel $(a)$, but for run DVI. $(c)$ Same as in panel $(a)$, but for run BIV.}
\label{fig:collsims_EBchar_AE}
\end{figure}

To calculate $E_{B,\mathrm{pl}}(k_{\|})$, we first apply a mask to the total (time-averaged) magnetic-energy spectrum $E_{B}(k_{\|},k_{\perp})$ to isolate the secondary parallel firehose modes; this mask covers the same region of $(k_{\|},k_{\perp})$-space as the one circumscribed by the white-dashed line in figure~\ref{fig:sims_dB_fig2}. We then integrate the masked spectra over all perpendicular wavenumbers to obtain $E_{B,\mathrm{pl}}(k_{\|})$. $E_{B,\mathrm{ob}}(k_{\|})$ is then calculated by subtracting $E_{B,\mathrm{pl}}(k_{\|})$ from the total parallel 1D magnetic-energy spectrum $E_{B}(k_{\|}) \equiv  \int \rmd k_{\perp} \,E_{B}(k_{\|},k_\perp)$. We show $E_{B,\mathrm{pl}}(k_{\|})$, $E_{B,\mathrm{ob}}(k_{\|})$ and $E_{B}(k_{\|})$ from three representative simulations in Alfv\'en-enabling states in figures~\ref{fig:collsims_EBchar_AE}$(a)$, $(b)$, and~$(c)$, respectively. 

Having calculated $E_{B,\mathrm{pl}}(k_{\|})$ and $E_{B,\mathrm{ob}}(k_{\|})$ numerically, we then fit both spectra with simple analytical functions of the form 
\begin{subeqnarray}
     E_{B,\mathrm{pl}}\!\left(k_{\|}\right) & \simeq & \frac{B_0^2}{8\upi}\left\{\frac{ \bar{E}_{B,\mathrm{pl}}}{\sqrt{\upi}\Delta {k}_{\|,\rm pl} \rho_i} \exp{\left[-\frac{(k_{\|}-{k}_{\rm \|, pl})^2 }{\Delta {k}_{\|,\rm pl}^{2}}\right]} + H(k_{\|} - {k}_{\rm \|, pl}) \frac{\bar{E}_{B,\mathrm{tail}}}{(k_{\|} \rho_i)^{p_{\rm tail}}} \right\}, \qquad \quad \\
     E_{B,\mathrm{ob}}\!\left(k_{\|}\right)
     & \simeq & \frac{B_0^2}{8\upi}\frac{\bar{E}_{B,\mathrm{ob}}}{\sqrt{\upi}\Delta {k}_{\|,\rm ob} \rho_i} \exp{\left[-\frac{(k_{\|} -{k}_{\rm \|, ob})^2}{\Delta {k}_{\|,\rm ob}^2}\right]} , \qquad \label{magnengspec}
\end{subeqnarray}
where ${k}_{\rm \|, pl}$ (${k}_{\rm \|, ob}$) is the wavenumber at which $E_{B,\mathrm{pl}}(k_{\|})$ ($E_{B,\mathrm{ob}}(k_{\|})$) attains its maximum, $\Delta {k}_{\|,\rm pl}$ ($\Delta {k}_{\|,\rm ob}$) is the characteristic width of the $k_{\|}$ interval over which $E_{B,\mathrm{pl}}(k_{\|})$ ($E_{B,\mathrm{ob}}(k_{\|})$) extends, and $\bar{E}_{B,\mathrm{pl}}$ ($\bar{E}_{B,\mathrm{ob}}$) is the total energy in the secondary parallel (oblique) firehose fluctuations. We also find it necessary to model the high-$k_{\|}$ wavenumber of the distribution of secondary parallel firehose modes with a power-law tail (of amplitude $\bar{E}_{B,\mathrm{tail}}$, and power-law index $p_{\rm tail}$); although the magnetic energy associated with modes of such high wavenumbers is much smaller than modes with $k_{\|} \rho_i \sim 1$, such modes nonetheless play a key role in determining the anisotropy of the ion distribution function in Alfv\'en-enabling states (see section~\ref{testcollop_anidistfunc}), and so cannot be disregarded. 

These particular functional fits are not derived analytically, but we find empirically that they describe the numerical spectra well. 
Practically, we first determine $\bar{E}_{B,\mathrm{pl}}$ and $\bar{E}_{B,\mathrm{ob}}$ by integrating each spectra, and then determine best-fit values to the other parameters, weighting the estimates by $E_{B,\mathrm{pl}}(k_{\|})$ and $E_{B,\mathrm{ob}}(k_{\|})$, respectively. For determining the fits for the oblique firehose spectrum, we exclude all parallel wavenumbers $k_{\|} < 0.4 \rho_i^{-1}$, because the spectrum of these longer-wavelength fluctuations is not well described by an analytic fit of the form (\ref{magnengspec}$b$), and such modes do not affect the anisotropy of thermal ions. We fit the power-law tail of the spectrum of parallel modes by first fitting the latter's peak with the Gaussian analytic form, then subtracting this fit and the spectrum of the noise from the total spectrum, and fitting the power law to what remains. In figures \ref{fig:collsims_EBchar_AE}$(a)$ and $(b)$, the good agreement between our fits of the form (\ref{magnengspec}) to the 1D magnetic-energy spectra of two representative simulations is illustrated. 

\begin{figure}
  \centering
  \includegraphics[width=0.9\linewidth]{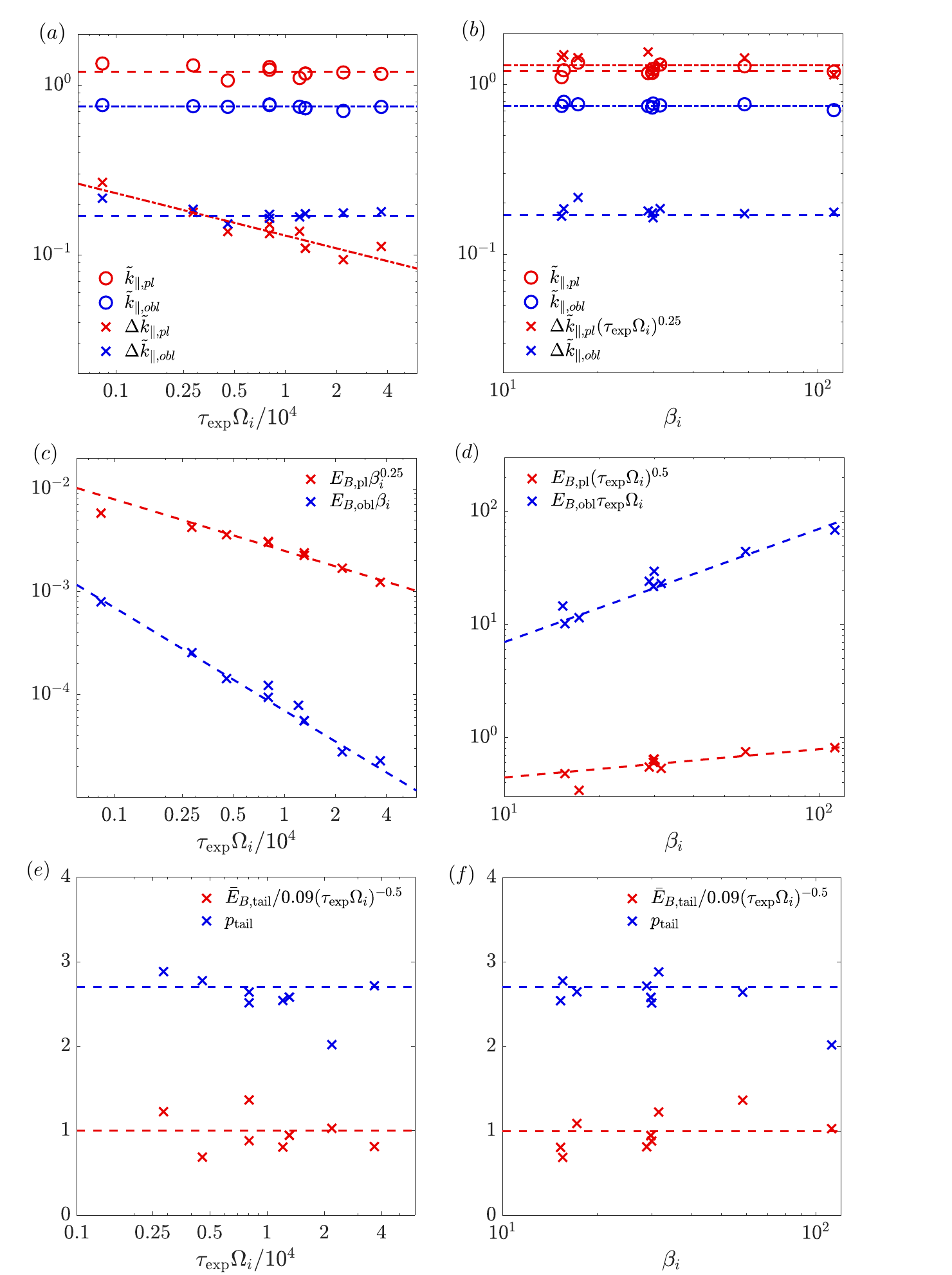}
  \caption{$(a)$ Best-fit estimates for wavenumber parameters introduced in (\ref{colloperator_scatter}) for all of our Alfv\'en-enabling simulations as a function of the expansion time. $(b)$ Same as in panel $(a)$, but as a function of $\beta_i$. $(c)$ Best-fit estimates for spectral amplitude parameters introduced in (\ref{colloperator_scatter}) for all of our Alfv\'en-enabling simulations as a function of the expansion time. $(d)$ Same as in panel $(e)$, but as a function of $\beta_i$. $(e)$ Best-fit estimates for high-wavenumber tail of parallel modes introduced in (\ref{colloperator_scatter}) for all of our Alfv\'en-enabling simulations as a function of the expansion time. $(f)$ Same as in panel $(e)$, but as a function of $\beta_i$.}
\label{fig:collsims_EBchar_AE_params}
\end{figure}

The wavenumber parameters of our best-quality fits for all of our simulations of Alfv\'en-enabling states as functions of $\tau_{\rm exp} \Omega_{i}$ and $\beta_i$ are presented in figure~\ref{fig:collsims_EBchar_AE_params}.
We find that all of the wavenumber parameters are approximately independent of both $\tau_{\rm exp} \Omega_i$ and $\beta_i$, save for $\Delta {k}_{\|,\rm pl}  \rho_i$, which has a weak dependence on $\tau_{\rm exp} \Omega_i$:\footnote{To avoid advocating for spuriously precise power-law fits based on our simulation data set, which, due to computational constraints, only consists of ten different runs in the Alfv\'en-enabling regime, we choose to specify power laws to the nearest quarter; this level of precision is chosen based on the size of the 95\% confidence intervals for the power-law indices of our fits, which is of characteristic magnitude ${\sim}$0.1--0.2.}
\begin{equation}
    {k}_{\rm \|, pl}  \rho_i \approx 1.2 , \quad {k}_{\rm \|, ob} \rho_i \approx 0.75 , \quad \Delta {k}_{\|,\rm pl}  \rho_i \approx \frac{1.3}{(\tau_{\rm exp}\Omega_i)^{0.25}} ,  \quad \Delta {k}_{\|,\rm ob}  \rho_i\approx 0.19 . \label{colloperatorparams_lengths}
\end{equation}
By contrast, both $\bar{E}_{B,\mathrm{ob}}$ and $\bar{E}_{B,\mathrm{pl}}$ do depend on $\tau_{\rm exp} \Omega_i$ and $\beta_i$, with those relationships being well approximated by the following scalings:
\begin{equation}
\bar{E}_{B,\mathrm{pl}} \approx 0.3 \frac{\beta_{i}^{0.25}}{(\tau_{\rm exp} \Omega_i)^{0.5}} , \quad \bar{E}_{B,\mathrm{ob}} \approx 0.7 \frac{\beta_{i}}{\tau_{\rm exp} \Omega_i} .  \label{colloperatorparams_energies}   
\end{equation}
Finally, for the high-wavenumber component of the secondary parallel firehose modes, we find that the  power-law index is approximately independent of both $\tau_{\rm exp} \Omega_i$ and $\beta_i$, but its amplitude has a comparable scaling to the peak amplitude of the secondary parallel firehose modes:
\begin{equation}
{p}_{\mathrm{tail}} \approx 2.7 , \quad \bar{E}_{B,\mathrm{tail}} \approx \frac{0.09}{(\tau_{\rm exp} \Omega_i)^{0.5}} . \label{colloperatorparams_powerlaw}
\end{equation}
We discuss possible theoretical justifications for these scalings in section~\ref{sec:interp_coll}. 

\begin{figure}
  \centering
  \includegraphics[width=\linewidth]{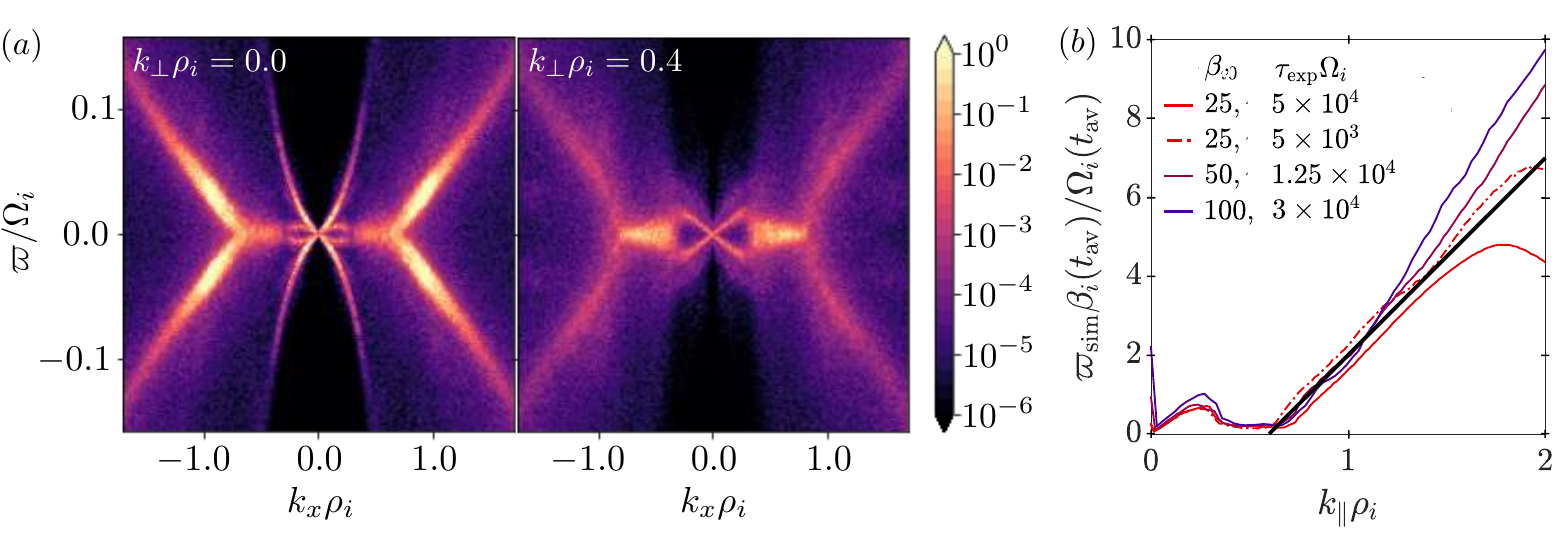}
  \caption{Panel $(a)$: slice plots at fixed $k_{\perp}$ of the frequency-dependent magnetic-energy spectrum $E_B(k_{\|},k_{\perp},\varpi)$ of the firehose fluctuations in run CV, averaged over the saturated state. Panel $(b)$: fluctuation-energy-weighted average value of real frequency $\varpi_{\rm sim}$ of the firehose fluctuations as a function of $k_{\|} \rho_i$ for several different runs that attain Alfv\'en-enabling states. The black line denotes the approximate fit \eqref{eqn:omegafit}.}
\label{fig:collsims_freqchar_AE}
\end{figure}

To calculate the dispersion relation $\varpi(k_{\|})$  of the secondary firehose modes in our simulation, as well as confirm that the oblique firehose modes are non-propagating, we compute the frequency-dependent magnetic-energy spectra $E_B(k_{\|},k_{\perp},\varpi)$ of the firehose fluctuations in saturation of our simulations of Alfv\'en-enabling states. Figure~\ref{fig:collsims_freqchar_AE}$(a)$ shows $E_B(k_{\|},k_{\perp},\varpi)$ computed for a representative simulation at two fixed values of $k_{\perp}$: for purely parallel modes ($k_{\perp} = 0$), and for oblique modes with $k_{\perp}$ comparable to that of the oblique firehose modes.    

For the parallel modes ($k_\perp\rho_i=0$; left panel), three distinct wave populations can be identified: at small wavenumbers ($k_{\|} \rho_i < 0.4$), both left- and right-handed Alfv\'en modes, while at ion-Larmor scales, secondary parallel firehose modes. As expected, the latter do indeed have a non-zero real frequency. For the oblique modes ($k_\perp\rho_i=0.4$; right panel), we also observe three distinct populations: at long-wavelengths ($k_{\|} \rho_i < 0.4$), shear Alfv\'en modes; just above ion-Larmor scales ($k_{\|} \rho_i \in [0.4,0.9]$), zero-frequency oblique firehose modes; and a weak population of (propagating) oblique secondary parallel firehose modes for $k_{\|} \rho_i > 0.9$. The presence of the long-wavelength modes in addition to the secondary parallel firehose and oblique firehose modes is perhaps surprising, because such long-wavelength modes are linearly damped at these levels of pressure anisotropy; we postulate that it is nonlinear coupling between secondary parallel firehose and oblique firehose modes that gives rise to them.

 From $E_B(k_{\|},k_{\perp},\varpi)$, we can then obtain a numerical  estimate of the dispersion relation of the parallel secondary firehose modes as a function of $k_{\|}$ by taking a weighted mean: 
 \begin{equation}
\langle \varpi \rangle(k_{\|}) \equiv \frac{\int_0^{k_{\perp,\rm max}} \rmd k_{\perp}  \int_0^{\varpi_{\rm max}} \rmd\varpi \, \varpi E_{B}(k_{\|},k_\perp, \varpi)}{\int_0^{k_{\perp,\rm max}} \rmd k_{\perp}  \int_0^{\varpi_{\rm max}} \rmd\varpi \, E_{B}(k_{\|},k_\perp, \varpi)} \, .
 \end{equation}
 Because the parallel secondary firehose modes are the dominant ones  at $k_{\|} \rho_i > 0.9$, the dependence of $\langle \varpi \rangle$ on $k_{\|}$ will correspond to their dispersion relation. 
 This numerical estimate for several representative simulations is shown in 
figure~\ref{fig:collsims_freqchar_AE}$(b)$. We find that, for $k_{\|} \rho_i \in [0.7,1.7]$, $\langle \varpi \rangle(k_{\|})$  is well approximated by the fit
\begin{equation}\label{eqn:omegafit}
\langle \varpi \rangle(k_{\|}) \approx \frac{\Omega_i}{\beta_i} (4.9 k_{\|} \rho_i - 2.9) \, .  
\end{equation}
The relatively narrow range of wavenumbers over which the spectrum of secondary parallel firehose modes exists mean that the clearly unphysical part of this fit to the dispersion relation (\emph{viz.}, $k_{\|} \rho_i \ll 1$, where $\varpi$ goes negative) are never used.

\subsection{Testing the model collision operator}

Having proposed an effective collision operator associated with firehose fluctuations in an Alfv\'en-enabling state, we now test whether this operator is consistent with two observables from our simulations: first, the velocity-dependent anisotropy of the distribution function for particles with speeds comparable to the thermal speed; secondly, Fokker--Planck coefficients calculated directly from the evolution of a sub-population of tracked (macro)particles.  

\subsubsection{Velocity-dependent anisotropy of the distribution function} \label{testcollop_anidistfunc}

Because collision operators describe how specific collisional processes affect the distribution function, confirming that our proposed firehose collision operator accounts for the observed distribution function's anisotropy is a natural test of our model. In the case of an expanding, high-$\beta_i$ plasma in an Alfv\'en-enabling state whose constituent particles have an effective collision rate $\nu_{\rm eff}$ that satisfies $\tau_{\rm exp}^{-1} \ll \nu_{\rm eff} \ll \Omega_i$, the relationship between the anisotropy of the distribution function and the collision operator takes a simple form. This condition is expected to hold for most particles in firehose-susceptible plasmas that attain Alfv\'en-enabling states, because the velocity-averaged collisionality $\langle \nu_{\rm eff} \rangle$ of particles satisfies $\langle \nu_{\rm eff} \rangle \sim \beta_i/\tau_{\rm exp} \gg 1/\tau_{\rm exp}$, while the pitch-angle scattering rate of even the most frequently scattered particles obeys the bound $\nu_{\rm eff} \ll \Omega_i$. 

To establish a relationship between the distribution function's anisotropy and the effective collision operator, we employ a modified version of a mathematical technique used in classical transport theory of plasmas: the Chapman--Enskog (CE) expansion~\citep[e.g.,][]{Yerger_2024}. This technique assumes that, in plasmas where the collision rate greatly exceeds the macroscopic evolution rate, the distribution function in the expanding plasma can be expanded in the form
\begin{equation}
f_i = f_{0i} + f_{1i} + \dots , \label{CE_exp}    
\end{equation}
where the first-order correction $f_{1i} \sim f_{0i}(\nu_{\rm eff} \tau_{\rm exp})^{-1}$ is asymptotically small compared to the leading-order term $f_{0i}$. Simultaneously, the condition that $\nu_{\rm eff} \ll \Omega_i$ means that, over the evolution timescales of interest, $f_i$ is approximately gyrotropic. If the gyroaveraged kinetic equation satisfied by the distribution function is also expanded in the small parameter $(\nu_{\rm eff} \tau_{\rm exp})^{-1}$, we find that, to leading order, $f_{0i}$ must satisfy $\mathfrak{C}_{\rm f}[f_{0i}] = 0$. Adopting our model firehose collision operator, and taking into account the Maxwellian initial condition of the distribution function in our simulations, this equation has the unique solution $f_{0i} = f_{\mathrm{M}i}$.\footnote{In general, our model firehose collision operator vanishes for any isotropic function $f_{0i}(\boldsymbol{v}) = f_{0i}(v)$. However, because the distribution function begins as Maxwellian in our simulations, and our collision operator does not directly generate a significant non-thermal population of particles, the zeroth-order solution remains Maxwellian. Our solution for the distribution function should be relevant to realistic plasmas, provided there is some process that pushes the plasma towards thermodynamic equilibrium -- for example, Coulomb collisions.} Considering the equation that arises to next order from the gyroaveraged kinetic equation, it follows that
\begin{equation}
    \mathfrak{C}_{\rm f}[f_{1i}] = \left[\left(\hat{\bb{b}}\hat{\bb{b}}-\frac{1}{3}\mathsfbi{I}\right)\bdbldot\mathsfbi{W}_i \right] \tilde{v}^2 P_2(\xi)
f_{\mathrm{M}i} \, , \label{CE_firstorder}
\end{equation}
where $\mathsfbi{W}_i$ is the (traceless, symmetric) rate-of-strain tensor of the ion bulk flow, $P_2(\xi)$ is the Legendre polynomial of second degree, and we remind the reader that $\tilde{v} \equiv v/v_{\mathrm{th}i}$. In the case of plasma that is linearly expanding on a timescale $\tau_{\rm exp}$ in one direction that is perpendicular to the background magnetic field, $\mathsfbi{W}_i = 2 \bnabla \bb{u} = -(2/\tau_{\rm exp})\ex\ex$, and so (\ref{CE_firstorder}) becomes
\begin{equation}
\mathfrak{C}_{\rm f}[f_{1i}] = -\frac{2}{3 \tau_{\rm exp}} \tilde{v}^2 P_2(\xi)
f_{\mathrm{M}i} \, . \label{CE_firstorder_v2}
\end{equation}
Now assuming that $\mathfrak{C}_{\rm f}[f_{1i}] = \mathfrak{C}_{\rm f}[f_{i}]$ takes the form given by (\ref{quasilinearscatteringrate_par}), and integrating (\ref{CE_firstorder_v2}) from $\xi = -1$ to $\xi_0 \rightarrow \xi$, we deduce that
\begin{equation}
 \frac{1}{2 \tilde{v}}\frac{\partial f_{i}}{\partial \xi} \approx f_{\mathrm{M}i}(v) \left\{\frac{\nu_{\rm eff,pl}(\tilde{v}_{\|})\tilde{v}_{\mathrm{wv}}(\tilde{v}_{\|})}{\nu_{\rm eff,pl}(\tilde{v}_{\|})+\nu_{\rm eff,ob}(\tilde{v}_{\|})} + \frac{\tilde{v}_{\|}}{3 \tau_{\rm exp}[\nu_{\rm eff,pl}(\tilde{v}_{\|})+\nu_{\rm eff,ob}(\tilde{v}_{\|})]}\right\} \, . \label{CE_funcaniso}
\end{equation}
Thus, we have established a simple relationship between the pitch-angle gradient of the distribution function, and the functions $\nu_{\rm eff,pl}(\tilde{v}_{\|})$, $\nu_{\rm eff,ob}(\tilde{v}_{\|})$, and $\tilde{v}_{\mathrm{wv}}(\tilde{v}_{\|})$ that characterise our model firehose collision operator. 

\begin{figure}
  \centering
  \includegraphics[width=\linewidth]{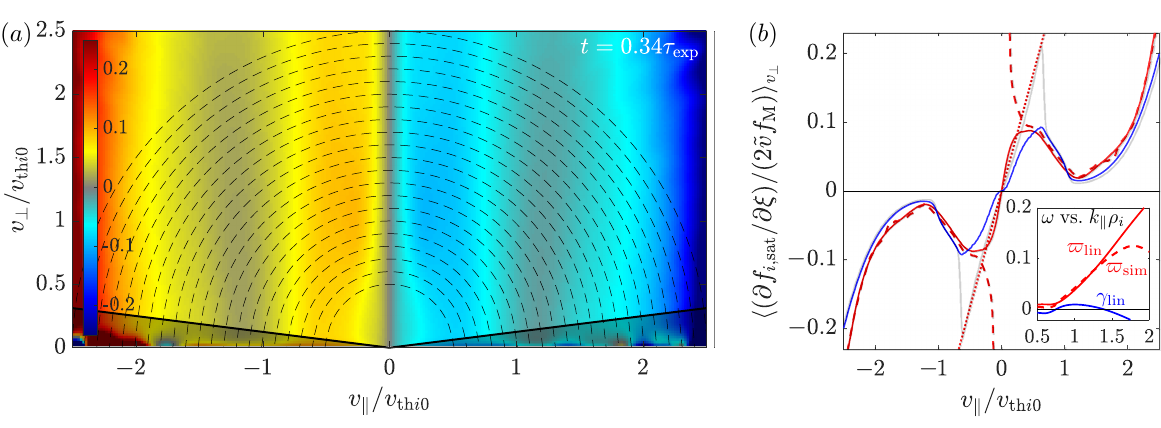}
  \caption{Panel $(a)$: slice plot of the pitch-angle gradient of the `saturated' ion distribution function $f_{i,\mathrm{sat}}$ divided by $2 \tilde{v} f_{\rm M}$ in an Alfv\'en-enabling state (run CV). Here, $f_{i,\mathrm{sat}}$ is the domain-averaged ion-distribution function $f(v_{\|},v_{\perp})$ time-averaged over the saturated period of the firehose instability. Panel $(b)$ shows the same quantity as $(a)$, but averaged over $v_{\perp}$ (solid red line); the average excludes the shaded region shown in panel $(a)$ which is negatively influenced by poor particle statistics. This is plotted with the theoretical prediction (\ref{CE_funcaniso}) for this quantity arising from our proposed collision operator (dashed line), and in the absence of any collisions (dotted line). The { dark blue (grey)} line shows the solutions of (\ref{CE_firstorder_timedep}) at $t = 0.34 \tau_{\rm exp}$, including (excluding) the high-wavenumber power-law tail of secondary parallel firehose modes. Inset:  complex frequency $\omega$ plotted against parallel wavenumber for linear low-frequency modes arising in a plasma with ion distribution function $f_{i,\mathrm{sat}}$. The real frequency $\varpi$ (growth rate $\gamma$) is shown in red (blue). Also plotted is the observed real frequency $\varpi_{\rm sim}$ of firehose fluctuations in the same run (red dashed).}
\label{fig:collsims_distfunctest_AE}
\end{figure}

Figure~\ref{fig:collsims_distfunctest_AE} provides a test of this relationship in the case of our asymptotic Alfv\'en-enabling run in its saturated phase.  
First, figure~\ref{fig:collsims_distfunctest_AE}$(a)$ illustrates a key feature of \eqref{CE_funcaniso}: that $({\partial f_{i}}/{\partial \xi})(1/\tilde{v}f_{\mathrm{M}i})$ is approximately independent of $v_{\perp}$, and primarily a function of $v_{\|}$. Secondly, figure~\ref{fig:collsims_distfunctest_AE}$(b)$ compares the time- and $v_{\perp}$-averaged value of $({\partial f_{i}}/{\partial \xi})(1/2\tilde{v}f_{\mathrm{M}i})$ (solid line) with the right-hand-side of (\ref{CE_funcaniso}) (dashed line), where we first compute $\nu_{\rm eff,pl}(\tilde{v}_{\|})$, $\nu_{\rm eff,ob}(\tilde{v}_{\|})$, and $\tilde{v}_{\mathrm{wv}}(\tilde{v}_{\|})$ assuming our quasilinear model applies instantaneously, and then time-average the entire expression. The agreement is very strong, save for $|\tilde{v}_{\|}| \ll 0.5$, supporting the claim that our model collision operator is appropriate for $|\tilde{v}_{\|}| \gtrsim 0.5$.  

To explain why reasonable agreement is not attained for comparatively small values of $\tilde{v}_{\|}$, we note that, in deriving (\ref{CE_funcaniso}), we have implicitly assumed that the rate of anomalous scattering is large enough that, at the specified time $t$ at which the comparison is made, either $\nu_{\rm eff,pl} t \gg 1$ or $\nu_{\rm eff,ob} t \gg 1$. As $\tilde{v}_{\|}$ is decreased from order-unity values to smaller ones, the amplitude of the increasingly high wavenumber firehose modes with which such particles are resonant decreases, leading to an ever-smaller scattering rate. Eventually, the collision rate decreases enough that $\nu_{\rm eff,ob} t \ll \nu_{\rm eff,pl} t \lesssim 1$, at which point there is no expectation for (\ref{CE_funcaniso}) to hold.  
Indeed, if $\nu_{\rm eff,pl} t \ll 1$, one should expect the distribution function's anisotropy to be consistent with continued double-adiabatic evolution; that is, in the absence of any collisions, the non-Maxwellian component of the distribution function should be given by
\begin{equation}
f_{1i} \approx \frac{2 t}{3 \tau_{\rm exp}} \tilde{v}^2 P_2(\xi)
f_{\mathrm{M}i} \, , \quad \frac{1}{2 \tilde{v}}\frac{\partial f_{i}}{\partial \xi} \approx \tilde{v}_{\|} \frac{t}{\tau_{\rm exp}} f_{\mathrm{M}i} \, .  \label{distfunc_DA}
\end{equation}
This expression is plotted in figure \ref{fig:collsims_distfunctest_AE}$(b)$ (dotted line) using the mean time of the `saturated' interval over which $f_i$ is averaged; good agreement is found for $\tilde{v}_\parallel\lesssim 0.2$. This implies that scattering of particles with small values of $v_{\|}$ compared to the ion thermal speed are indeed too infrequent to impact the distribution function anisotropy at such velocities.

We can further test this hypothesis by considering the evolution equation of the first-order correction $f_{1i}$ under the ordering $t \sim \tau_{\rm exp}/\beta_{i} \sim \langle \nu_{\rm eff} \rangle^{-1}$:
\begin{equation}
\frac{\partial f_{1i}}{\partial t} - \mathfrak{C}_{\rm f}[f_{1i}] = \frac{2}{3 \tau_{\rm exp}} \tilde{v}^2 P_2(\xi) 
f_{\mathrm{M}i} \label{CE_firstorder_timedep} .
\end{equation}
It is clear that, taking the subsidiary limit $t \langle \nu_{\rm eff} \rangle \gg 1$ recovers the steady-state solution (\ref{CE_firstorder_v2}), while the opposite limit $t \langle \nu_{\rm eff} \rangle \ll 1$ returns adiabatic evolution, with $f_{1i}$ given by (\ref{distfunc_DA}). We then solve (\ref{CE_firstorder_timedep}) numerically, with the effective collision operator given by (\ref{quasilinearscatteringrate_par_intro}) and (\ref{colloperator_scatter_intro}) when $\Delta_i < -1.35/\beta_i$ (i.e., when the oblique firehose is first destabilised). We integrate forward in time for the same duration as in our {\tt Pegasus++} runs and compute the pitch-angle gradients. A illustrative comparison of the two results (dark blue vs red line) for run CV is shown in  figure~\ref{fig:collsims_distfunctest_AE}$(b)$; in this (and other simulations) we find quantitative agreement, supporting our hypothesis. 

We can also use numerical solutions of (\ref{CE_firstorder_timedep}) to investigate the importance (or possible lack thereof) of the high-wavenumber power-law tail of secondary parallel firehose modes. If we remove the contribution of these modes from 
(\ref{colloperator_scatter_intro}$a$), and re-run our numerical solution of (\ref{CE_firstorder_timedep}), we obtain the light-blue line in figure~\ref{fig:collsims_distfunctest_AE}$(b)$. The resulting pitch-angle derivative of $f_{i,{\rm sat}}$ matches the {\tt Pegasus++} results well for $|v_{\|}| \gtrsim 0.6 v_{\mathrm{th}i}$ and for $|v_{\|}| \lesssim 0.2 v_{\mathrm{th}i}$. For intermediate values of $v_{\|}$, the numerical solution implies (erroneously) that the pitch-angle gradient of the distribution function should, for such values of $v_{\|}$, be given by the double-adiabatic result (\ref{distfunc_DA}). The reason that the double-adiabatic prediction is incorrect is simply that, if the high-wavenumber power-law tail of secondary parallel firehose modes is not modelled, then the scattering rate due to modes with $k_{\|} \rho_i \gtrsim 2$ implied by (\ref{colloperator_scatter_intro}) is insufficient for the distribution function's anisotropy to have been regulated in any meaningful way. We conclude that the high-wavenumber secondary firehose modes -- which, as we argue in section \ref{sec:sims:res:satfirehose_parfirehose_subthermal}, should be present physically -- play a non-trivial role in determining the velocity-dependent anisotropy of the ion distribution function in saturation.

\subsubsection{Increment method}

Another approach for testing our proposed model for the firehose collision operator is to try to characterise drag and diffusion of particles in our simulation directly, and compare such measurements to predictions from our model. Under two quite general assumptions -- specifically, that collisions are a near-Markovian process, and that individual scattering events do not lead to large-angle scattering -- it can be shown that any operator characterising those collisions must be to a good approximation a Fokker--Planck operator:
\begin{equation}
\mathfrak{C}[f] \approx -\frac{\partial}{\partial \bb{v}} \bcdot \left(\bb{A}f\right)+\frac{1}{2}\frac{\partial}{\partial \bb{v}} \frac{\partial}{\partial \bb{v}} \bdbldot \left(\mathsfbi{B} f\right) \, .
\end{equation}
Here, the (vector) drag coefficient $\bb{A}$ and the (rank-two tensor) diffusion coefficient $\mathsfbi{B}$ are given by 
\begin{equation}
 \bb{A} \equiv \lim_{\Delta t \rightarrow\zeroquote} \frac{\langle \Delta \boldsymbol{v} \rangle}{\Delta t} , \quad \mathsfbi{B} \equiv \lim_{\Delta t \rightarrow\zeroquote}\frac{\langle \Delta \boldsymbol{v} \Delta \boldsymbol{v} \rangle}{\Delta t} \,  ,
\end{equation}
where $\langle \Delta \boldsymbol{v} \rangle$ and $\langle \Delta \boldsymbol{v} \Delta \boldsymbol{v} \rangle$ are the first- and second-order jump moments, and the limit $\Delta t \rightarrow\zeroquote$ is to be interpreted as a time interval $\Delta t$ that satisfies $2 \pi \Omega_i^{-1} \ll \Delta t \ll 2 \pi \nu_{\rm c}^{-1}$ (where $\nu_{\rm c}$ is rate of scattering). This result gives us a general approach for estimating drag and diffusion due to a collisional process occurring in a PIC simulation: consider a time increment $\Delta t_{\rm inc}$ satisfying $2 \pi \Omega_i^{-1} \ll \Delta t_{\rm inc} \ll 2 \pi \nu_{\rm c}^{-1}$, calculate the jump moments associated with that time interval, and then estimate $\bb{A}$ and $\mathsfbi{B}$ via
\begin{equation}
 \bb{A} \approx \frac{\langle \Delta \boldsymbol{v} \rangle(\Delta t_{\rm inc})}{\Delta t_{\rm inc}} , \quad \mathsfbi{B} \approx \frac{\langle \Delta \boldsymbol{v} \Delta \boldsymbol{v} \rangle(\Delta t_{\rm inc})}{\Delta t_{\rm inc}} \,  .
 \label{eqn:jumpmoments}
\end{equation}
If the estimate is a good one, then different increment sizes satisfying $2 \pi \Omega_i^{-1} \ll \Delta t_{\rm inc} \ll 2 \pi \nu_{\rm c}^{-1}$ should give similar results. For simplicity's sake, we assume that the effective firehose collision operator is a function of pitch-angle only, and is therefore given by
\begin{equation}
\mathfrak{C}_{\rm f}[f]=-\frac{\partial}{\partial \xi} \left[A(v,\xi)f\right]+\frac{1}{2}\frac{\partial^2}{\partial \xi^2} \left[B(v,\xi)f\right] \, ,\label{eqn:FPpitch}
\end{equation}
where
\begin{equation}
 A(v,\xi) \equiv \lim_{\Delta t \rightarrow \zeroquote} \frac{\langle \Delta \xi \rangle}{\Delta t} , \quad B(v,\xi) \equiv \lim_{\Delta t \rightarrow \zeroquote}\frac{\langle \Delta \xi^2 \rangle-\langle \Delta \xi \rangle^2}{\Delta t} \, , 
\end{equation}
are the scalar pitch-angle drag and diffusion coefficients, respectively.  

\begin{figure}
  \centering
  \includegraphics[width=\linewidth]{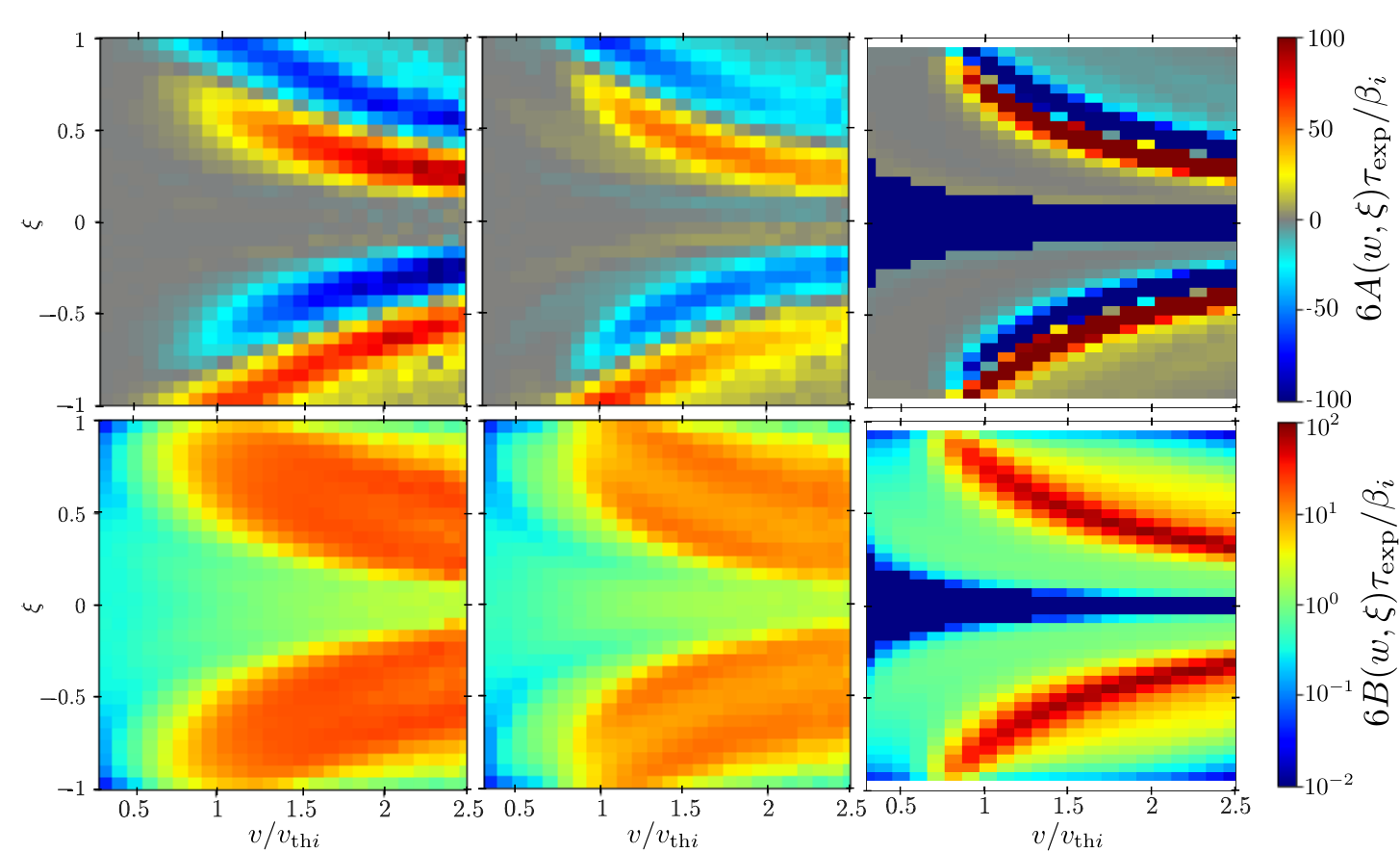}
  \caption{Fokker--Planck coefficients $A(v,\xi)$ (top row) and $B(v,\xi)$ (bottom row) obtained two different ways: using tracked-particle data from run~CV to calculate the jump moments \eqref{eqn:jumpmoments} assuming either $\Delta t = 4 \upi \Omega_i^{-1}$ (left column) or $\Delta t = 8 \upi \Omega_i^{-1}$ (middle column); and comparing our quasilinear pitch-angle scattering operator \eqref{colloperator_scatter_intro} with \eqref{eqn:FPpitch} to read off $A$ and $B$ (right column). The coefficients are normalised such that order-unity values are comparable to the velocity-averaged scattering rate.}
\label{fig:collsims_incrementtest_AE}
\end{figure}

Figure~\ref{fig:collsims_incrementtest_AE} presents the $A$ and $B$ coefficients calculated using tracked-particle data from our asymptotic Alfv\'en-enabling simulation (run CV, which has $\langle \nu_{\rm eff}\rangle \simeq 0.21 \beta_{\|i}/\tau_{\rm exp} \simeq 1.1 \times 10^{-4} \Omega_i$). We use two different increments: $\Delta t = 4 \pi \Omega_i^{-1}$ (left column) and $8 \pi \Omega_i^{-1}$ (middle column). The right column displays the coefficients associated with our model collision operator \eqref{colloperator_scatter_intro}.  
The comparison demonstrates reasonable qualitative agreement between the two models. $A(v,\xi)$ and $B(v,\xi)$ do not vary significantly along the resonant contours $w \xi = \mathrm{const}$ for both values of $\Delta t_{\rm inc}$, which implies that they are primarily functions of $v_{\|}$ only. Further, the drag coefficient changes sign in the same manner at a particular parallel velocity $v_{\|} \lesssim v_{\mathrm{th}i}$, and the magnitudes of both the drag and diffusion coefficients peak in the vicinity of this value of $v_{\|}$. 

However, the quantitative agreement is less convincing: compared to our quasilinear model, the peak values of $A$ and $B$ inferred using the increment method are reduced and features are noticeably broadened. Investigating the cause of this discrepancy, we find that one of the key assumptions underlying the increment method -- that particles undergo local jumps in phase space -- is violated by our data. Particles starting with pitch angles corresponding to regions of $(v,\xi)$-space in which there is strong scattering quickly move to other regions in which $\nu_{\rm eff,pl}(v,\xi)$ is smaller, and so sample a range of scattering rates during the chosen time increment. By contrast, the increment method assumes that just the initial scattering rate is sampled. This can be seen numerically by examining the root-mean-square change in pitch angle over the chosen increment; we find that even for $\Delta t_{\rm inc} = 4 \upi \Omega_i^{-1} = 1.3 \times 10^{-3} \langle \nu_{\rm eff} \rangle^{-1}$, particles starting near $v_{\|} \sim v_{\mathrm{th}i}$ experience changes $\Delta \xi$ to their pitch angle of order $\Delta \xi \sim 0.1$--$0.2$ (not shown). While these changes can be attributed partially to the direct effect of scattering, they are also due to fluctuations in the pitch angle of particles with $v_{\|} \sim v_{\mathrm{th}i}$ on timescales ${\sim}\Omega_i^{-1}$ that naturally arise as the particles traverse larger-scale oblique firehose fluctuations. This implies that the results of the increment method should not be regarded as being quantitative; indeed, the fact that caution is warranted is also evidenced by the discrepant results obtained using different increment sizes (see figure~\ref{fig:collsims_incrementtest_AE}). That being said, the broadening of resonances by both scattering and via non-resonant interactions is a physical effect, and one that is not currently accounted for in our quasilinear model. An extended discussion of this phenomenon -- considered when constructing a collision operator for the whistler heat-flux instability -- is given by~\citet[][\S 6.4]{Yerger_2024}.

\subsection{The perturbed magnetic energy of firehose fluctuations in saturation: part II} \label{sec:interp_coll}

In this section, we consider the parameter dependence of the functions $\nu_{\rm eff,ob}$ and $\nu_{\rm eff,pl}$ that we have deduced numerically, and offer possible explanations for them. The essence of these explanations is that these functions take the observed time-averaged forms in order to maintain a state of marginal linear stability with respect to both the oblique firehose and parallel secondary firehose instability.  However, because these arguments are somewhat speculative, yet technical, the impatient reader may wish to pass over them and move straight onto our conclusions in section~\ref{sec:applications}. 

\subsubsection{A qualitative theory of scattering by oblique firehose fluctuations} \label{sec:sims:res:satfirehose_obliquefirehose} 

When constructing our model collision operator, we assumed that a (resonant) quasilinear scattering operator was a reasonable simplification to adopt. For this assumption to be a logically consistent one, we must also assume that the growth rate of oblique firehose modes remains accurately modelled by linear kinetic theory in the saturated state of the instability. If this is to be the case, the growth rate $\gamma_{\rm ob}$ must satisfy $\gamma_{\rm ob} \sim \tau_{\rm exp}^{-1} \ll \Omega_i/\beta_{i}$, and therefore we require that the oblique firehose instability is approximately marginalised (in a time-averaged sense) over all wavevectors at which oblique firehose modes are detected in our simulations. Because the oblique firehose threshold is sensitive to the anisotropy of the distribution function, this requirement provides a significant constraint on the magnitude of the distribution function's anisotropy.  

While we are unable to write down a simple mathematical expression for the threshold condition of oblique firehose modes at arbitrary wavevectors, it is 
shown in Appendix~\ref{append_firehose_threshold_condition} that the threshold of quasi-parallel ($k_{\perp} \ll k_{\|}$) oblique firehoses with wavelengths that are not too small ($k_{\|} \rho_i \lesssim 0.5$) -- a subset of the unstable oblique firehose modes -- is given by [cf. (\ref{obliquefirethres_quasipar_append})]
\begin{equation}
    \frac{\upi}{n_{i0}} \mathcal{P} \! \int_{-\infty}^{\infty} \mathrm{d} \tilde{v}_{\|} \,\frac{\tilde{v}_{\|} }{1- k_{\|}^2 {\rho}_i^2 \tilde{v}_{\|}^2} \int_0^{\infty} \mathrm{d} \tilde{v}_{i\bot} \, \tilde{v}_{i\bot}^3 \frac{1}{\tilde{v}}\frac{\partial {f}_{i0}}{\partial \xi} \approx \frac{1}{\beta_{\|i}} \, . \label{obliquefirethres_quasipar}    
\end{equation}
Now substituting in (\ref{CE_funcaniso}) for the distribution function anisotropy, and assuming that the contribution to the principal value integral is dominated by parallel wavenumbers near those of the oblique firehose modes themselves, we deduce that
\begin{equation}
    \frac{1}{3\sqrt{\upi}} \mathcal{P} \! \int_{-\infty}^{\infty} \mathrm{d} \tilde{v}_{\|}\, \frac{\tilde{v}_{\|}^{2} }{1- k_{\|}^2 \tilde{\rho}_i^2 \tilde{v}_{\|}^2} \frac{1}{\nu_{\rm eff,ob}(\tilde{v}_{\|})} \exp{(-\tilde{v}_{\|}^2)} \approx \frac{\tau_{\rm exp}}{\beta_{\|i}} \, . \label{obliquefirethres_quasipar_simp}    
\end{equation}
If the integral equation~(\ref{obliquefirethres_quasipar_simp}) is to hold over a range of different values of  $k_{\|} \rho_i$, it follows that $\nu_{\rm eff,ob} \sim \beta_i/\tau_{\rm exp}$, and so $\bar{E}_{B,\mathrm{ob}} \sim \beta_i/\tau_{\rm exp} \Omega_i$, as we  have indeed observed numerically [cf. (\ref{colloperatorparams_energies})]. Beyond that, there is no obvious dependence of $\nu_{\rm eff,ob}(v_{\|i})$ on any other parameters -- which is consistent with the numerical observation [cf. (\ref{colloperatorparams_lengths})] that the fitting parameters $k_{\|,\mathrm{ob}}$ and $\Delta k_{\|,\mathrm{ob}}$ that characterise the mean and spread of parallel wavenumbers, respectively, are numbers, and not dependent on any other parameters. Equation~(\ref{obliquefirethres_quasipar_simp}) presumably also places a constraint on the functional form of $\nu_{\rm eff,ob}(v_{\|})$; however, because inverting (\ref{obliquefirethres_quasipar_simp}) is a non-trivial mathematical problem whose well-posedness is unclear, we do not attempt to pursue this further.  

\subsubsection{A qualitative theory of scattering of thermal particles by parallel secondary firehose fluctuations} \label{sec:sims:res:satfirehose_parfirehose} 

For the secondary parallel firehose modes that emerge in the Alfv\'en-enabling state, the saturation mechanism cannot be the same one as for the oblique firehose modes, because the secondary parallel firehose modes are propagating. This means that (quasilinear) pitch-angle scattering regulates the ion distribution function's anisotropy towards isotropy in the wave frame that is co-moving with the secondary parallel firehose modes. It can, in fact, be shown that the gradient of $f_i$ with respect to the pitch-angle $\xi'$ in the wave frame is related to ${\partial \tilde{f}_i}/{\partial \xi}$ and the parallel phase velocity ${v}_{\rm wv}$ of the waves by
\begin{equation}
\frac{\partial {f}_i}{\partial \xi'} \bigg|_{v'} = \frac{\partial {f}_i}{\partial \xi} \bigg|_{v} - \frac{2 {v}_{\rm wv} v}{v_{\mathrm{th}i}^2} {f}_{\rm M} ,     
\end{equation}
where $v'$ is the speed of particles in the wave frame~\citep{Yerger_2024}. Assuming that the amplitude of the secondary parallel firehose modes is sufficiently small that their growth rate $\gamma_{\rm pl}$ can still be modelled by (\ref{gamma_prlmodes}$b$), the magnitude of ${\partial \tilde{f}_i}/{\partial \xi'}$ around the parallel velocities $|v_{\|}| \sim v_{\mathrm{th}i}$ that are associated with cyclotron resonance is related to $\gamma_{\rm pl}$ by [cf. (\ref{pitchangle_gradient_biMax})]
\begin{equation}
\frac{\partial {f}_i}{\partial \xi'}\bigg|_{v_{\|}=v_{\|},\mathrm{res}} \sim v \mathcal{I}_{\pm} \sim \frac{\gamma_{\rm pl}}{\Omega_i} {f}_{\rm M} \, .
\end{equation}
Such an assumption also implies that the effective rate $\nu_{\rm eff,pl}$ of pitch-angle scattering by the secondary parallel firehose modes is related to their characteristic saturation amplitude $\delta B_{\rm pl}/B_0$ by $\nu_{\rm eff,pl} \sim \Omega_i \delta B_{\rm pl}^2/B_0^2$. Next, in saturation, the rate of change of the equilibrium distribution in saturation must balance the rate at which the secondary parallel firehose fluctuations cause pitch-angle diffusion of the distribution function, {\it viz.}
\begin{equation}
\nu_{\rm eff,pl} \frac{\partial {f}_i}{\partial \xi'}\bigg|_{v_{\|}=v_{\|},\mathrm{res}} \sim \frac{1}{\tau_{\rm exp}} {f}_{\rm M} \sim \nu_{\rm eff,pl} \frac{\gamma_{\rm pl}}{\Omega_i} {f}_{\rm M} .
\end{equation}
It therefore follows that 
\begin{equation}
 \frac{\delta B_{\rm pl}^2}{B_0^2} \sim \frac{\nu_{\rm eff,pl}}{\Omega_i} \sim \frac{1}{\gamma_{\rm pl}\tau_{\rm exp}} .  \label{plfirehose_dB_scaleA}
\end{equation}
This scaling has one particularly notable consequence. {In the saturated state, the growth rate of the secondary parallel firehose modes must be much smaller than their real frequency. If this were not the case -- that is, if unstable secondary parallel firehose modes grew at the same rate at which they propagated -- then their growth rate would be comparable to their phase speed, which is generally much faster than the macroscopic evolution rate. It follows from this that $\gamma_{\rm pl} \ll \Omega_i/\beta_i$, and so the amplitude of the saturated secondary modes, ${\delta B_{\rm pl}^2}/{B_0^2}$, must greatly exceed the value ${\sim}$$\beta_i/\tau_{\rm exp} \Omega_i$ that might be inferred from a naive quaslinear scattering model [cf.~(\ref{obliquefirehose_dB_scale})]. A simple physical explanation of this phenomenon is that particle scattering by secondary firehose modes acts to isotropise the distribution in the wave frame, not the laboratory frame. As a result, these modes must attain a larger-than-anticipated amplitude for this particle scattering to regulate the macroscopic generation of anisotropy.}

Determining a correct estimate of $\gamma_{\rm pl}$, and thereby ${\delta B_{\rm pl}^2}/{B_0^2}$ and $\nu_{\rm eff,pl}$, is a more challenging question. Making the na\"{i}ve presumption that, in order for saturation to occur, $\gamma_{\rm pl} \sim 1/\tau_{\rm exp}$, it follows from (\ref{plfirehose_dB_scaleA}) that ${\delta B_{\rm pl}^2}/{B_0^2} \sim 1$. This is inconsistent with the measured amplitude of parallel secondary firehose modes in our simulations [cf. (\ref{colloperatorparams_energies})], implying that a different mechanism must cause saturation to occur more efficiently. The condition that scattering of resonant particles by the secondary parallel firehose modes should not exceed the rate at which those modes grow ({\it viz.}, $\nu_{\rm eff,pl} \lesssim \gamma_{\rm pl}$) places a more stringent condition on ${\delta B_{\rm pl}^2}/{B_0^2}$, with the predicted saturation amplitude being 
\begin{equation}
\frac{\delta B_{\rm pl}^2}{B_0^2} \sim \frac{1}{\nu_{\rm eff,pl}\tau_{\rm exp}} \sim \frac{1}{\left(\tau_{\rm exp} \Omega_i\right)^{1/2}} , \label{plfirehose_dB_scaleB}
\end{equation}
and $\nu_{\rm eff,pl} \sim \Omega_i^{1/2} \tau_{\rm exp}^{-1/2}$. 
The scaling (\ref{plfirehose_dB_scaleB}) of ${\delta B_{\rm pl}^2}/{B_0^2}$ is almost consistent with (\ref{colloperatorparams_energies}), save for the $\beta_i^{1/4}$ dependence. Where this $\beta_i$-dependence arises from -- as well as the weak dependence of the breadth of the $k_{\|}$-interval over which firehose modes are detected on $(\tau_{\rm exp} \Omega_i)^{1/4}$ -- remains unclear to the authors, but could indicate that other possible saturation mechanisms (e.g., wave--wave interactions) could play some role. Establishing the precise mechanism of saturation would require the development of additional tools for analyzing our simulation results -- in particular, a full quasilinear code that solves for the evolution of the distribution function and the magnetic perturbations self-consistently -- so we defer this to future study.   

\subsubsection{A theory of scattering of subthermal particles by sub-ion-Larmor-scale secondary parallel firehose fluctuations} \label{sec:sims:res:satfirehose_parfirehose_subthermal} 

Finally, we motivate the inclusion of a high-wavenumber power-law tail in our model of the magnetic-energy spectrum of secondary parallel firehose modes. 

As shown in section~\ref{sec:sims:res:secondparfirehose}, the secondary parallel firehose modes that are initially destabilised have a characteristic number that is smaller than the reciprocal of the ion Larmor radius ($k_{\|} \rho_i \approx 0.7$), and modes with $k_{\|} \rho_i \gg 1$ are not destabilised. However, as the expansion proceeds, subthermal particles whose parallel velocity is initially too small to interact resonantly with the secondary parallel firehose modes continue to evolve according to the double-adiabatic conservation laws. As a result, the pitch-angle anisotropy of the distribution function at parallel velocities satisfying $|v_{\|}| \lesssim 0.5 v_{\mathrm{th}i}$ continues to grow. This has the consequence that, as the expansion proceeds, secondary parallel firehose modes with increasingly large wavenumbers become destabilised.  

This claim can be  proven explicitly for modes with wavelengths that are much smaller than $\rho_i$ (or, equivalently, $k_{\|} \rho_i \gg 1$). In this limit, the real frequency (\ref{gamma_prlmodes}$a$) of modes is  given approximately by $\varpi \approx \pm k_{\|}^2 d_i^2 \Omega_i$, and their growth rate (\ref{gamma_prlmodes}$b$) is\footnote{These modes are, in fact, just whistler waves -- such modes are the only parallel-propagating modes at sub-ion-Larmor scales that satisfy the ordering $\omega/k_{\|} v_{\mathrm{th}i} \sim \beta_i^{-1}$.}
\begin{equation}
\gamma \approx \pm \Omega_i \frac{\upi^2}{n_{i0}} \int_0^{\infty} \mathrm{d}v_{\perp} v_{\perp}^{3} \left(\frac{1}{w} \frac{\partial {f}_{i0}}{\partial \xi}-\frac{2 v_{\rm wv}}{v_{\mathrm{th}i}^2} {f}_{\rm M}\right)\Bigg|_{v_{\|} = \pm  v_{\mathrm{th}i}/k_{\|} \rho_i } . \label{gamma_prlmodes_subionLarmorscale}
\end{equation}
Then, assuming that the evolution of the ion distribution function's anisotropy is given by (\ref{distfunc_DA}) (i.e., it is approximately double-adiabatic), the growth rate (\ref{gamma_prlmodes_subionLarmorscale}) can be evaluated, giving
\begin{equation}
\gamma = \sqrt{\upi} \left(\frac{1}{k_{\|} \rho_i}\frac{t}{2 \tau_{\rm exp}} - \frac{k_{\|}\rho_i}{\beta_i} \right) \Omega_i . \label{gamma_prlmodes_subionLarmorscale_init}
\end{equation}
Thus, modes with $k_{\|} \rho_i < (t \beta_i/2 \tau_{\rm exp})^{1/2}$ are unstable at time $t$. Note that, for this calculation to be self-consistent, it must be the case that $t \gg \tau_{\rm exp}/\beta_i$, which implies that these sub-ion-Larmor scales modes will only be destabilised at times much later than the onset time of both oblique firehose modes and ion-Larmor-scale secondary parallel firehose modes.  

Once these sub-ion-Larmor-scale modes are destabilised, it is reasonable to propose that they will grow until they too scatter the particles with which they are resonant, regulating the anisotropy of particles with $v_{\|} \approx v_{\mathrm{th}i}/k_{\|} \rho_i \ll v_{\mathrm{th}i}$. Irrespective of the precise saturation mechanism, this regulation generically gives rise to magnetic-energy spectra at $k_{\|} \rho_i \gg 1$ that satisfy a power law. To show this, we posit that, in saturation, the distribution function's anisotropy evaluated at such $v_{\|}$ would satisfy (\ref{CE_funcaniso}) if scattering were to regulate the distribution function's anisotropy. Assuming that there are no non-propagating sub-ion-Larmor-scale oblique firehose modes, this implies that  
\begin{equation}
\frac{1}{2 \tilde{v}}\frac{\partial f_{i}}{\partial \xi} \approx f_{\mathrm{M}i}(v) \left[\tilde{v}_{\mathrm{wv}}(\tilde{v}_{\|}) + \frac{\tilde{v}_{\|}}{3 \tau_{\rm exp}\nu_{\rm eff,pl}(\tilde{v}_{\|})}\right] \, . \label{CE_funcaniso_pl}
\end{equation}
Further assuming that the behaviour of these modes is correctly described by quasilinear theory, even in the saturated state, their growth rate will still be given by (\ref{gamma_prlmodes_subionLarmorscale}). Substituting (\ref{CE_funcaniso_pl}) gives that the growth rate $\gamma(k_{\|} \rho_i)$ of modes with parallel wavenumber $k_{\|}$ is
\begin{equation}
\gamma(k_{\|} \rho_i) = \frac{2\sqrt{\upi}}{3}\frac{1}{k_{\|} \rho_i}\frac{1}{\tau_{\rm exp}\nu_{\rm eff,pl}(1/k_{\|} \rho_i)} \Omega_i .     
\end{equation}
It follows from the relation (\ref{colloperator_scatter}$a$) between the magnetic-energy spectrum $E_{B,\mathrm{pl}}\!\left(k_{\|} \rho_i\right)$ of parallel modes and $\nu_{\rm eff,pl}$ that
\begin{equation}
E_{B,\mathrm{pl}}\!\left(k_{\|} \rho_i\right) \approx \frac{B_0^2}{8\upi}\frac{4}{3\sqrt{\upi}} \frac{1}{(k_{\|} \rho_i)^2}\frac{1}{\tau_{\rm exp}\gamma(k_{\|} \rho_i)} . 
\end{equation}
Thus, if $\gamma(k_{\|} \rho_i) \propto k_{\|}^\alpha$, for some index $\alpha$ {(as would be expected due to scale invariance of sub-ion-Larmor-scale modes)}, then $E_{B,\mathrm{pl}}\!\left(k_{\|} \rho_i\right) \propto k_{\|}^{-\alpha -2}$.

It is clear that the specific index of the power law depends on exactly why the sub-ion-Larmor-scale secondary parallel firehose modes saturate -- the analogous question that we discussed in section \ref{sec:sims:res:satfirehose_parfirehose} for the ion-Larmor-scale modes. If, for example, $\gamma(k_{\|} \rho_i) \sim 1/\tau_{\rm exp}$ at all wavenumbers in saturation, then it would follow that $E_{B,\mathrm{pl}}\!\left(k_{\|} \rho_i\right) \sim (B_0^2/8 \upi) (k_{\|} \rho_i)^{-2}$; if, instead, $\gamma(k_{\|} \rho_i) \sim \nu_{\rm eff,pl}(1/k_{\|} \rho_i)$, then $E_{B,\mathrm{pl}}\!\left(k_{\|} \rho_i\right) \sim (B_0^2/8 \upi) (\tau_{\rm exp} \Omega_i)^{-1/2} (k_{\|} \rho_i)^{-1.5}$. Comparing to our numerical results [cf. (\ref{colloperatorparams_powerlaw})], we find that the scaling of the amplitude of the power-law with $\tau_{\rm exp} \Omega_i$ is consistent with this latter result. However, the slope of the power law we observe is significantly more negative. One plausible explanation for this discrepancy is the comparatively large amplitude of the spectrum of the noise in our simulations [$E_{B,\mathrm{noise}}(k_{\|}) \sim 2\times 10^{-5} (B_0^2/8\upi)$] compared with the measured amplitude of the power-law tail [$E_B(k_{\|}) \sim 2$--$5\times 10^{-5} (B_0^2/8\upi)$ at $k_{\|} \rho_i \approx 3$]. It may therefore be the case that the effect of numerical collisionality on subthermal particles steepens the observed spectral slope of the magnetic field compared with what might be observed if the numerical collisionality were lower, { because grid-scale electric fields supplant sub-ion-Larmor-scale magnetic perturbations in their role of scattering sub-thermal particles}. Unfortunately, due to the high computational cost of running simulations with an even larger number of particles per cell, we are unable to explore this possibility further at this time. 

\section{Discussion and applications} \label{sec:applications}

Our theory of firehose-instability saturation and its effect on thermodynamics and collisionality has implications both for prior simulation studies and for astrophysical applications. In the former arena, our theory explains the apparent differences between the value $\Delta_i \simeq -2/\beta_{\|i}$ of the pressure anisotropy that was obtained in the saturated state of previous high-$\beta$ shearing-box simulations of firehose-susceptible plasmas \citep{Kunz2014_b}, and the value $\Delta_i \simeq -1.4/\beta_{\|i}$ obtained in $\beta \gtrsim 1$ expanding-box simulations~\citep{Hellinger_2008,Hellinger_2015a,Hellinger_2019,Bott_2021}. Specifically, in the shearing-box simulations, $\beta_{i} \geq 200$, and the simulation with the longest characteristic shearing time $\tau_{\rm sh}$ satisfied $\tau_{\rm sh} \Omega_i \leq 10^{4}$, so $\tau_{\rm sh} \Omega_i/\beta_{i}^{1.6} \leq 2$. Taking numerical prefactors into account, all of these simulations can be described as being in the Alfv\'en-inhibiting state. By contrast, previous expanding-box simulations of the firehose instability all have $\beta_i \lesssim 8$, and $\tau_{\rm exp} \Omega_i \gtrsim 10^3$, so 
$\tau_{\rm exp} \Omega_i/\beta_i^{1.6} \gtrsim 36$; these simulations therefore describe plasmas in the Alfv\'en-enabling state.    

Having characterised three possible states, one obvious question to ask is whether the collisionless astrophysical plasmas in which the firehose instability is thought to operate (e.g., the ICM, black-hole accretion flows, the near-Earth solar wind) end up in ultra-high-beta, Alfv\'en-inhibiting, or Alfv\'en-enabling states. Generically, one of the key features of these astrophysical systems is that they exhibit huge scale separations between the timescales $\tau$ on which they evolve macroscopically, and plasma timescales such as the ion-Larmor period $2 \upi \Omega_i^{-1}$. We illustrate this by calculating $\Omega_i^{-1}$ using characteristic values of $B$ in three specific environments, and comparing it to their macroscopic evolution timescale $\tau$: the ICM at the cooling radius ($B \sim 1~\mu \mathrm{G}$, $\tau \sim 10^{14}~\mathrm{s}$, $\tau \Omega_i \sim 10^{12}$), Sgr~A$^{*}$ at the Bondi radius ($B \sim 1~\mathrm{mG}$, $\tau \sim 10^{6}~\mathrm{s}$, $\tau \Omega_i \sim 10^{7}$), and the solar wind at one astronomical unit ($B \sim 30~\mu \mathrm{G}$, $\tau \sim 10^{5}~\mathrm{s}$, $\tau \Omega_i \sim 4 \times 10^{5}$). Another key feature of these three particular systems is that their characteristic values of $\beta_i$ are larger than unity, but not by many orders of magnitude: for the ICM, $\beta_i \sim 10^{2-3}$; for Sgr~A$^{*}$, $\beta_i \sim 10$; for the near-Earth solar wind, $\beta_i \sim 1$. Considering these two features together, we find that $\tau \Omega_i/\beta_i^{1.6} \gg 1$ in all three of these systems (for the ICM, $\tau \Omega_i/\beta_i^{1.6} \sim 10^{9}$; for Sagittarius A$^{*}$, $\tau \Omega_i/\beta_i^{1.6} \sim 10^{5}$; for the solar wind, $\tau \Omega_i/\beta_i^{1.6} \sim 4 \times 10^{5}$). In short, these three examples of astrophysical firehose-susceptible plasmas should all saturate in Alfv\'en-enabling states.

{ Given the relevance of our findings to the solar wind, it is pertinent to compare our results with various theoretical and numerical studies of expanding plasmas completed in that context~\citep{Matteini_2006,Hellinger_2008,Hellinger_2017b}.  These studies have tended to focus on plasma with $\beta_i \sim 1$, and employed a spherical (rather than uni-directional) expansion; however, some have considered larger values of $\beta_i$ up to $\beta_i \approx 10$, which overlaps with the lowest $\beta_i$ that we have simulated, and there are areas of significant commonality. One such example is the quasi-periodic (as opposed to quasi-static) nature of the `saturated' state in the Alfv\'en-enabling regime. Such behaviour has also been seen in numerous two-dimensional simulations of expanding solar wind~\citep{Hellinger_2008,Hellinger_2017b}, and has been attributed to the tendency of the oblique firehose instability to be `self-destructive'. Another commonality concerns the interplay between oblique firehose modes and a second population of parallel firehose modes. For our run with the largest value of $\tau \Omega_i/\beta_i^{1.6}$ (run CV; $\tau \Omega_i = 5 \times 10^{4}$, $\beta_{i0} = 25$) -- viz., our `asymptotic' Alfv\'en-enabling run -- the evolution of the perturbed magnetic energy in parallel and oblique modes (cf. figure \ref{fig:sims_dB_fig3}) is quite similar to the results of two-dimensional simulations using a spherical expansion with $\tau_{\rm exp} \Omega_{i0} = 10^{4}$, and $\beta_{i0} = 0.5$~\citep[][, figure 5]{Hellinger_2008}. On this point, the inverse relationship between $\delta B_{\rm f}^2/B_0^2$ and $\tau_{\rm exp}$ presented in figure \ref{fig:sims_dB_fig2} is consistent with the findings of \citet[][figure 6]{Matteini_2006}. Finally, the departure from a bi-Maxwellian distribution caused by the interaction of thermal particles with firehose modes in our simulations is consistent with the `butterfly'-shaped contours of the ion distribution function routinely observed in prior simulations in the solar-wind context~\citep{Matteini_2006,Hellinger_2008,Hellinger_2015a,Hellinger_2019}. 

That being said, there are some differences worth noting. Firstly, there is a small difference between the particular value of $\Delta_i \simeq -1.4/\beta_{\|i}$ obtained in the saturated state of these previous expanding-box simulations, and the value $\Delta_i \simeq -1.6/\beta_{\|i}$ that we observed in our `asymptotic' Alfv\'en-enabling runs. The most plausible explanation for this difference is the larger values of $\beta_i$ that were used in our expanding-box simulations compared to the previous ones. We believe that, because of  differences in the linear physics of the firehose instability when $\beta_i \gtrsim 1$ and $\beta_i \gg 1$ -- specifically, the resonant parallel firehose instability can no longer be disregarded when $\beta_i \gtrsim 1$ -- it is likely that the secondary parallel firehose instability does not emerge in the same way when $\beta_i \gtrsim 1$ as when $\beta_i \gg 1$. Because these modes push the ion distribution function away from a bi-Maxwellian form, thereby altering the linear stability threshold of the oblique firehose instability, it is likely that this could affect the precise saturated value of the pressure anisotropy. This difference in the linear physics also presents itself in the comparative evolution of the perturbed magnetic energy in parallel and oblique firehose fluctuations; for example, in the $\beta_{i} \sim 1$ simulations of \citet{Hellinger_2008}, the energy in oblique firehose modes is always subdominant to that in (primary) resonant parallel firehose modes, whereas, initially, the opposite holds for our simulations of the Alfv\'en-enabling regime. Finally, we do not find evidence of significant interactions with suprathermal particles in our simulations of expanding high-$\beta_i$ plasmas, contrasting with the power-law tails observed by~\citet{Matteini_2006}.}

The conclusion that the firehose-unstable collisionless plasma present in astrophysical systems is typically in an Alfv\'en-enabling state has several important consequences for various physical phenomena. First of these concerns the behaviour of Alfv\'en waves, and in particular the phenomenon of Alfv\'en-wave interruption. It was recently shown that, in collisionless plasma, long-wavelength, linearly polarised, parallel-propagating Alfv\'en waves -- that is, modes with $k_{\|\mathrm{A}} \rho_i \ll 1$ -- with a sufficiently large amplitude ($\delta B_{\perp}/B_0 \gtrsim 2 \beta_i^{-1/2}$) could generate sufficient pressure anisotropy to  remove the Alfv\'enic restoring force on the wave and to trigger the firehose instability in local regions of plasma, leading to efficient damping of the wave \citep{Squire_2016,Squire_2017}. The implication of this work initially seemed to be that collisionless plasmas could not support Alfv\'enic perturbations above a critical amplitude that decreased with increasing $\beta_i$. However, this conclusion implicitly assumed that the regions of plasma in which firehose modes were produced attain an Alfv\'en-inhibiting state; this is, in fact, only the case whenever $\Omega_i/\beta_i^{1.6} \lesssim \omega_{\rm A}$, where $\omega_{\rm A} \equiv k_{\|\mathrm{A}} v_{\rm A}$ is the frequency of the long-wavelength Alfv\'en wave. This can be rearranged to give a lower bound on the parallel wavenumber at which the assumption holds: $k_{\|\mathrm{A}} \rho_i \gtrsim \beta_i^{-1.1}$. The simulations with the largest scale separation reported in \citet{Squire_2017} have $\beta_i = 100$ and $k_{\|\mathrm{A}} \rho_i = 2 \upi/1000 \approx \beta_i^{-1.1}$, which was consistent with expectations. If, however, $k_{\|\mathrm{A}} \rho_i \ll \beta_i^{-1.1}$ while $\delta B_{\perp}/B_0 \gtrsim 2 \beta_i^{-1/2}$, regions driven unstable to the firehose instability would instead attain Alfv\'en-enabling states, and so would not lead to the Alfv\'en wave's interruption (because the Alfv\'enic restoring force would not be completely negated). This has the implication that only mesoscale Alfv\'en waves with amplitudes $\delta B_{\perp}/B_0 \gtrsim 2 \beta_i^{-1/2}$ will experience interruption, while macroscales ones will not. Generating mesoscale Alfv\'en waves with such large amplitudes in astrophysical environments requires highly localised, intense energy injection; mesoscale waves generated in less extreme ways -- such as those forming a turbulent Alfv\'enic cascade -- will typically have much smaller amplitudes. 

The tendency of collisionless astrophysical plasma to arrive at an Alfv\'en-enabling state when the firehose instability is triggered also has significant ramifications for the nature of magnetised turbulence {in plasma that, on account of its macroscopic evolution (e.g., global expansion), acquires a background negative pressure anisotropy}. Specifically, it causes such turbulence to be similar to magnetohydrodynamical (MHD) turbulence. MHD turbulence with a strong guide magnetic field $B_0$ has two key
features. First, at length scales well below the outer scale $L$ at which the turbulence is driven but well above the ion-Larmor scale, a conservative cascade of Alfv\'enic fluctuations (with amplitude $\delta B_{\perp} \ll B_0$) is established via localised nonlinear interactions. Secondly, the fluctuations themselves are spatially anisotropic, with that anisotropy being determined by critical balance: $\tau_{\rm A} \sim l_{\|}/v_{\rm A} \sim \tau_{\rm nl} \sim l_{\perp}/u_{\perp}$, where $l_{\perp}$ and $l_{\|}$ are the characteristic scales of Alfv\'enic fluctuations in the directions parallel and perpendicular to the local background magnetic field, $\tau_{\rm A}$ is the fluctuation's characteristic linear evolution period, $\tau_{\rm nl}$ is the nonlinear interaction time and $u_{\perp}$ is the fluctuation's velocity perturbation \citep{gs95}. It follows that $l_{\perp}/l_{\|} \sim u_{\perp}/v_{\rm A} \sim \delta B_{\perp}/B_0 \sim (l_\perp/L)^{1/3} \ll 1$. {Hybrid-kinetic simulations have recently confirmed theoretical expectations~\citep{Schekochihin_2009} that pressure-isotropic collisionless $\beta \sim 1$ plasma would share these characteristics and be MHD-like~\citep[e.g.,][]{Arzamasskiy_2019}. However, it is unclear, \textit{a priori},  whether this resemblance persists in high-$\beta$ collisionless plasmas that are simultaneously developing a background pressure anisotropy $\Delta_{i0} < 0$. } If $\Delta_{i0}$ exceeds either the mirror or firehose instability thresholds, a turbulent cascade could be bypassed by the nonlocal transfer of magnetic energy from such large-scale fluctuations to small-scale ones. Indeed, recent hybrid-kinetic simulations of high-$\beta$, large-amplitude Alfv\'enic turbulence in collisionless plasma provide evidence of this~\citep{Arzamasskiy_2022}. In addition to nonlocality, the negation of Alfv\'enic restoring forces in plasma with $\Delta_{i0} \leq -2/\beta_{\|i}$ and $v_{\rm A,eff}^2 \leq 0$ would prevent critical balance from being established and thereby render the turbulence quasi-hydrodynamic. However, if the plasma attains an Alfv\'en-enabling state, then $v_{\rm A,eff}$ is simply a finite fraction of $v_{\rm A}$, and so should be qualitatively the same. \citet{Bott_2021} found the latter outcome in hybrid-kinetic simulations of $\beta_i \gtrsim 1$ Alfv\'enic turbulence in a collisionless plasma that generated a negative value of $\Delta_{i0}$ via a macroscopic expansion.  A developed cascade of MHD-like Alfv\'enic turbulence -- from inertial-range scales down through the ion-Larmor scale -- coexisted with firehose fluctuations that supported an Alfvén-enabling state, with critical balance being maintained via adaptation of the nonlinear turbulent decorrelation time to the modified linear timescale of the Alfv\'enic fluctuations. The existence of saturation in the Alfv\'en-enabling regime is, therefore, crucial to such a system being able to support a standard Alfv\'enic turbulent cascade (albeit with a modified wave speed).

By contrast, for astrophysical plasmas in which pressure anisotropies are generated by the Alfv\'enic fluctuations themselves, the fact that such plasmas tend to attain Alfv\'en-enabling states is of less importance for determining the nature of the turbulence itself. That is not to say that turbulent Alfv\'enic fluctuations in such systems will generate Alfv\'en-inhibiting regions of plasma. Indeed, the condition $\tau_{\rm A} \Omega_i \lesssim \beta_i^{1.6}$ required for Alfv\'en-inhibiting regions to be created implies an upper bound on the perpendicular scale $l_{\perp}$ of fluctuations required for those fluctuations to give rise to Alfv\'en-inhibiting regions that is seldom attained: assuming the Goldreich--Sridhar scaling $l_\| \sim l_{\perp}^{2/3} L^{1/3}$ for the anisotropy of the turbulent fluctuations~\citep{gs95}, it follows that this bound on $l_\perp$ is $l_{\perp}/\rho_i \lesssim \beta_i^{1.65} (\rho_i/L)^{1/2}$. For all astrophysical systems (except those having exceptionally high $\beta_i$), the right-hand side of this inequality is typically very small.\footnote{The simulations described in \citet{Arzamasskiy_2022} do produce local regions that are in Alfv\'en-inhibiting states. In that simulation, $L/\rho_i \approx 120$ and $\beta_{i0} = 16$, so according to our theoretical estimates, turbulent fluctuations with perpendicular scales $l_{\perp}/\rho_i \lesssim \beta_i^{1.65} (\rho_i/L)^{1/2} \approx 10$ might be expected to drive pressure anisotropies $\Delta_i \leq -2/\beta_{\|i}$. This is consistent (to within order-unity factors) with the scale of the Alfv\'en-inhibiting regions that are observed in the simulations [see Figure 6, panel (f), of~\citet{Arzamasskiy_2022}].} Instead, another recently discovered phenomenon in high-$\beta$ Alfv\'enic turbulence -- magnetoimmutability~\citep{Squire_2019} -- means that the volume-filling fraction of the plasma that approaches even the (less restrictive) threshold for the oblique firehose instability is much smaller than would be anticipated naively based on the Goldreich--Sridhar scaling. \citet{Squire_2023} and \citet{Majeski_2024} showed explicitly in simulations that, as a result, it makes little difference to the turbulence which firehose threshold is reached. The suppression of pressure-anisotropy fluctuations by magnetoimmutability also renders high-$\beta$ Alfv\'enic turbulence MHD-like, but this conclusion is not dependent upon microphysical changes induced by the firehose instability (or the mirror instability, for that matter). The fundamental difference between this case, with turbulently driven pressure anisotropy, and the case discussed above, with globally forced pressure anisotropy, relates to the ability of the plasma to respond dynamically via the pressure-anisotropy stress, which is driven by pressure-anisotropy gradients and can be comparable to Maxwell stresses in a high-beta plasma. In the turbulent setting, this pressure-anisotropy stress suppresses motions that generate significant pressure anisotropies, leaving little of the plasma at the firehose (or mirror) thresholds; in a globally forced case, this is not possible, and the whole plasma can attain a threshold together.

A third consequence is that firehose fluctuations are unlikely to have any significant direct effect on the acceleration and propagation of cosmic rays through astrophysical plasmas such as the ICM. In the conventional picture, scattering of cosmic rays is typically thought to be either due to resonant interactions with inertial-range turbulent fluctuations, or due to the excitation of MHD waves via resonant streaming instabilities. However, it had also been argued that ion-Larmor-scale modes excited by pressure anisotropies can give rise to particle acceleration \citep{Ley_2019}, and more recently been proposed that mirror fluctuations in the ICM scatter sub-TeV cosmic rays much more efficiently than other mechanisms~\citep{Reichherzer_2024, Ewart2024}. So, we consider here whether the firehose fluctuations present in Alfv\'en-enabling firehose plasma could give rise to non-negligible degrees of scattering in the ICM. For cosmic rays whose Larmor radius $\rho_{\rm CR}$ greatly exceeds the characteristic scale {$\sim$}$\rho_i$ of the firehose fluctuations, such scattering would have to be non-resonant and quasi-unmagnetised. By analogy to the arguments presented in~\citet[][section 1.1]{Reichherzer_2024}, we conclude that the spatial diffusion coefficent $\kappa_{\rm CR}$ of such cosmic rays due to scattering by firehoses is given by $\kappa_{\rm CR} \sim c (\rho_{\rm CR}^2/\rho_{i}) (\delta B_{\rm f}/B_0)^{-2}$. If the plasma through which the cosmic rays are passing is in an Alfv\'en-enabling state, then our theory predicts that $\delta B_{\rm f}^2/B_0^2 \sim \beta_i^{1/4} (\tau \Omega_i)^{-1/2}$, and so $\kappa_{\rm CR} \sim c (\rho_{\rm CR}^2/\rho_{i}) \beta_i^{-1/4} (\tau \Omega_i)^{1/2}$. For TeV cosmic-ray protons passing through the ICM, it is the case that $\rho_{\rm CR} \sim 3 \times 10^{5} \rho_i$ and $\tau \Omega_i \sim 10^{12}$, so $\kappa_{\rm CR} \sim 10^{38} \, \mathrm{cm}^2 \, \mathrm{s}^{-1}$. This is eight orders of magnitude larger than the spatial diffusion coefficients arising due to other scattering mechanisms~\citep{Reichherzer_2024}, implying that scattering of cosmic rays by firehoses is a negligible effect. We note, however, that the firehose instability could still have indirect effects on cosmic-ray dynamics. Our estimates have implicitly treated the saturation of the cosmic-ray streaming and firehose instabilities separately, which may not be reasonable. Further, cosmic-ray-streaming-instability-driven Alfv\'enic modes in a plasma that has attained an Alfv\'en-enabling state will maintain a phase velocity that is a finite fraction of $v_{\rm A}$, with the consequence that the classical picture of the cosmic-ray streaming instability should still apply; this would not be the case in a plasma in an Alfv\'en-inhibiting state. 

Although we have argued that most astrophysical plasmas of interest will attain 
Alfv\'en-enabling states if they become susceptible to the firehose instability, we note that there are a few circumstances in which the Alfvén-inhibiting or ultra-high-beta states could still be relevant. One such circumstance is plasma with very large $\beta_i$. The plasma created during the reionization epoch, which is thought to be only very weakly magnetised, with the ion-Larmor radius initially comparable to the mean free path $\lambda_i$, is a good example of this: in such plasma, $\beta_i \sim 10^{20}$, which would certainly be large enough to put any firehose-susceptible plasma into the ultra-high-beta state. This conclusion implies that recent~\citep{St-Onge_2020} and future studies of the action of the fluctuation dynamo inside weakly collisional plasmas in the early universe cannot simply assume that the plasmas they are modelling are in Alfv\'en-enabling states. Another example is that of local regions of ICM plasma in which its tangled stochastic field is reversing sign. The ICM is not observed to have an ordered component to its magnetic field, implying that locally, the ICM will have regions in which $\beta_i$ is much larger than its typical value and therefore Alfv\'en-inhibiting states (or even ultra-high-beta states) could be realised locally. Finally, collisionless or weakly collisional magnetised plasmas with much smaller separations between macroscopic and plasma timescales require comparatively smaller values of $\beta_i$ in order to attain Alfv\'en-inhibiting states or ultra-high-beta states. This is particularly pertinent for any future laser-plasma experiments that might investigate the firehose instability, because state-of-the-art laboratory astrophysics experiments that have investigated weakly collisional magnetised plasmas on the world's highest-energy laser facilities such as the National Ignition Facility only achieved a timescale separation $\tau \Omega_i \sim 30$ \citep{Meinecke_2022}. 

{ A natural question about this study concerns the extent to which our findings generalise beyond the specific set-up explored in this work (a uni-directional expansion in a two-dimensional plane) to a broader set of macroscopic motions that generate pressure anisotropy (e.g., spherical expansions, or shearing motions). While there are several physical systems of interest for which our simulations can be interpreted formally as a local model, we consider understanding the extent to which our results apply to firehose instabilities driven by arbitrary macroscopic motions to be more pertinent. It is plausible that some aspects of firehose instability saturation may depend on the precise details of the geometrical expansion for macroscopic evolution times that only exceed the Larmor period by a few orders of magnitude (a condition that covers some of our simulations, particularly those in the Alfvén-inhibiting regime). These differences could lead to, for example, distinct saturation amplitudes, depending on the macroscopic motion in question; indeed, this is a plausible explanation for why $\delta B_{\rm f}/B_0$ is larger by an order-unity factor in our expanding-box simulations than in the prior shearing-box simulations of \citet{Kunz2014_b} and \citet{Melville_2016} at analogous values of $\tau$ (see section \ref{sec:AIoverview}). Nevertheless, the numerous areas of consistency with the previous shearing-box simulations and also other simulations that employed a quasi-spherical expansion suggest that many features of firehose instabilities are not sensitive to the precise nature of the macroscopic motion that causes their instability. Indeed, when the timescale of macroscopic evolution of the plasma greatly exceeds the saturation timescale of the firehose instability -- as we observe at sufficiently large expansion times -- we expect that the precise details of the macroscopic evolution should become increasingly unimportant. Preliminary results from 3D simulations  that we have performed recently of high-$\beta$ collisionless plasma undergoing quasi-spherical expansion support this claim; at sufficiently large values of the parameter $\tau \Omega_i/\beta_i^{1.6}$, we recover Alfv\'en-enabling states with characteristics -- such as magnetic-field morphology -- that closely resemble those seen in the 2.5D simulations reported here.}

{ Another aspect of this problem that merits further study pertains to the assumption of fluid-like, pressure-isotropic electrons in our hybrid-kinetic simulations. While this assumption is appropriate for some astrophysical plasmas (e.g., the ICM), in other plasma systems (e.g., black-hole accretion flows) where the collisionality is sufficiently weak, the assumption of isotropic electrons may not be a good one. Specifically, if $\tau \nu_{e} \ll \beta_e$, where $\nu_e$ is the electron collision frequency and $\beta_e$ the electron plasma beta, the macroscopic evolution of the plasma will naturally give rise to both ion and electron pressure anisotropies of sufficient magnitude to drive various kinetic instabilities. For example, electron pressure anisotropies satisfying $\Delta_e \lesssim -1.4/\beta_e$ will drive electron-Larmor-scale modes unstable~\citep[see, e.g.,][]{Hollweg_1970,Li_2000}; if $\Delta_e < -2/\beta_i$, ion firehose modes can be destabilised even in the absence of ion pressure anisotropy~\citep[e.g.,][]{Kunz_2018}. In the case of purely collisionless, magnetised plasma, in which both electrons and ions satisfy the CGL equations (\ref{CGL_withcoll}) and both  $\Delta_e$ and $\Delta_i$ are generated at the same rate by the plasma's macroscopic evolution, we expect changes in the evolution and saturation of the ion firehose instability as compared with its evolution with pressure-isotropic electrons. Indeed, such differences have already been reported by \citet{Riquelme_2018}, who found that the regulated ion pressure anisotropy was less negative if electron pressure anisotropy was not fixed but instead allowed to evolve dynamically. While fully kinetic simulations that resolve both ion and electron pressure anisotropies are an important direction for future work, we note that incorporating such physics is computationally demanding and would have made it significantly more difficult to isolate the processes we have explored here. For this reason, our focus on a hybrid-kinetic framework -- in which electron pressure is assumed isotropic -- is both pragmatic and physically motivated. Even so, two conceptual aspects of our results should be relevant to future studies of the firehose instability that incorporate electron pressure anisotropy. The first is the possible existence of distinct thermodynamic states depending on the macroscopic evolution rate, the Larmor frequencies, and $\beta_i$ and $\beta_e$; this follows from linear theory, which suggests that firehose modes at different scales can still have distinct thresholds~\citep[see, e.g.,][]{Bott_2023}. Secondly, secondary firehose instabilities arising from the interaction of primary firehose modes with both electrons and ions is a plausible phenomenon worth further investigation, possibly using the analytic framework developed in this work. For example, \citet{Ley_2024} report secondary ion-cyclotron and whistler instabilities driven by primary mirror modes, suggesting that secondary kinetic instabilities could be a ubiquitous phenomenon in high-$\beta$ collisionless plasmas.}

\section{Summary} \label{sec:summary}

In this paper, we have argued that high-$\beta$, collisionless (or weakly collisional) plasmas that become susceptible to the firehose instability due to their macroscopic evolution attain one of three qualitatively distinct states once the instability has saturated -- ultra-high-beta, Alfv\'en-inhibiting, or Alfv\'en-enabling. Which state is realized depends on whether a critical parameter dependent on the plasma's macroscopic evolution time $\tau$, the ion-Larmor frequency $\Omega_i$, and the ion plasma beta parameter $\beta_i$, is large or small. For plasmas with $\beta_i \ll 10^{5}$, this condition takes a particularly simple form: whenever $\tau \Omega_{i} \gtrsim 10 \beta_i^{1.6}$, an Alfv\'en-enabling state is attained; plasmas with $\beta_i \ll \tau \Omega_{i} \lesssim \beta_i^{1.6}$ will settle into Alfv\'en-inhibiting states; and plasmas with $\tau \Omega_{i} \lesssim \beta_i$ reside in the ultra-high-beta regime. The key macroscopic difference between Alfv\'en-enabling or Alfv\'en-inhibiting states is the value of the steady-state regulated pressure anisotropy $\Delta_i$, and thereby the effective Alfv\'en velocity $v_{\rm A,eff}$. In Alfv\'en-inhibiting states, $\Delta_i \simeq -2/\beta_{\|i}$ and $v_{\rm A,eff}^2/v_{\rm A}^2 \simeq 0$, and so Alfv\'en waves are unable to propagate; in Alfv\'en-enabling states, $\Delta_i \simeq -1.6/\beta_{\|i}$ and $v_{\rm A,eff}^2/v_{\rm A}^2 \simeq 0.2$, and so Alfv\'en waves can propagate (albeit at a reduced phase speed). The two states are also qualitatively distinct microphysically. The magnetic-energy spectrum of firehose fluctuations in the Alfv\'en-inhibiting state is broad, including modes with characteristic wavelengths that are much larger that the ion Larmor radius $\rho_i$; in the Alfv\'en-enabling state, firehose fluctuations are predominantly at ion-Larmor scales, and are of two distinct types (oblique firehose modes, and the newly identified secondary parallel firehose modes) that are separable in wavenumber space. The distinct characteristics of the firehose modes give rise to ion distribution functions with subtly different characteristics: specifically, in the Alfv\'en-inhibiting state, the distribution function is quasi-isotropic for particles with parallel velocities $v_{\|} \gtrsim v_{\mathrm{th}i}$, while (at any one time) only a subset of such particles are isotropic in the Alfv\'en-enabling state. In both instances, the distribution function is not well described as a bi-Maxwellian, with pitch-angle anisotropy being concentrated at smaller characteristic velocities.  

In addition to uncovering the distinction between the Alfv\'en-enabling and Alfv\'en-inhibiting states, we have also characterised the effective collisionality that emerges in firehose-susceptible plasmas. We first computed the particle-averaged collisionality $\nu_{\rm eff}$, finding qualitatively that $\nu_{\rm eff} \sim \beta_i/\tau$ in both states (in agreement with previous work). More quantitatively, we have proposed a precise value for the effective collisionality in firehose-unstable, high-$\beta$ plasmas that attain Alfv\'en-enabling ($\nu_{\rm eff} \approx 0.21 \beta_i/\tau_{\rm exp}$) and Alfv\'en-inhibiting ($\nu_{\rm eff} \approx 0.17 \beta_i/\tau_{\rm exp}$) states respectively. Computing this effective collisionality allowed us to in turn determine the effective parallel Braginskii viscosity $\mu_{\rm B}$ in such plasmas: $\mu_{\rm B} \approx 0.4 (B^2/4 \upi) \tau_{\rm exp}$ in Alfv\'en-enabling states, and $\mu_{\rm B} \approx 0.5 (B^2/4 \upi)\tau_{\rm exp}$ in Alfv\'en-inhibiting states. Finally, we proposed a quasilinear pitch-angle scattering model (with parallel-velocity-dependent scattering rates) for the effective collision operator associated with firehose fluctuations in Alfv\'en-enabling states. We found that this model was consistent with data from two specialised simulation diagnostics including the anisotropy of the distribution function. The scattering model proposed here may be useful for kinetic simulations that average out cyclotron motion for computational efficiency (e.g., gyrokinetic or drift-kinetic simulations); in this case velocity-space instabilities could be included via an imposed collision operator such as that described here.

We hope that this work provides a helpful model for future investigations of kinetic instabilities in collisionless (and weakly collisional) plasmas. Judicious use of specialised numerical techniques such as the HEB method in PIC simulations has the benefit of maximising the achievable separation between macroscopic and microscopic scales at fixed computational cost; as we have shown here, this can be essential for accessing the parameter regimes that are relevant to astrophysics. Furthermore, performing scans over key parameters (such as $\beta_i$) is often helpful for identifying the physical mechanisms that cause the saturation of kinetic instabilities. As for this paper's key results -- in particular, our computation of the effective collisionality ($\nu_{\rm eff} \simeq 0.4 \beta_i/\tau$) and parallel Braginskii viscosity ($\mu_{\rm B} \simeq 0.8 \tau B^2/4 \upi$) in astrophysically relevant plasma -- we believe that using local kinetic simulations to compute effective transport coefficients, which could then be subsequently implemented into global fluid simulations, provides a promising route towards building successful models of macroscopic collisionless plasma environments.

\vspace{2ex} A.F.A.B., M.W.K., and E.Q.~were supported for this research by funding from DOE awards DE-SC0019046 and DE-SC0019047 through the NSF/DOE Partnership in Basic Plasma Science and Engineering. A.F.A.B.~received additional support during the latter part of this work from the UKRI (grant number MR/W006723/1). {J.S. was supported by Rutherford Discovery Fellowship RDF-U001804 and Marsden Fund grant MFP-U002221, which are managed through the Royal Society Te Aparangi. } The simulations were performed using the Stellar cluster at the PICSciE-OIT TIGRESS High Performance Computing Center and Visualization Laboratory of Princeton University. The authors are grateful to the two
anonymous reviewers, whose recommendations have improved the paper. Declaration of Interests: the authors report no conflict of interest.

\appendix

\section{Supporting linear theory for the firehose instability}

In this appendix, we write out explicitly the dispersion relation of linear modes in a hot collisionless plasma. We then use this expression for a few analytical 
calculations pertaining to the linear theory of the firehose instability that support the results outlined in the main text. 

\subsection{The hot-plasma dispersion relation} \label{append:hotplasmadispreldef}

The hot-plasma dispersion relation is given by
\begin{equation}
   \mbox{Det}\left[\frac{c^2 k^2}{\omega^2} \left(\hat{\bb{k}}\hat{\bb{k}}-\mathsfbi{I}\right)+ \bb{\mathfrak{E}} \right]=0 
   ,
   \label{hotplasmadisprel}
\end{equation}
where $\hat{\bb{k}} \equiv \bb{k}/k$ is the direction of the 
perturbation's wavevector, 
\begin{equation}
  \bb{\mathfrak{E}} \equiv \mathsfbi{I} + \frac{4 \upi \mathrm{i}}{\omega} 
  \bb{\sigma} \label{dielecttensfull}
\end{equation}
is the plasma dielectric tensor, and $\bb{\sigma}$ is the plasma 
conductivity tensor. The conductivity tensor is a function of the 
equilibrium distribution functions $f_{i0}(v_{\|},v_{\bot})$ and $f_{e0}(v_{\|},v_{\bot})$ of constituent ions and 
electrons, respectively; 
it is explicitly given by
\begin{eqnarray}
\bb{\sigma} = \sum_s \bb{\sigma} _s & = & - \frac{\mathrm{i}}{4 \upi \omega} \sum_s \omega_{\mathrm{p}s}^2 \bigg[ \frac{2}{\sqrt{\upi}} \frac{k_{\|}}{|k_{\|}|}  \int_{-\infty}^{\infty} \mathrm{d} \tilde{w}_{\|s} \, \tilde{w}_{\|s} \int_0^{\infty} \mathrm{d} \tilde{v}_{s\bot} \,\Lambda_s(\tilde{w}_{\|s},\tilde{v}_{s\bot}) \hat{\bb{z}} \hat{\bb{z}} \nonumber \\*
&& \mbox{} + \tilde{\omega}_{\|s} \frac{2}{\sqrt{\upi}} \int_{C_{\rm L}} \mathrm{d} \tilde{w}_{\|s} \int_0^{\infty} \mathrm{d} \tilde{v}_{s\bot} \,\tilde{v}_{s\bot}^2 \Xi_s(\tilde{w}_{\|s},\tilde{v}_{s\bot}) \sum_{n = -\infty}^{\infty} \frac{\mathsfbi{R}_{sn}}{\zeta_{sn} -\tilde{w}_{\|s}}  \bigg] \, . \qquad \label{conductivity}
\end{eqnarray}
Here, $\left\{\hat{\bb{x}},\hat{\bb{y}},\hat{\bb{z}}\right\}$ 
are the basis vectors of an orthogonal coordinate system defined in terms of $\bb{B}_0$ and 
$\bb{k}$ by 
\begin{equation}
     \hat{\bb{z}} \equiv \frac{\bb{B}_0}{B_0}, \quad 
     \hat{\bb{x}}  \equiv \frac{\bb{k}_{\perp}}{k_{\perp}} \equiv \frac{\bb{k}- k_{\|} \hat{\bb{z}}}{k_{\bot}} , 
     \quad
     \hat{\bb{y}}  \equiv \hat{\bb{z}} \times \hat{\bb{x}} 
     , \label{xyzcoordinatebasis}
\end{equation}
where $B_0 \equiv |\bb{B}_0|$, $k_{\|} \equiv \bb{k}\bcdot \hat{\bb{z}}$, and $k_{\bot} \equiv 
\left|\bb{k}_{\perp}\right|$, $q_s$ is the charge of particles of species $s$, $m_s$ their masses, $n_{s0}$ their densities, $T_{s0}$ their temperatures, $v_{\mathrm{th}s} \equiv \sqrt{2 T_{s}/m_s}$ their thermal velocities, $\tilde{v}_{\|s} \equiv v_{\|}/v_{\mathrm{th}s}$, $\tilde{v}_{\perp s} \equiv v_{\perp}/v_{\mathrm{th}s}$,
and 
\begin{equation}
  \omega_{\mathrm{p}s} \equiv \sqrt{\frac{4 \upi q_s^2 n_{s0}}{m_s}} , \label{plasmafrequency_def}
\end{equation} 
\begin{equation}
\tilde{w}_{\|s} \equiv \frac{k_{\|} \tilde{v}_{\|s}}{|k_{\|}|},   \label{w_pl_def}
\end{equation}
\begin{equation}
  \tilde{\rho}_s \equiv \frac{m_s c v_{\mathrm{th}s}}{q_{s} B_0} =  \frac{|q_{s}|}{q_{s}} \rho_s,
\end{equation}
\begin{equation}
  \tilde{\omega}_{\|s} \equiv \frac{\omega}{|k_{\|}| v_{\mathrm{th}s}} , \label{normfrequency}
\end{equation}
\begin{equation}
  \zeta_{sn} \equiv \tilde{\omega}_{\|s} - \frac{n}{|k_{\|}| \tilde{\rho}_s} , \label{zeta_def}
\end{equation}
\begin{equation}
  \tilde{f}_{s0}(\tilde{v}_{\|s},\tilde{v}_{s\bot}) \equiv \frac{\upi^{3/2} v_{\mathrm{th}s}^3}{n_{s0}} f_{s0}\left(\frac{k_{\|}}{|k_{\|}|} v_{\mathrm{th}s} \tilde{w}_{\|s},v_{\mathrm{th}s} \tilde{v}_{s\bot}\right)
  , \label{renorm_distfunc_def}
\end{equation}
\begin{equation}
  \Lambda_s(\tilde{w}_{\|s},\tilde{v}_{s\bot}) \equiv \tilde{v}_{s\bot} \frac{\p \tilde{f}_{s0}}{\p \tilde{w}_{\|s}}-\tilde{w}_{\|s} \frac{\p \tilde{f}_{s0}}{\p \tilde{v}_{s\bot}} 
  , \label{anisotropyfunc}
\end{equation}
\begin{equation}
  \Xi_s(\tilde{w}_{\|s},\tilde{v}_{s\bot}) \equiv \frac{\p \tilde{f}_{s0}}{\p \tilde{v}_{s\bot}}
  + \frac{\Lambda_s(\tilde{w}_{\|s},\tilde{v}_{s\bot})}{\tilde{\omega}_{\|s}} , \label{IntgradCond}
\end{equation}
\normalsize
\begin{subeqnarray}
 (\mathsfbi{R}_{sn} )_{xx} & \equiv & \frac{n^2 \besselJ_n(k_{\bot} \tilde{\rho}_s \tilde{v}_{s\bot})^2}{k_{\bot}^2 \tilde{\rho}_s^2 \tilde{v}_{s\bot}^2} , \\
 (\mathsfbi{R}_{sn} )_{xy} & \equiv & \frac{\mathrm{i} n \besselJ_n(k_{\bot} \tilde{\rho}_s \tilde{v}_{s\bot}) \besselJ_n'(k_{\bot} \tilde{\rho}_s \tilde{v}_{s\bot})}{k_{\bot} \tilde{\rho}_s \tilde{v}_{s\bot}} , \\
 (\mathsfbi{R}_{sn} )_{xz} & \equiv & \frac{n \bigl[\besselJ_n(k_{\bot} \tilde{\rho}_s \tilde{v}_{s\bot})\bigr]^2}{k_{\bot} \tilde{\rho}_s \tilde{v}_{s\bot}} \frac{k_{\|} \tilde{w}_{\|s}}{|k_{\|}| \tilde{v}_{s\bot}}  ,\\
 (\mathsfbi{R}_{sn} )_{yx} & = & - (\mathsfbi{R}_{sn} )_{xy} ,\\
 (\mathsfbi{R}_{sn} )_{yy} & \equiv & \bigl[\besselJ_n'(k_{\bot} \tilde{\rho}_s \tilde{v}_{s\bot})\bigr]^2 ,\\
 (\mathsfbi{R}_{sn} )_{yz} & \equiv & -\mathrm{i} n \besselJ_n(k_{\bot} \tilde{\rho}_s \tilde{v}_{s\bot}) \besselJ_n'(k_{\bot} \tilde{\rho}_s \tilde{v}_{s\bot}) \frac{k_{\|} \tilde{w}_{\|s}}{|k_{\|}| \tilde{v}_{s\bot}} , \\
 (\mathsfbi{R}_{sn} )_{zx} & = & (\mathsfbi{R}_{sn} )_{xz} ,\\
 (\mathsfbi{R}_{sn} )_{zy} & = & -(\mathsfbi{R}_{sn} )_{yz} ,\\
 (\mathsfbi{R}_{sn} )_{zz} & \equiv & \frac{\tilde{w}_{\|s}^2}{\tilde{v}_{s\bot}^2} \bigl[\besselJ_n(k_{\bot} \tilde{\rho}_s \tilde{v}_{s\bot})\bigr]^2 . \label{defMsn}
\end{subeqnarray}
For bi-Maxwellian ions and Maxwellian electrons, 
\begin{subeqnarray}
f_{i0}(v_{\|},v_{\bot}) & = & \frac{n_{i0}}{\upi^{3/2} v_{\mathrm{th}\|i} v_{\mathrm{th}\perp i}^2} \exp \left(-\frac{v_{\|}^2}{v_{\mathrm{th}\|i}^2}-\frac{v_{\perp}^2}{v_{\mathrm{th}\perp i}^2}\right) 
, \\ f_{e0}(v_{\|},v_{\bot}) & = & \frac{n_{e0}}{\upi^{3/2} v_{\mathrm{th}e}^3} \exp \left(-\frac{v^2}{v_{\mathrm{th}e}^2}\right) 
,
\end{subeqnarray}
where $v_{\mathrm{th}\|i} \equiv \sqrt{2 T_{\|i}/m_i}$ the parallel thermal ion velocity, $v_{\mathrm{th}\perp i} \equiv \sqrt{2 T_{\perp i}/m_i}$ the perpendicular thermal ion velocity, and $n_{e0} = n_{i0}$ the electron number density, the integrals in (\ref{conductivity}) can be evaluated in terms of the plasma 
dispersion function and modified Bessel functions~\citep{Davidson_1983}. 

\subsection{The growth rate of the resonant parallel firehose instability in $\beta_i \gg 1$ plasma with a weak anisotropy} \label{append_resparfiregrowthrate}

Next, we derive an analytic expression for the linear growth rate of resonant parallel firehose modes in magnetised, $\beta_i \gg 1$ plasma. As discussed in the main text, these modes are hydromagnetic waves that become resonantly unstable in a plasma with 
a bi-Maxwellian ion distribution (and Maxwellian electron distribution) whose parallel ion temperature is greater than its perpendicular temperature. We focus on modes whose 
wavevector is exactly parallel to the background magnetic field $\bb{B}_0$, which previous numerical work indicates are the fastest growing resonant parallel firehose modes \citep{Gary_1998}. Also motivated by the findings of this prior research, we assume that the real frequency $\varpi$ of these modes is much greater than the growth rate $\gamma$ (an assumption we confirm \textit{a posteriori}). Consistent with previous analytic results \citep{Sagdeev_1960}, we find that right-handed, circularly polarised modes are unstable at arbitrarily small negative pressure anisotropy $\Delta_i$. However, for plasmas in which $\Delta_i \sim -1/\beta_i \ll 1$, we find that the fastest growing resonant parallel firehose modes have a characteristic scale that is much larger than the ion Larmor scale ($k_{\|} \rho_i \sim \Delta_i^{1/2} \ll 1$), and a growth rate that is exponentially small 
in $|\Delta_i| \ll 1$. 

\subsubsection{Dispersion relation}

We start from the hot-plasma dispersion relation for parallel modes in a bi-Maxwellian, non-relativistic plasma~\citep{Davidson_1983}:
\begin{equation}
  \frac{c^2 k_{\|}^2}{\omega^2} - 1 = \sum_s \frac{\omega_{\mathrm{p}s}^2}{\omega^2} 
  \left\{\frac{\omega}{k_{\|} v_{\mathrm{th}s\|}} Z(\xi_s^{\pm}) + \Delta_s \left[1+\xi_s^{\pm}Z(\xi_s^{\pm})\right]\right\} 
  \, , \label{parallelhotplasmadisprel}
\end{equation}
where we remind the reader that $\omega \equiv \varpi + \mathrm{i} \gamma$ 
is the complex frequency, $Z(x)$ the plasma dispersion function (Fried and Conte 1960), 
\begin{equation}
  \xi_s^{\pm} = \frac{\omega\pm\Omega_s}{k_{\|} v_{\mathrm{th}s\|}} \, ,
\end{equation}
and $\Omega_s \equiv q_s B_0/m_s c$ is the Larmor frequency of species $s$; the sum is taken over all particle species in the plasma. For forward-propagating modes ({\it viz.}, $k_{\|} > 0$), right/left-handed circularly polarised modes are described by the $+/-$ branch of 
(\ref{parallelhotplasmadisprel}), respectively. 

To characterise resonant parallel firehose modes, we specialise to the $+$ branch, and 
then assume a two-species plasma with Maxwellian electrons. Equation~\eqref{parallelhotplasmadisprel} then simplifies to 
\begin{equation}
  c^2 k_{\|}^2 =\omega^2 + \omega_{\mathrm{p}i}^2 \frac{\omega}{k_{\|} v_{\mathrm{th}i\|}} Z(\xi_i^{+}) 
  + \omega_{\mathrm{p}i}^2 \Delta_i \left[1+\xi_i^{+}Z(\xi_i^{+})\right] + \omega_{\mathrm{p}e}^2 \frac{\omega}{k_{\|} v_{\mathrm{th}e}} Z(\xi_e^{+}) 
\, . \label{disprelgen}
\end{equation}
To proceed further analytically, we must adopt an ordering.

\subsubsection{Ordering and simplifications}

In order to characterise near-marginal modes, we assume that $\varpi \gg \gamma$, and 
\begin{equation}
  \frac{\varpi}{\Omega_i} \sim |\Delta_i| \sim \frac{1}{\beta_{\|i}} \sim k_{\|}^2 \rho_i^2  \ll 1 \, . 
  \label{ordering_highbeta}
\end{equation}
This ordering implicitly assumes that the wavelength of the resonant parallel firehose modes is much longer than the ion Larmor scale (an assumption that we will verify \textit{a posteriori}). 
Under this ordering, we can neglect 
the displacement current term on the left-hand side of  (\ref{disprelgen}), because $\omega^2/c^2 k_{\|}^2 \sim \beta_{i}^{-1} v_{\mathrm{th}i\|}^2/c^2 \ll 1$. We also have that
\begin{equation}
  \xi_s^{+} \approx \frac{\Omega_s}{k_{\|} v_{\mathrm{th}s\|}} = \frac{1}{k_{\|} \rho_{\|s}} 
  \gg 1 
\end{equation}
for both species, where $\rho_{\|s} \equiv v_{\mathrm{th}s\|}/\Omega_s$ (note that, due to our assumed ordering, ${|\rho_{\|s}/\rho_{s} - 1| \ll 1}$). 
We can therefore use the large-argument expansion of the plasma dispersion 
function: 
\begin{equation}
  Z(\xi_s^{+}) = \left[-\frac{1}{\xi_s^{+}}-\frac{1}{2(\xi_s^{+})^3} + \mathcal{O}\left({k_{\|}^{5/2} |\rho_{\|s}|^{5/2}}\right)\right] 
  + \mathrm{i} \sqrt{\upi} \exp{\left[-(\xi_s^{+})^2\right]} .
\end{equation} 
To expand in $\gamma/\varpi \ll 1$, we use
\begin{equation}
\frac{1}{(\xi_s^{+})^n} = (k_{\|} \rho_{\|s})^n \left[1 - \frac{n \omega}{\Omega_s} + \mathcal{O}\left(\frac{\varpi^2}{\Omega_s^2}\right) \right] 
\, .
\end{equation}
Then, it can be shown that  
\begin{eqnarray}
\frac{\omega}{k_{\|} v_{\mathrm{th}s\|}} Z(\xi_s^{+}) & = & \left[ - \frac{\varpi}{\Omega_s} 
+ \frac{\varpi^2}{\Omega_s^2}  - \frac{1}{2} k_{\|}^2 \rho_{\|s}^2 \frac{\varpi}{\Omega_s} 
+ \mathcal{O}\left(\frac{\varpi^3}{\Omega_s^3}\right) \right] \nonumber \\*
&& \mbox{} + \mathrm{i} \left\{\frac{\gamma}{\Omega_s} \left(\frac{2 \varpi}{\Omega_s} - 1\right)+ \frac{\varpi}{\Omega_s} \frac{\sqrt{\upi}}{k_{\|} \rho_{\|s}} \exp{\left[-\frac{\left(1-\varpi/\Omega_s\right)^2}{k_{\|}^2 \rho_{\|s}^2}\right]} \right\} 
\, , \\
1 + \xi_s^{+} Z(\xi_s^{+}) & = & \left[-\frac{1}{2} k_{\|}^2 \rho_{\|s}^2 + \mathcal{O}\left({k_{\|}^{4} |\rho_s|^{4}}\right) \right] 
+ \frac{\mathrm{i} \sqrt{\upi}}{k_{\|} \rho_{\|s}} \exp{\left[-\frac{\left(1-\varpi/\Omega_s\right)^2}{k_{\|}^2 
\rho_{\|s}^2}\right]} \, .
\end{eqnarray}

\subsubsection{Real frequency}

Assuming that $T_e = T_{\|i}$, the final term in \eqref{disprelgen} associated with the electrons becomes
\begin{equation}
\omega_{\mathrm{p}e}^2\frac{\omega}{k_{\|} v_{\mathrm{th}e}} Z(\xi_e^{+}) 
= \omega_{\mathrm{p}i}^2 \frac{\varpi}{\Omega_i} \left[1+ \mathcal{O}\left(\frac{m_e}{m_i}\right)\right] 
\, .
\end{equation}
The real part of the dispersion relation (\ref{disprelgen}) then gives (to leading-order in $\Delta_i \ll 
1$), 
\begin{equation}
 \frac{k_{\|}^2 \rho_i^2}{\beta_{\|i}} \approx \frac{\varpi^2}{\Omega_i^2} - \frac{1}{2} k_{\|}^2 \rho_i^2 \frac{\varpi}{\Omega_i} 
 - \frac{1}{2} \Delta_i k_{\|}^2 \rho_i^2 \, , \label{realfreqdisprel}
\end{equation}
where we have used the result $\rho_i = \beta_{\|i}^{1/2} d_i$ 
to relate the ion Larmor radius to the ion skin depth $d_i =  
c/\omega_{\mathrm{p}i}$. The (positive) roots of (\ref{realfreqdisprel}) are given by
\begin{equation}
  \frac{\varpi}{\Omega_i} \approx \frac{1}{4} k_{\|}^2 \rho_i^2 + \sqrt{\frac{1}{16} k_{\|}^4 \rho_i^4 + \left(\frac{1}{\beta_{\|i}}+\frac{\Delta_i}{2}\right) k_{\|}^2 \rho_i^2} 
  \, . \label{realfreqresult}
\end{equation}

\subsubsection{Growth rate}

The imaginary part of (\ref{disprelgen}) 
is
\begin{equation}
 \frac{2 \gamma \varpi}{\Omega_i^2} +  \frac{\varpi}{\Omega_i} \frac{\sqrt{\upi}}{k_{\|} \rho_{\|i}} \exp{\left[-\frac{\left(1-\varpi/\Omega_i\right)^2}{k_{\|}^2 
\rho_{\|i}^2}\right]} +  \frac{\sqrt{\upi}}{k_{\|} \rho_{\|i}} \Delta_i \exp{\left[-\frac{\left(1-\varpi/\Omega_i\right)^2}{k_{\|}^2 \rho_{\|i}^2} \right]}
= 0 ,
\end{equation}
which can be rearranged to give
\begin{equation}
 \frac{\gamma}{\Omega_i} \approx \frac{\sqrt{\upi}}{2 k_{\|} \rho_{\|i}} \left(-1 - \frac{\Delta_i}{\varpi/\Omega_i}\right) 
 \exp{\left[-\frac{\left(1-\varpi/\Omega_i\right)^2}{k_{\|}^2 \rho_{\|i}^2} \right]} \, . 
 \label{growthrateresult}
\end{equation}
It is clear from (\ref{growthrateresult}) that right-handed modes can be 
unstable for $\Delta_i < 0$ if
\begin{equation}
  \frac{\varpi}{\Omega_i} < |\Delta_i|  \, . \label{thresholdA}
\end{equation}
Substituting \eqref{realfreqresult} in for $\varpi/\Omega_i$, the inequality \eqref{thresholdA} is equivalent to 
\begin{equation}
  k_{\|} \rho_{\|i} < |\Delta_i| \beta_{\|i}^{1/2} \, .
\end{equation}
Thus, for $\Delta_i$ arbitrarily small, right-handed modes are unstable at 
sufficiently large parallel wavelengths; for $|\Delta_i| \sim \beta_{\|i}^{-1}$, we have $k \rho_{\|i} \sim |\Delta_i|^{1/2} \ll 1$.

It can be shown by considering the magnitude of the neglected higher-order terms in the expansion of $\omega/\Omega_i$ in $|\Delta_i|\sim\beta_{\|i}^{-1} \ll 1$ that the growth rate $\gamma_{\rm peak}$ of the fastest-growing modes 
satisfies the asymptotic scaling
\begin{equation}
 \gamma_{\rm peak} \sim |\Delta_i|^{1/2} \exp{\left(-\frac{1}{|\Delta_i|} \right)} 
 \Omega_i \sim \beta_{\|i}^{-1/2} \exp{\left(-\beta_{\|i} \right)} \Omega_i
 \, , \label{peakgrowthrate}
\end{equation}
at the wavenumber
\begin{equation}
 (k_{\|} \rho_{\|i})_{\rm peak} \approx |\Delta_i| \beta_{\|i}^{1/2} - \alpha \Delta_i^{3/2} \, ,  
\end{equation}
 where $\alpha$ is some order-unity, positive number. The associated real frequency is given by
 \begin{equation}
 \frac{\varpi}{\Omega_i} \approx |\Delta_i| + \mathcal{O}\left(\Delta_i^{2}\right) \, ,
 \end{equation}
 which validates our assumption that $\varpi \gg \gamma$. Equation (\ref{peakgrowthrate}) is the main result of this section, which is used in \S\ref{linfiretheory_resparfire} of the main text. Determining exact expressions 
 requires going to next order in the expansion (an algebraically involved exercise, which we leave to the reader). 
 
\subsection{Calculating the threshold of the oblique firehose instability} \label{append_firehose_threshold_condition}

Numerical calculations for a plasma with bi-Maxwellian ions and Maxwellian electrons
indicate that the resonant oblique firehose instability is non-propagating. Therefore, we assume that 
at the threshold for the instability, $\omega = 0$. Under this assumption, the threshold for the resonant oblique firehose
instability with arbitrary ion and electron distribution functions follows from (\ref{hotplasmadisprel}): 
\begin{equation}
\mbox{Det}\left\{\frac{k^2 \rho_i^2}{\beta_i} \left(\hat{\bb{k}}\hat{\bb{k}}-\mathsfbi{I}\right) + \tilde{\bb{\sigma}}_0 \right\} = 0 
, \label{Det_thresholdeqn}
\end{equation}
where 
\begin{eqnarray}
\tilde{\bb{\sigma}}_0 & = & \sum_s \frac{\omega_{\mathrm{p}s}^2}{\omega_{\mathrm{p}e}^2} \bigg[ \frac{2}{\sqrt{\upi}} \frac{k_{\|}}{|k_{\|}|}  \int_{-\infty}^{\infty} \mathrm{d} \tilde{w}_{\|s} \, \tilde{w}_{\|s} \int_0^{\infty} \mathrm{d} \tilde{v}_{s\bot} \, \Lambda_s(\tilde{w}_{\|s},\tilde{v}_{s\bot}) \hat{\bb{z}} \hat{\bb{z}} \nonumber \\*
&& \mbox{} - \frac{2}{\sqrt{\upi}} \int_{C_{\rm L}} \mathrm{d} \tilde{w}_{\|s} \int_0^{\infty} \mathrm{d} \tilde{v}_{s\bot} \,\tilde{v}_{s\bot}^2 \Lambda_s(\tilde{w}_{\|s},\tilde{v}_{s\bot}) \sum_{n = -\infty}^{\infty} \frac{\mathsfbi{R}_{sn}}{n/|k_{\|}|\tilde{\rho}_s +\tilde{w}_{\|s}}  \bigg] \, . \qquad \label{conductivity_thres}
\end{eqnarray}
For any species with an isotropic distribution function, $\Lambda_s = 
0$; so only anisotropic species provide a non-zero contribution to $\tilde{\bb{\sigma}}_0$. 
It can be shown that, for any set of distribution functions that are even in $\tilde{w}_{\|s}$, 
there exists a solution of (\ref{Det_thresholdeqn}) with $k_{\|} < 0$ if and 
only if there exists a solution with $k_{\|} > 0$; we therefore, without loss of 
generality, take $k_{\|} > 0$, and so $\tilde{w}_{\|s} = \tilde{v}_{\|s}$. For 
the same set of distribution functions, 
the tensor $\tilde{\bb{\sigma}}_0$ has the following exact symmetries:
\begin{subeqnarray}
  (\tilde{\bb{\sigma}}_0)_{xz} & = & (\tilde{\bb{\sigma}}_0)_{zx} = - \frac{k_{\bot}}{k_{\|}} (\tilde{\bb{\sigma}}_0)_{xx} \, , \\
  (\tilde{\bb{\sigma}}_0)_{yz} & = & -(\tilde{\bb{\sigma}}_0)_{zy} = \frac{k_{\bot}}{k_{\|}} (\tilde{\bb{\sigma}}_0)_{xy} = - \frac{k_{\bot}}{k_{\|}} (\tilde{\bb{\sigma}}_0)_{yx} \, , \\
  (\tilde{\bb{\sigma}}_0)_{zz} & = & \frac{k_{\bot}^2}{k_{\|}^2} (\tilde{\bb{\sigma}}_0)_{xx} \, , 
  \label{dielectricsyms}
\end{subeqnarray}
where the three independent components of $\tilde{\bb{\sigma}}_0$ are
\begin{subeqnarray}
 (\tilde{\bb{\sigma}}_0)_{xx} & \equiv &  -\frac{2}{\sqrt{\upi}} \sum_{s} \frac{\omega_{\mathrm{p}s}^2}{\omega_{\mathrm{p}e}^2}  \sum_{n=-\infty}^{\infty} \Bigg\{ \frac{n^2}{k_{\bot}^2 \tilde{\rho}_s^2}\int_{C_{\rm L}} \mathrm{d} \tilde{v}_{\|s}\, \frac{1}{\tilde{v}_{\|s}+n/k_{\|} \tilde{\rho}_s} \nonumber \\*
 && \mbox{} \qquad \qquad \qquad \times \int_0^{\infty} \mathrm{d} \tilde{v}_{s\bot} \,\Lambda_s(\tilde{v}_{\|s},\tilde{v}_{s\bot}) \bigl[\besselJ_n(k_{\bot} \tilde{\rho}_s \tilde{v}_{s\bot})\bigr]^2 \Bigg\} ,  \\
 (\tilde{\bb{\sigma}}_0)_{xy} & \equiv & -\frac{2 \mathrm{i}}{\sqrt{\upi}} \sum_{s} \frac{\omega_{\mathrm{p}s}^2}{\omega_{\mathrm{p}e}^2}  \sum_{n=-\infty}^{\infty} \left[ \frac{n}{k_{\bot} \tilde{\rho}_s} \int_{C_{\rm L}} \mathrm{d} \tilde{v}_{\|s}\, \frac{1}{\tilde{v}_{\|s}+n/k_{\|} \tilde{\rho}_s} \right. \nonumber \\*
 && \mbox{} \left. \qquad \qquad \times \int_0^{\infty} \mathrm{d} \tilde{v}_{s\bot} \, \tilde{v}_{s\bot} \Lambda_s(\tilde{v}_{\|s},\tilde{v}_{s\bot}) \besselJ_n(k_{\bot} \tilde{\rho}_s \tilde{v}_{s\bot}) \besselJ_n'(k_{\bot} \tilde{\rho}_s \tilde{v}_{s\bot}) \right] , \quad \\
 (\tilde{\bb{\sigma}}_0)_{yy} & \equiv & -\frac{2}{\sqrt{\upi}} \sum_{s} \frac{\omega_{\mathrm{p}s}^2}{\omega_{\mathrm{p}e}^2} \sum_{n=-\infty}^{\infty} \Bigg\{ \int_{C_{\rm L}} \mathrm{d} \tilde{v}_{\|s}\,\frac{1}{\tilde{v}_{\|s}+n/k_{\|} \tilde{\rho}_s} \nonumber \\*
 && \mbox{} \qquad \qquad \qquad \times \int_0^{\infty} \mathrm{d} \tilde{v}_{s\bot} \, \tilde{v}_{s\bot}^2 \Lambda_s(\tilde{v}_{\|s},\tilde{v}_{s\bot}) \bigl[\besselJ_n'(k_{\bot} \tilde{\rho}_s \tilde{v}_{s\bot})\bigr]^2 \Bigg\} . 
  \label{dielectric0elements}
\end{subeqnarray}
A corollary of this
useful property is that $\tilde{\bb{\sigma}}_0$ is orthogonal to $\hat{\bb{k}}$, 
and can be written in the form 
\begin{equation}
\tilde{\bb{\sigma}}_0 =   \frac{k^2}{k_{\|}^2} (\tilde{\bb{\sigma}}_0)_{xx} \bb{e}_1 \bb{e}_1 + \frac{k}{k_{\|}} (\tilde{\bb{\sigma}}_0)_{xy}  \left(\bb{e}_1 
  \bb{e}_2 - \bb{e}_2 \bb{e}_1 \right) + (\tilde{\bb{\sigma}}_0)_{yy}
  \bb{e}_2 \bb{e}_2  , 
\end{equation} 
where $\{\bb{e}_1,\bb{e}_2,\bb{e}_3\}$ is a coordinate 
basis defined by 
\begin{equation}
\bb{e}_1 \equiv \hat{\bb{y}} \times \hat{\bb{k}} , \quad 
\bb{e}_2 \equiv \hat{\bb{y}} , \quad 
\bb{e}_3 \equiv \hat{\bb{k}} . 
\end{equation}
Because $\mathsfbi{I}-\hat{\bb{k}}\hat{\bb{k}} = \bb{e}_1 \bb{e}_1 +\bb{e}_2 \bb{e}_2 
$, it follows that (\ref{Det_thresholdeqn}) becomes
\begin{equation}
\mbox{Det}\left\{\left[(\tilde{\bb{\sigma}}_0)_{xx} -\frac{k_{\|}^2 \rho_e^2}{\beta_e}\right] \bb{e}_1 \bb{e}_1 + \frac{k_{\|}}{k} (\tilde{\bb{\sigma}}_0)_{xy}  \left(\bb{e}_1 
  \bb{e}_2 - \bb{e}_2 \bb{e}_1 \right)+\left[(\tilde{\bb{\sigma}}_0)_{yy} -\frac{k^2 \rho_e^2}{\beta_e}\right] \bb{e}_2 \bb{e}_2 \right\} = 0 
. \label{Det_thresholdeqn_v2}
\end{equation}

Next, we note some identities that will prove to be useful for simplifying (\ref{Det_thresholdeqn_v2}). First, for any function $\mathcal{G}(\tilde{v}_{\|s})$,
\begin{eqnarray}
\sum_{n = -\infty}^{\infty} n \int_{C_{\rm L}} \mathrm{d} \tilde{v}_{\|s}\,\frac{1}{\tilde{v}_{\|s}+n/k_{\|} \tilde{\rho}_s} \mathcal{G}(\tilde{v}_{\|s}) 
& = & 2 \mathcal{P} \! \int \mathrm{d} \tilde{v}_{\|s} \sum_{n = 1}^{\infty} \frac{n^2 k_{\|} \tilde{\rho}_s}{n^2- k_{\|}^2 \tilde{\rho}_s^2 \tilde{v}_{\|s}^2} \tilde{v}_{\|s} \mathcal{G}(\tilde{v}_{\|s}) \nonumber \\
&&\mbox{} - \mathrm{i} \upi \sum_{n = -\infty}^{\infty} n \mathcal{G}(n/k_{\|} \tilde{\rho}_s) 
\, .
\end{eqnarray}
It follows that, if $\mathcal{G}(\tilde{v}_{\|s})$ is odd 
in $\tilde{v}_{\|s}$, then 
\begin{equation}
\sum_{n = -\infty}^{\infty} n \int_{C_{\rm L}} \mathrm{d} \tilde{v}_{\|s}\,\frac{1}{\tilde{v}_{\|s}+n/k_{\|} \tilde{\rho}_s} \mathcal{G}(\tilde{v}_{\|s}) 
 =  - \mathrm{i} \upi \sum_{n = -\infty}^{\infty} n \mathcal{G}(n/k_{\|} \tilde{\rho}_s) 
\, .
\end{equation}
If we then choose
\begin{equation}
\mathcal{G}(\tilde{v}_{\|s}) = \frac{1}{k_{\bot} \tilde{\rho}_s} 
\int_0^{\infty} \mathrm{d} \tilde{v}_{s\bot} \, \tilde{v}_{s\bot} \Lambda_s(\tilde{v}_{\|s},\tilde{v}_{s\bot}) \besselJ_n(k_{\bot} \tilde{\rho}_s \tilde{v}_{s\bot}) \besselJ_n'(k_{\bot} \tilde{\rho}_s \tilde{v}_{s\bot}) ,  
\end{equation}
it follows that, if the distribution functions $\tilde{f}_{s0}$ are even in 
$\tilde{v}_{\|s}$, then $\Lambda_s$ is odd in $\tilde{v}_{\|s}$, and so 
\begin{eqnarray}
(\tilde{\bb{\sigma}}_0)_{xy} & = & -2 \sqrt{\upi} \sum_{s} \frac{\omega_{\mathrm{p}s}^2}{\omega_{\mathrm{p}e}^2} 
\nonumber \\*
&& \mbox{} \times \sum_{n=-\infty}^{\infty} \frac{n}{k_{\bot} \tilde{\rho}_s} \int_0^{\infty} \mathrm{d} \tilde{v}_{s\bot} \, \tilde{v}_{s\bot} \Lambda_s(n/k_{\|} \tilde{\rho}_s,\tilde{v}_{s\bot}) \besselJ_n(k_{\bot} \tilde{\rho}_s \tilde{v}_{s\bot}) \besselJ_n'(k_{\bot} \tilde{\rho}_s \tilde{v}_{s\bot}) . \quad \quad \quad \label{conductivity_xy_identity}
\end{eqnarray}
It can be shown similarly that 
\begin{eqnarray}
\sum_{n = -\infty}^{\infty} n^2 \int_{C_{\rm L}} \mathrm{d} \tilde{v}_{\|s}\, \frac{1}{\tilde{v}_{\|s}+n/k_{\|} \tilde{\rho}_s} \mathcal{G}(\tilde{v}_{\|s}) 
& = & -2 \mathcal{P} \! \int \mathrm{d} \tilde{v}_{\|s} \sum_{n = 1}^{\infty} \frac{n^2 k_{\|}^2 \tilde{\rho}_s^2}{n^2- k_{\|}^2 \tilde{\rho}_s^2 \tilde{v}_{\|s}^2} \tilde{v}_{\|s} \mathcal{G}(\tilde{v}_{\|s}) \nonumber \\*
&& \mbox{} + \mathrm{i} \upi \sum_{n = -\infty}^{\infty} n^2 \mathcal{G}(n/k_{\|} \tilde{\rho}_s) 
\, ,
\end{eqnarray}
and so if $\mathcal{G}(\tilde{v}_{\|s})$ is an odd function, then 
\begin{equation}
\sum_{n = -\infty}^{\infty} n^2 \int_{C_{\rm L}} \mathrm{d} \tilde{v}_{\|s}\,\frac{1}{\tilde{v}_{\|s}+n/k_{\|} \tilde{\rho}_s} \mathcal{G}(\tilde{v}_{\|s}) 
 = -2 \mathcal{P} \! \int \mathrm{d} \tilde{v}_{\|s} \sum_{n = 1}^{\infty} \frac{n^2 k_{\|}^2 \tilde{\rho}_s^2}{n^2- k_{\|}^2 \tilde{\rho}_s^2 \tilde{v}_{\|s}^2} \tilde{v}_{\|s} \mathcal{G}(\tilde{v}_{\|s}) 
 .
\end{equation}
Now choosing 
\begin{equation}
\mathcal{G}(\tilde{v}_{\|s}) = \frac{1}{k_{\bot}^2 \tilde{\rho}_s^2}
\int_0^{\infty} \mathrm{d} \tilde{v}_{s\bot} \, \Lambda_s(\tilde{v}_{\|s},\tilde{v}_{s\bot}) \bigl[ \besselJ_n(k_{\bot} \tilde{\rho}_s \tilde{v}_{s\bot})\bigr]^2 ,  
\end{equation}
and again assuming that the distribution functions $\tilde{f}_{s0}$ are even in 
$\tilde{v}_{\|s}$, we deduce that
\begin{eqnarray}
 (\tilde{\bb{\sigma}}_0)_{xx} & \equiv &  \frac{4}{\sqrt{\upi}} \sum_{s} \frac{\omega_{\mathrm{p}s}^2}{\omega_{\mathrm{p}e}^2}  \sum_{n=1}^{\infty} \Bigg\{  \frac{k_{\|}^2 \tilde{\rho}_s^2}{k_{\perp}^2 \tilde{\rho}_s^2} \,\mathcal{P} \! \int \mathrm{d} \tilde{v}_{\|s}\,\frac{\tilde{v}_{\|s}}{1- k_{\|}^2 \tilde{\rho}_s^2 \tilde{v}_{\|s}^2/n^2} \nonumber \\*
 && \mbox{} \qquad \qquad \qquad \times \int_0^{\infty} \mathrm{d} \tilde{v}_{s\bot} \, \Lambda_s(\tilde{v}_{\|s},\tilde{v}_{s\bot}) \bigl[\besselJ_n(k_{\bot} \tilde{\rho}_s \tilde{v}_{s\bot})\bigr]^2 \Bigg\}
 . \label{conductivity_xx_identity}
\end{eqnarray}

\subsubsection{The long-wavelength, oblique $(k_{\perp} \sim k_{\|} \ll \rho_i^{-1})$ limit}

We can now carry out one possible secondary subsidiary expansion of (\ref{Det_thresholdeqn_v2}): we assume $k_{\|} \tilde{\rho}_s \sim k_{\perp} \tilde{\rho}_s \ll 
1$. For a plasma with bi-Maxwellian distribution functions for all species (or any distribution which 
does not have anisotropic power-law tails), then $\Lambda_s$ is exponentially small in $k_{\|} \tilde{\rho}_s \ll 
1$, and so therefore is $(\tilde{\bb{\sigma}}_0)_{xy}$. It follows 
that, if $k_{\|} \tilde{\rho}_s \ll 1$, equation~(\ref{Det_thresholdeqn_v2}) simplifies 
to
\begin{equation}
\mbox{Det}\left\{\left[(\tilde{\bb{\sigma}}_0)_{xx} -\frac{k_{\|}^2 \rho_e^2}{\beta_e}\right] \bb{e}_1 \bb{e}_1 +\left[(\tilde{\bb{\sigma}}_0)_{yy} -\frac{k^2 \rho_e^2}{\beta_e}\right] \bb{e}_2 \bb{e}_2 \right\} = 0  
. \label{Det_thresholdeqn_v3}
\end{equation}
The electric field eigenvector of oblique firehose modes is parallel to 
$\bb{e}_1$, so the threshold condition for oblique firehose modes is 
\begin{equation}
(\tilde{\bb{\sigma}}_0)_{xx} -\frac{k_{\|}^2 \rho_e^2}{\beta_e} = 0 \, .  
\end{equation}
We now expand $(\tilde{\bb{\sigma}}_0)_{xx}$ in $k_{\|} \tilde{\rho}_s \sim k_{\perp} \tilde{\rho}_s \ll 
1$ using the summation identity
\begin{equation}
 \frac{1}{1- k_{\|}^2 \tilde{\rho}_s^2 \tilde{v}_{\|s}^2/n^2} = 1 + \frac{1}{n^2} k_{\|}^2 \tilde{\rho}_s^2 \tilde{v}_{\|s}^2 + \frac{1}{n^4} k_{\|}^4 \tilde{\rho}_s^4 \tilde{v}_{\|s}^4 + \dots  
\, ,
\end{equation}
and also
\begin{subeqnarray}
\bigl[ \besselJ_1(k_{\bot} \tilde{\rho}_s \tilde{v}_{s\bot})\bigr]^2 & = & \frac{1}{4}k_{\bot}^2 \tilde{\rho}_s^2 \tilde{v}_{s\bot}^2-\frac{1}{16} k_{\bot}^4 \tilde{\rho}_s^4 \tilde{v}_{s\bot}^4 + \dots \, , 
\\
\bigl[ \besselJ_2(k_{\bot} \tilde{\rho}_s \tilde{v}_{s\bot})\bigr]^2 & = & \frac{1}{64} k_{\bot}^4 \tilde{\rho}_s^4 \tilde{v}_{s\bot}^4 + \dots \, . \label{longwavelengthBessel}
\end{subeqnarray}
Equation~(\ref{conductivity_xx_identity}) then becomes
\begin{eqnarray}
 (\tilde{\bb{\sigma}}_0)_{xx} & \equiv &  \frac{1}{\sqrt{\upi}} \sum_{s} k_{\|}^2 \tilde{\rho}_s^2 \frac{\omega_{\mathrm{p}s}^2}{\omega_{\mathrm{p}e}^2} \Bigg[ \int_{-\infty}^{\infty} \mathrm{d} \tilde{v}_{\|s} \int_0^{\infty} \mathrm{d} \tilde{v}_{s\bot} \, \tilde{v}_{\|s} \tilde{v}_{s\bot}^2 \Lambda_s(\tilde{v}_{\|s},\tilde{v}_{s\bot}) \nonumber \\*
 && \mbox{}\quad + k_{\|}^2 \tilde{\rho}_s^2 \int_{-\infty}^{\infty} \mathrm{d} \tilde{v}_{\|s} \int_0^{\infty} \mathrm{d} \tilde{v}_{s\bot} \, \tilde{v}_{\|s}^3 \tilde{v}_{s\bot}^2 \Lambda_s(\tilde{v}_{\|s},\tilde{v}_{s\bot})  
 \\
  && \mbox{}\quad - \frac{3}{16} k_{\perp}^2 \tilde{\rho}_s^2 \int_{-\infty}^{\infty} \mathrm{d} \tilde{v}_{\|s} \int_0^{\infty} \mathrm{d} \tilde{v}_{s\bot} \, \tilde{v}_{\|s} \tilde{v}_{s\bot}^4 \Lambda_s(\tilde{v}_{\|s},\tilde{v}_{s\bot})  + \mathcal{O}(k^4 \tilde{\rho}_s^4) \Bigg] 
 . \label{conductivity_xx_expansion}
\end{eqnarray}
Now we use the identities
\begin{subeqnarray}
\int_{-\infty}^{\infty} \mathrm{d} \tilde{v}_{\|s} \int_0^{\infty} \mathrm{d} \tilde{v}_{s\bot} \, \tilde{v}_{\|s} \tilde{v}_{s\bot}^2 \Lambda_s & = & 2 \int_{-\infty}^{\infty} \mathrm{d} \tilde{v}_{\|s} \int_0^{\infty} \mathrm{d} \tilde{v}_{s\bot} \, \tilde{v}_{s\bot} \left(\tilde{v}_{\|s}^2-\frac{1}{2} \tilde{v}_{\perp s}^2 \right) \tilde{f}_{s0} \nonumber \\
& = & -\frac{\sqrt{\upi}}{2}  \frac{T_{\|s}}{T_{s}} \Delta_s , \\
\int_{-\infty}^{\infty} \mathrm{d} \tilde{v}_{\|s} \int_0^{\infty} \mathrm{d} \tilde{v}_{s\bot} \, \tilde{v}_{\|s}^3 \tilde{v}_{s\bot}^2 \Lambda_s & = & 4 \int_{-\infty}^{\infty} \mathrm{d} \tilde{v}_{\|s} \int_0^{\infty} \mathrm{d} \tilde{v}_{s\bot} \, \tilde{v}_{s\bot} \tilde{v}_{\|s}^2 \left(\tilde{v}_{\|s}^2-\frac{1}{4} \tilde{v}_{\perp s}^2 \right) \tilde{f}_{s0}  \qquad \quad \\
& = & -\frac{\sqrt{\upi}}{2}  \frac{T_{\|s}}{T_{s}} \mathcal{A}_{4s} , \\
\int_{-\infty}^{\infty} \mathrm{d} \tilde{v}_{\|s} \int_0^{\infty} \mathrm{d} \tilde{v}_{s\bot} \, \tilde{v}_{\|s} \tilde{v}_{s\bot}^4 \Lambda_s & = & 2 \int_{-\infty}^{\infty} \mathrm{d} \tilde{v}_{\|s} \int_0^{\infty} \mathrm{d} \tilde{v}_{s\bot} \, \tilde{v}_{s\bot}^3 \left(\tilde{v}_{\|s}^2-\frac{1}{2} \tilde{v}_{\perp s}^2 \right) \tilde{f}_{s0} 
 \\
& = & -\frac{\sqrt{\upi}}{2}  \frac{T_{\|s}}{T_{s}} \mathcal{B}_{4s} , \label{fourthordermoments}
\end{subeqnarray}
where $\Delta_s = T_{\perp s}/T_{\|s}-1$ is the pressure anisotropy of species $s$, 
and $\mathcal{A}_{4s}$ and $\mathcal{B}_{4s}$ are constants that are 
proportional to fourth-order non-dimensionalised moments of the plasma's 
distribution functions. We deduce that 
\begin{equation}
 (\tilde{\bb{\sigma}}_0)_{xx} = -\frac{1}{2} \sum_{s} k_{\|}^2 \tilde{\rho}_s^2 \frac{\omega_{\mathrm{p}s}^2}{\omega_{\mathrm{p}e}^2}  \frac{T_{\|s}}{T_{s}}\Bigg[ 
 \Delta_s + k_{\|}^2 \tilde{\rho}_s^2 \mathcal{A}_{4s} - \frac{3}{16} k_{\perp}^2 \tilde{\rho}_s^2 \mathcal{B}_{4s} + \mathcal{O}(\Delta_s k^4 \tilde{\rho}_s^4) \Bigg] 
 . \label{conductivity_xx_expansion_outcome}
\end{equation} 
For a plasma with bi-Maxwellian distribution functions, 
\begin{equation}
    \Lambda_s(\tilde{v}_{\|s},\tilde{v}_{s\bot}) = -2 \Delta_s \tilde{v}_{\|s} \tilde{v}_{s\bot} \exp \left(-\tilde{v}_{s}^2\right) 
    ,
\end{equation}
so 
\begin{equation}
\mathcal{A}_{4s} = \frac{3}{2} \Delta_s , \quad 
\mathcal{B}_{4s} = 2 \Delta_s \, .
\end{equation}
  
We can now write down the threshold condition of the resonant oblique firehose instability for the special case 
of a two-species plasma with isotropic electrons and anisotropic ions:
\begin{equation}
\Delta_i + \frac{2}{\beta_{\|i}} +k_{\|}^2 {\rho}_i^2 \mathcal{A}_{4i} - \frac{3}{16} k_{\perp}^2 {\rho}_i^2 \mathcal{B}_{4i} = \mathcal{O}(\Delta_i k^4 
{\rho}_i^4). \label{threshold_nonbimax_Append}
\end{equation}
For a plasma with bi-Maxwellian ions, this becomes
\begin{equation}
\Delta_i \left(1+\frac{3}{2} k_{\|}^2 {\rho}_i^2 - \frac{3}{8} k_{\perp}^2 {\rho}_i^2\right)+ \frac{2}{\beta_{\|i}} = \mathcal{O}(\Delta_i k^4 
{\rho}_i^4), \label{threshold_bimax_Append}
\end{equation}
reproducing (\ref{threshold_bimax_main}).  

\subsubsection{The ion-Larmor-scale, quasi-parallel $(k_{\perp} \ll k_{\|} \lesssim 0.5 \rho_i^{-1})$ subsidiary limit}

As an alternative to the oblique, long-wavelength limit, we instead consider firehose modes with $k_{\perp} \rho_i \ll k_{\|} \rho_i \lesssim 0.5$. In this particular subsidiary limit, we again use the identities (\ref{longwavelengthBessel}) to simplify the dependence of $(\tilde{\bb{\sigma}}_0)_{xx}$ on the Bessel functions, but now only neglect terms that are exponentially small in $k_{\|} \rho_i \ll 1$, not algebraically small. In this case, we have from (\ref{conductivity_xx_expansion}) that for distribution functions $\tilde{f}_{s0}$ that are even in $v_{\|}$, 
\begin{equation}
 (\tilde{\bb{\sigma}}_0)_{xx} \approx \frac{1}{\sqrt{\upi}} \sum_{s} \frac{\omega_{\mathrm{p}s}^2}{\omega_{\mathrm{p}e}^2} k_{\|}^2 \tilde{\rho}_s^2 \,\mathcal{P} \! \int_{-\infty}^{\infty} \mathrm{d} \tilde{v}_{\|s}\, \frac{\tilde{v}_{\|s} }{1- k_{\|}^2 \tilde{\rho}_s^2 \tilde{v}_{\|s}^2} \int_0^{\infty} \mathrm{d} \tilde{v}_{s\bot} \, \tilde{v}_{s\bot}^2 \Lambda_s(\tilde{v}_{\|s},\tilde{v}_{s\bot})
 . \label{conductivity_xx_identity_quasiparsimp}
\end{equation}
For the special case 
of a two-species plasma with isotropic electrons and anisotropic ions, the threshold condition of quasi-parallel oblique firehoses with $k_{\|} \rho_i \lesssim 0.5$ is given by
\begin{equation}
    \frac{1}{\sqrt{\upi}} \, \mathcal{P} \! \int_{-\infty}^{\infty} \mathrm{d} \tilde{v}_{\|i}\, \frac{\tilde{v}_{\|i} }{1- k_{\|}^2 \tilde{\rho}_i^2 \tilde{v}_{\|i}^2} \int_0^{\infty} \mathrm{d} \tilde{v}_{i\bot} \, \tilde{v}_{i\bot}^2 \Lambda_i(\tilde{v}_{\|i},\tilde{v}_{i\bot}) \approx \frac{1}{\beta_{\|i}} \, . \label{obliquefirethres_quasipar_append}
\end{equation}
This condition is used in \S\ref{sec:sims:res:satfirehose_obliquefirehose} of the main text.

\section{Numerical collisionality} \label{append:collisionality}

As with all hybrid-kinetic PIC simulations with a finite number of particles per cell, our {\tt Pegasus++} simulations are affected by random noise, which in turn gives rise to an effective numerical collisionality $\nu_{\rm num}$. In our simulations runs, we attempted to mitigate this affect by using a large number of particles per cell, which reduces the thermal noise and thereby $\nu_{\rm num}$. However, due to the large scale separation in some of our runs between the expansion time $\tau_{\rm exp}$ and the ion-Larmor period $2 \upi \Omega_i^{-1}$, we observed indirect evidence in some of runs that numerical collisionality could be playing a role: specifically, the initial evolution of the pressure anisotropy departing from that of a purely collisionless plasma. Here, we therefore characterise the box-averaged collisionality, demonstrate that it can account for the observed evolution of $\Delta_i$, and provide a simple estimate of its expected impact on key physical quantities.

\begin{figure}
  \centering
  \includegraphics[width=\linewidth]{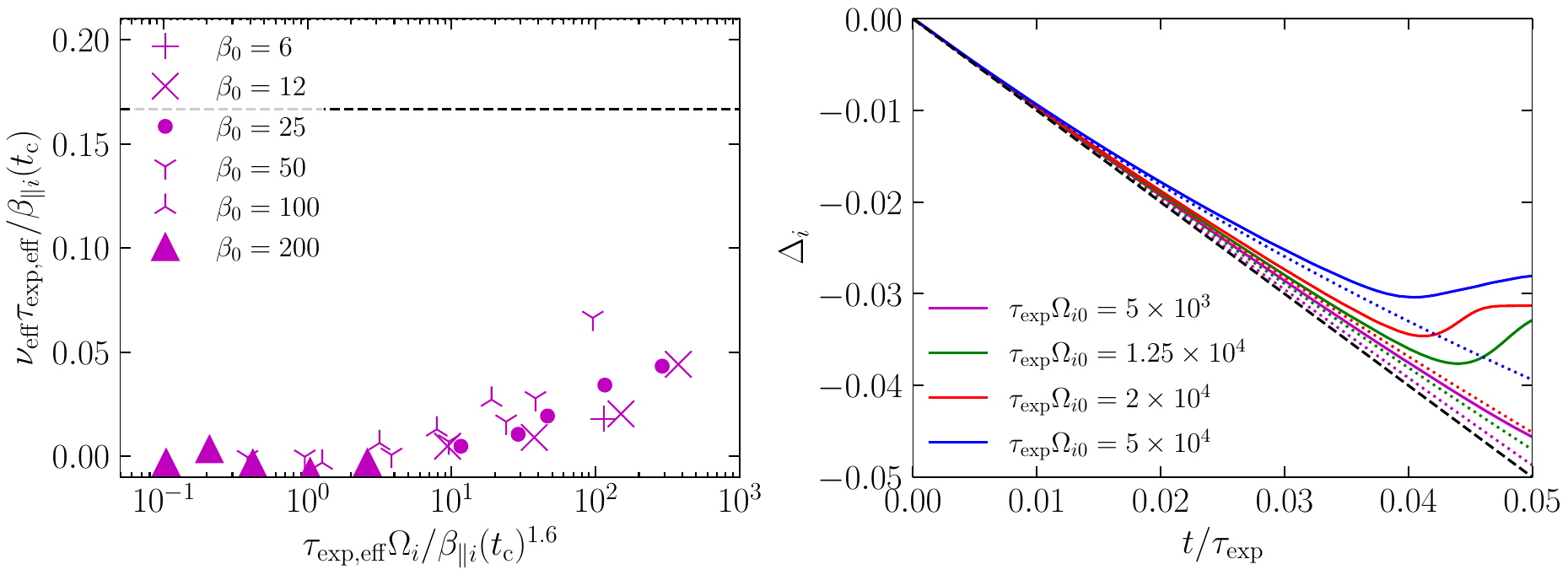}
  \caption{$(a)$ Values of the effective collisionality $\nu_{\rm eff}$ measured directly in all simulations at the time $t_{\rm c}$ at which the oblique firehose threshold is reached. The dashed line indicates the effective (time-averaged) value $\nu_{\rm eff} = \beta_{\|i}/6 \tau_{\rm exp,eff}$ of the collisionality predicted in asymptotic Alfv\'en-inhibiting states. $(b)$ Time evolution of the pressure anisotropy $\Delta_i$ for runs DIV, DV, DVI and DVIII ($\beta_{i0} = 50$). The dotted lines denotes the model (\ref{pressaniso_evolv_coll}) for the evolution of the pressure anisotropy in the presence of the numerical collisionalities for these runs given in panel $(a)$. The dashed black line shows the evolution of the pressure anisotropy in the absence of numerical collisionality.}
\label{fig:numcoll_boxav_allplots}
\end{figure}

To measure directly the numerical collisionality, we adopt the same approach used to measure the box-averaged effective collisionality that was employed in section \ref{sec:boxavcollisionality}, but now apply it at the time $t_{\rm c}$ at which the oblique firehose threshold is surpassed. We choose this specific time because measurements of $\nu_{\rm num}$ using this method will not be distorted by firehose-induced collisionality, but the thermal noise will be as similar as possible to that present during the growth and saturation of the firehoses. The results of this analysis for all of our simulations is shown in figure \ref{fig:numcoll_boxav_allplots}$(a)$. 
We find that for all of our simulations, $\nu_{\rm num} \tau_{\rm exp,eff}/\beta_{\|i} \lesssim 0.07$, decreasing significantly below this at smaller values of the parameter $\tau_{\rm exp,eff} \Omega_i/\beta_{\|i}^{1.6}$. Therefore, in all of our simulations, numerical collisionality should only have a small effect on the pressure anisotropy's evolution; further, $\nu_{\rm num} \lesssim 0.3 \nu_{\rm eff}$, with the implication that the effect of numerical collisionality on the induced-firehose collisionality should be a small correction as opposed to an order-unity effect. 

To characterise the effect of the numerical collisionality on the initial evolution of $\Delta_i$ more quantitatively, we construct a simple model based on the assumption that, prior to the emergence of firehose fluctuations, the only processes that can affect the pressure anisotropy are the expansion and numerical collisionality. Under this assumption, $\Delta_i$ evolves according to
\begin{equation}
    \frac{\mathrm{d} \Delta_i}{\mathrm{d}t} = \frac{\mathrm{d}}{\mathrm{d}t} \log{B} - 3 \nu_{\rm num} \Delta_i \simeq -\frac{1}{\tau_{\rm exp}} - 3 \nu_{\rm num} \Delta_i , 
\end{equation}
where the latter approximation follows whenever $t \ll \tau_{\rm exp}$. Solving for $\Delta_i$ with initial condition $\Delta_i(t = 0) = 0$, we find that
\begin{equation}
\Delta_i(t) = -\frac{1}{3 \tau_{\rm exp} \nu_{\rm num}} \left[1-\exp{\left(-3 \nu_{\rm num} t\right)}\right] \simeq -\frac{t}{\tau_{\rm exp}}\left(1-\frac{3}{2}\nu_{\rm num}t + \ldots\right) \, , \label{pressaniso_evolv_coll}
\end{equation}
where the final result is derived in the subsidiary limit $\nu_{\rm eff} t \ll 1$. We compare the actual evolution of $\Delta_i$ with the model (\ref{pressaniso_evolv_coll}) combined with the numerical collisionality in figure \ref{fig:numcoll_boxav_allplots}$(b)$; reasonable agreement is obtained, implying that the discrepancy of the evolution $\Delta_i$ from the collisionless prediction $\Delta_i = -t/\tau_{\rm exp}$ is most plausibly explained by numerical collisionality.

As for how we can estimate the effect of numerical collisionality of key quantities of interest in our simulations, we note that we can incorporate the effect of numerical collisionality into our interpretation of our results at a fixed value of $\Delta_i$ by revising our definition of the expansion time:
\begin{equation}
\tau_{\rm exp,num} \equiv \frac{\tau_{\rm exp}}{1+3 \tau_{\rm exp} \nu_{\rm num} \Delta_i} \, .
\end{equation}
In the saturated Alfv\'en-enabling states that we have simulated -- which generically have the largest values of $\nu_{\rm num} \tau_{\rm exp,eff}/\beta_{\|i}$ -- we use $\Delta_i \approx -1.6/\beta_{\|i}$ to estimate that $\tau_{\rm exp,num} \lesssim 1.3 \tau_{\rm exp}$. Thus, in short, numerical collisionality might be expected to decrease the effective collisionality associated with the firehoses by a small but finite factor, as well as somewhat suppress the observed values of $\delta B_{\rm f}^2/B_0^2$ that we observed in our Alfv\'en-enabling simulations.

\bibliographystyle{jpp}
\bibliography{main}

\end{document}